\documentclass[12pt]{article}
\pdfoutput=1
\usepackage[a4paper]{geometry}
\usepackage{jheppub, amsmath,amssymb,amsfonts,amsxtra, mathrsfs, makeidx,graphics,graphicx,amsthm,epsfig, ytableau,bm,longtable,float, color,tikz,mathtools,xfrac,footnote,rotating, lscape, makecell, environ,mathtools, empheq, colortbl, xcolor,enumerate}

\usetikzlibrary{positioning}
\usetikzlibrary{chains}
\usetikzlibrary{arrows, arrows.meta ,fit,decorations.pathreplacing}
\tikzstyle{every picture}+=[remember picture]
\tikzstyle{na} = [baseline]

\usetikzlibrary{arrows, decorations.markings, calc, fadings, decorations.pathreplacing, patterns, decorations.pathmorphing, positioning}
\tikzset{>={Latex[width=1.5mm,length=1.5mm]}}

\usepackage{mdframed}
\newenvironment{claim}{  \begin{mdframed}[linecolor=black!0,backgroundcolor=brown!25]\noindent\itshape\ignorespaces}{\end{mdframed}}

\def\node#1#2{\overset{#1}{\underset{#2}{{\color{gray} \bullet}}}}

\def\node#1#2{\overset{#1}{\underset{#2}{\circ}}}

\tikzstyle{every picture}+=[remember picture]
\tikzstyle{na} = [baseline=-.5ex]

\usepackage{bbm}
\usepackage{subfigure}

\newcommand{\eg}{\textit{e.g.}}

\newcommand{\ie}{\textit{i.e.}}

\numberwithin{equation}{section}

\newcommand{\nn}{\nonumber}

\newcommand{\be}{\begin{equation}} \newcommand{\ee}{\end{equation}}
\newcommand{\bea}{\begin{equation} \begin{aligned}} \newcommand{\eea}{\end{aligned} \end{equation}}

\def\tilde{\widetilde}

\def\hat{\widehat}

\def\rt2{\sqrt{2}}

\def\Tr{\mathop{\rm Tr}}
\def\tr{\mathop{\rm tr}}

\def\CI{{\cal I}}

\def\CN{{\cal N}}

\def\CT{{\cal T}}

\def\1{{\ds 1}}

\newcommand{\cI}{\mathcal{I}}

\newcommand{\fm}{\mathfrak{m}}

\def\repa{\raise4pt\hbox{$\square$}\mkern-14mu\raise-4pt\hbox{$\square$}}
\def\repab{\overline{\raise4pt\hbox{$\square$}\mkern-14mu\raise-4pt\hbox{$\square$}\mkern-1mu}}

\def\smileface{\ensuremath{\hbox{\large$\bigcirc$}\mkern-15mu\raise-1pt\hbox{\scriptsize$\smallsmile$}%
\mkern-10mu\raise4pt\hbox{..}\mkern4mu}}
\def\frownface{\ensuremath{\hbox{\large$\bigcirc$}\mkern-15mu\raise-1pt\hbox{\scriptsize$\smallfrown$}%
\mkern-10mu\raise4pt\hbox{..}\mkern4mu}}

\newcommand{\ba}{\begin{array}}
\newcommand{\ea}{\end{array}}
\newcommand{\bi}{\begin{itemize}}
\newcommand{\ei}{\end{itemize}}
\def\vec#1{\bm{#1}}
\def\bea#1\eea{\allowdisplaybreaks \begin{align}#1\end{align}}
 \newcommand{\ben}{\begin{enumerate}}
\newcommand{\een}{\end{enumerate}}
\newcommand{\bean}{\begin{eqnarray*}}
\newcommand{\eean}{\end{eqnarray*}}
\newcommand{\eref}[1]{(\ref{#1})}

\newcommand{\BR}{\mathbb{R}}

\newcommand{\BZ}{\mathbb{Z}}

\newcommand{\comment}[1]{}

\definecolor{light-gray}{gray}{0.7}

\newcommand{\blue}{\color{blue}}

\newcommand{\red}{\color{red}}

\def\aup#1 {\overset{#1}{\uparrow} \, \overset{\tilde{#1}}{\downarrow}}

\tikzset{snake it/.style={decorate, decoration={snake, amplitude=.4mm, segment length=2mm,
                       post length=0mm,pre length=0mm}}}
                       
\newcommand{\udl}[1]{\mathrm{d} #1 \,}

\newcommand{\xfac}[1]{\left( #1; x^2 \right)_\infty}

\def\gd{\delta}

\title{Duality walls in the 4d $\CN=2$ $SU(N)$ gauge theory with $2N$ flavours}
\author[a,b]{Ivan Garozzo,} 
\author[b,c]{Noppadol Mekareeya,}
\author[a,b]{and Matteo Sacchi}
\affiliation[a]{Dipartimento di Fisica, Universit\`a di Milano-Bicocca, \\ Piazza della indexenza 3, I-20126 Milano, Italy}
\affiliation[b]{INFN, sezione di Milano-Bicocca, \\Piazza della indexenza 3, I-20126 Milano, Italy}
\affiliation[c]{Department of Physics, Faculty of indexence, \\
Chulalongkorn University, Phayathai Road, \\
Pathumwan, Bangkok 10330, Thailand}
\emailAdd{ivangarozzo@gmail.com}
\emailAdd{n.mekareeya@gmail.com}
\emailAdd{m.sacchi13@campus.unimib.it}
\abstract{$S$-duality domain walls are extended objects in supersymmetric gauge theories with several rich physical properties. This paper focuses on 3d $\CN=2$ gauge theories associated with $S$-duality walls in the 4d $\CN=2$ $SU(N)$ gauge theory with $2N$ flavours. The theories associated with multiple duality walls are constructed by gluing together a basic building block, which is the theory associated with a single duality wall. We propose the prescription for gluing many copies of such a basic building block together as well as present the prescription for self-gluing. A number of dualities between such theories are discovered and studied using the supersymmetric index.  This work generalises the notion of the $S$-fold theory, which has been so far  studied extensively in the context of duality walls in the 4d super-Yang-Mills theory, to the theory with lower amounts of supersymmetry.}
\begin{document}
\maketitle

\section{Introduction}
$S$-duality domain walls are interesting extended objects in four dimensional supersymmetric gauge theories. Such a domain wall is an interface that allows the gauge coupling to vary in the vicinity of the wall from one constant value to another as one crosses the wall. The theories on opposite sides of the wall are related by $S$-duality in the sense that when $S$-duality is applied to one side of the wall, the gauge couplings of the two theories become equal.  The boundary condition can be chosen such that the duality wall is half-BPS, \ie~ preserving half of the amounts of supersymmetry of the four dimensional theory in question.  

The $S$-duality wall in the 4d $\CN=4$ super--Yang--Mills (SYM) with gauge group $SU(N)$ has been studied in detail in \cite{Gaiotto:2008sa, Gaiotto:2008sd, Gaiotto:2008ak}.  The theory associated with the wall is a 3d superconformal field theory with $\CN=4$ supersymmetry known as $T(SU(N))$.  The description of such a theory can be determined when the gauge coupling of the 4d SYM approaches zero.  Let us briefly summarise how this works.  Suppose that one takes the gauge coupling of the 4d theory on one side of the wall to be close to zero.  On this side of the wall the degrees of freedom decouple, and on the other side one has a Dirichlet boundary condition.  The latter can be realised in Type IIB string theory as $N$ D3 branes, each ending on an NS5 brane.  $S$-duality in 4d $\CN=4$ SYM can be viewed as inherited from $S$-duality in Type IIB string theory, which gives the dual configuration consisting of $N$ D3 branes, each ending on a D5 brane.  The 4d theory on the worldvolume of D3 branes in the latter configuration is weakly coupled, and one can decouple their degrees of freedom by ending the other side of each D3 brane on a D5 branes.  From this brane system, the quiver description can be read off as follows
\be
\begin{tikzpicture}[baseline]
\tikzstyle{every node}=[font=\footnotesize,minimum size=1cm]
\node[draw, circle] (1) at (0,0) {$1$};
\node[draw, circle] (2) at (2,0) {$2$};
\node[draw=none] (3) at (4,0) {$\cdots$};
\node[draw, circle] (4) at (6,0) {\tiny $N-1$};
\node[draw, rectangle] (5) at (8,0) {$N$};
\draw (1)--(2)--(3)--(4)--(5); 
\end{tikzpicture}
\ee 
where each circular node denotes an $U(n)$ gauge group, each line denotes bifundamental hypermultiplets and the square node denotes a flavour symmetry.

An interesting generalisation is to investigate 3d theories associated with duality walls in other 4d theories, possibly with lower amounts of supersymmetry.  One of the obvious candidates is the 4d $\CN=2$ $SU(N)$ gauge theory with $2N$ flavours.  This has an exactly marginal gauge coupling, with an interesting $S$-duality group being $SL(2,\BZ)$ for $N=2$ \cite{Seiberg:1994aj} and $\Gamma^0(2) \subset SL(2,\BZ)$ for $N \geq 3$ \cite{Argyres:1995wt, Hanany:1995na}. This theory can also be realised as a twisted compactification of 6d $(2,0)$ theory of type $A_{N-1}$ on a punctured Riemann surface \cite{Gaiotto:2009we}.  The 3d theory associated with the duality wall in this 4d theory can be determined by utilising the AGT correspondence \cite{Alday:2009aq, Wyllard:2009hg}, which relates the partition function of the 4d theory on the squashed four-sphere to an observable in the Liouville or Toda theory on the Riemann surface \cite{Hama:2012bg, Nosaka:2013cpa}.  As pointed out in \cite{Drukker:2010jp}, the partition function of the 3d theory associated with $S$-duality wall placed along the squashed three-sphere, which is the equator of the aformentioned four-sphere, corresponds to a collection of the duality transformation coefficients of conformal blocks of the Liouville or Toda theory.  From such a partition function, one may extract the gauge group and matter content of the 3d theory in question \cite{Hosomichi:2010vh, Teschner:2012em, Floch:2015hwo}. In fact, this technique has been successfully applied to determine the 3d theory associated with the $S$-duality wall in the 4d $\CN=2^*$ gauge theory \cite{Hosomichi:2010vh}.  For the 4d $\CN=2$ $SU(N)$ gauge theory with $2N$ flavours, this method was applied by the authors of \cite{Teschner:2012em, Floch:2015hwo} (see also \cite{Gang:2012ff} for the superconformal index).  In \cite{Floch:2015hwo}, the theory associated with the duality wall was then identified as the 3d $\CN=2$ $U(N-1)$ gauge theory with $2N$ flavours, with the $R$-charges of the chiral fields fixed to certain values.  Later, it was pointed out by the authors of \cite{Benini:2017dud} that the superpotential of such a theory should be $W= V_+ + V_-$, where $V_\pm$ are the basic monopole operators of the $U(N-1)$ gauge group. For convenience, following \cite{Benini:2017dud}, we refer to this 3d theory as $\CT_{\mathfrak{M}}$ and it will be discussed in more detail in section \ref{sec:3dtheorywall} of this paper.  It should be remarked that this approach that is used to identify the 3d theory is different from that used by Gaiotto and Witten \cite{Gaiotto:2008ak}, mentioned in the previous paragraph.  Although the Type IIA brane configuration of the 4d theory \cite{Witten:1997sc} and the Type IIB brane configuration of the 3d theory with the monopole superpotential \cite{Amariti:2015yea,Amariti:2015mva,Amariti:2016kat,Amariti:2017gsm, Amariti:2019rhc} are known, to the best of our knowledge, it is not clear how to identify the latter as the theory associated with the duality wall in the former.

In this paper, we consider the 3d theory associated with one and higher number of duality walls.  We use $\CT_{\fm}$ as a basic building block to construct the other theories.  Such a construction involves gluing the basic building block to each other in various possible ways, described in section \ref{sec:glue}.  We, in fact, adopt the gluing prescription from \cite{Kim:2017toz,Kim:2018lfo,Pasquetti:2019hxf}, and discuss the motivation in doing so in section \ref{sec:motivationglue}.  The number of duality walls in question is equal to the number of basic building block involved in the gluing procedure.  Moreover, we discuss the procedure of self-gluing, where by some or all of the flavour symmetries are commonly gauged.  Along the way, we find several of interesting dualities relating a number of theories we construct.  Supersymmetric index is used as a main tool to study the operators at the superconformal fixed point, as well as to check the duality proposed in this paper.

Another important motivation of this paper is to provide a generalisation of the 3d $S$-fold theory, previously studied in \cite{Terashima:2011qi, Gang:2015wya, Gang:2018wek, Gang:2018huc} and \cite{Assel:2018vtq, Garozzo:2018kra, Garozzo:2019hbf, Garozzo:2019ejm}, to the set-up with lower amounts of supersymmetry.  The $S$-fold theory is a  3d superconformal field theory associated with duality walls, each of which gives rise to a local $SL(2,\BZ)$ transformation, in the 4d $\CN=4$ super--Yang--Mills theory.  In the field theory description, the building block of the $S$-fold theory is $T(U(N))$, which is the aforementioned $T(SU(N))$ theory along with the mixed CS coupling between two $U(1)$ symmetries that gives rise to a $U(N) \times U(N)$ global symmetry for the $T(U(N))$ theory.   Such basic building blocks can then be glued together (or self-glued) to obtain new 3d superconformal theories.  As discussed extensively in the above references, $S$-fold theories have many interesting features.  For example, from the naive field theory description, the $S$-fold theory generically has an $\CN=3$ supersymmetry; however, in many cases, a more detailed study reveals that supersymmetry can be enhanced to $\CN=4$ or even to $\CN=5$.  In this paper, we study the analog of the $S$-fold theories in the context of 4d $\CN=2$ $SU(N)$ gauge theory with $2N$ flavours.  Although we do not find any supersymmetry enhancement in this paper, the theories we discuss still have many interesting properties.

The paper is organised as follows.  In section \ref{sec:3dtheorywall}, we introduce the $\CT_{\fm}$ theory as the basic building block that will be used to construct the other theories.  We mention how to couple the 4d fields to $\CT_{\fm}$ as well as examine various duality frames.  In section \ref{sec:glue}, we present the prescription for gluing many copies of the basic building blocks together as well as propose the prescription for self-gluing.  The concept of the ``skeleton diagram'', which is the analog of the Riemann surface with punctures (used extensively in \cite{Kim:2017toz} to construct a large class of theories) and gives rise to a geometric interpretation of the gluing, is introduced in sections \ref{sec:3dtheorywall} and \ref{sec:glue}. In section \ref{sec:singlewall}, we discuss two classes of theories associated with a single wall, whose skeleton diagram contains (1) two external legs and genus one and (2) zero external leg and genus two.  The quadralities between such theories are discussed.  In sections \ref{sec:twowallsclassic} and \ref{sec:twowallsalternative}, theories associated with two duality walls, using two different types of the basic building block, are constructed and discussed.  We finally conclude the paper in section \ref{sec:conclude}.  In appendix \ref{app:3dindex}, we briefly summarise some generalities about 3d supersymmetric index.

\section{The 3d gauge theory with a monopole superpotential} \label{sec:3dtheorywall}
The theory associated with the $S$-duality wall of the 4d $\CN=2$ $SU(N)$ gauge theory with $2N$ flavours is the 3d $\CN=2$ $U(N-1)$ gauge theory with $2N$ flavours and superpotential $W = V_+ + V_-$, where $V_\pm$ are the basic monopole operators of the latter theory \cite{Benini:2017dud}.  For the sake of brevity, following \cite{Benini:2017dud}, we refer to the aforementioned 3d theory as $\CT_{\mathfrak{M}}$, where $\mathfrak{M}$ stands for the monopole superpotential.  The identification of the theory on the $S$-duality wall of the 4d theory and the $\CT_{\mathfrak{M}}$ theory\footnote{On the other hand, the 3d theory associated with the $S$-duality wall of the 4d $\CN=2^*$ $SU(N)$ gauge theory has been identified as the axial mass-deformed $T(SU(N))$ gauge theory by \cite{Hosomichi:2010vh}.} had been attempted by several authors, \eg~ \cite{Teschner:2012em, Gang:2012ff, Floch:2015hwo}.  The main technique was to study a collection of the duality transformation coefficients of conformal blocks (also known as the kernel) of the Liouville or Toda theory, which are in the AGT correspondence \cite{Alday:2009aq, Wyllard:2009hg} with the 4d theory.  The kernel was then interpreted as the partition function of the 3d theory associated with the duality wall \cite{Drukker:2010jp}.  Knowing the former allows one to identify the matter content of the 3d theory associated with the duality wall \cite{Teschner:2012em, Floch:2015hwo}.  In \cite{Floch:2015hwo} it was observed that the $R$-charges of the chiral fields in the 3d theory were fixed to certain particular values.  This was later interpreted in \cite{Benini:2017dud} as due to the monopole superpotential.

The $\CT_{\mathfrak{M}}$ theory has a global symmetry $SU(2N) \times SU(2N)$.  We represent this theory by the following quiver diagram:
\be \label{UNm1w2N}
\CT_{\mathfrak{M}}: \qquad
\begin{tikzpicture}[baseline]
\tikzstyle{every node}=[font=\footnotesize]
\node[draw, circle, fill=yellow] (node1) at (0,0) {$N-1$};
\node[draw, rectangle] (sqnode1) at (2,0) {$2N$};
\node[draw, rectangle] (sqnode2) at (-2,0) {$2N$};
\draw[draw=black,solid,->]  (sqnode2) to (node1) ; 
\draw[draw=black,solid,->]  (node1) to (sqnode1) ; 
\end{tikzpicture}
\ee
where we denoted the gauge node in yellow in order to indicate the monopole superpotential $W=V_+ + V_-$.  Due to the monopole superpotential, the topological and axial symmetries are broken, and the $R$-charge $r$ of the chiral fields is fixed to be $r=1/2$ due to the relation $2N(1-r)-(N-1-1)=2$.

In fact, as pointed out in \cite{Benini:2017dud}\footnote{More precisely, in \cite{Benini:2017dud}, a more general duality relating the $U(N_c)$ gauge theory with $N_f$ flavors and $W= V_+ + V_-$ and the $U(N_c-N_f)$ gauge theory with $N_f$ flavors,  $N_f^2$ singlets $M$ and $W= V_+ + V_-+Mq\tilde{q}$ was proposed.}, theory \eref{UNm1w2N} is dual to another theory with the same gauge group $U(N-1)$, also with $2N$ flavours and $4N^2$ singlets $M$
\be \label{dualUNm1w2N}
\begin{tikzpicture}[baseline]
\tikzstyle{every node}=[font=\footnotesize]
\node[draw, circle, fill=yellow] (node1) at (0,0) {$N-1$};
\node[draw, rectangle] (sqnode1) at (2,0) {$2N$};
\node[draw, rectangle] (sqnode2) at (-2,0) {$2N$};
\draw[draw=black,solid,<-]  (sqnode2) to node[midway,above]{$q$} (node1) ; 
\draw[draw=black,solid,<-]  (node1) to node[midway,above]{$\tilde{q}$} (sqnode1) ; 
\draw[draw=black,solid,<-]  (sqnode1) to [bend left=60] node[midway,below]{$M$} (sqnode2) ; 
\end{tikzpicture}
\ee
and superpotential $W = V_+ + V_- + M q \tilde{q}$.  In other words, we have duality
\be \label{mainduality}
\eref{UNm1w2N} ~\longleftrightarrow~ \eref{dualUNm1w2N}
\ee

\subsection{Indices of theories \eref{UNm1w2N} and \eref{dualUNm1w2N}}
Our main tool to study the theories in this paper is the supersymmetric index, which we shall refer to as {\it index} for the sake of brevity. It can be computed as the partition function on $\mathbb{S}^2\times\mathbb{S}^1$. We summarise the necessary detail in appendix \ref{app:3dindex}.

In order to write the supersymmetric index of a theory with monopole superpotential one has to take into account suitable contributions of BF couplings with the global symmetries and the $R$-symmetry that make the monopole operators uncharged and exactly marginal. In the case of theory \eref{UNm1w2N} we are considering the monopole superpotential breaks the topological as well as the axial symmetries. Hence, the index can be easily obtained from that of the $U(N-1)$ gauge theory with $2N$ flavours and zero superpotential, turning off the fugacities for the axial and the topological symmetries, as well as setting the $R$-charge of the chiral fields to $r=\frac{1}{2}$
\be
\begin{split}
&\mathcal{I}_{\eref{UNm1w2N}}(x; \{\bm{\mu},\bm{n}\},\{\bm{\tau}, \bm{p}\}) \\
&=\sum_{\bm{m}\in\mathbb{Z}^{N-1}}\frac{1}{(N-1)!}\oint\prod_{a=1}^{N-1}\frac{\udl{u_a}}{2\pi i\,u_a}Z_{vec}
(x; \{\bm{u},\bm{m}\})
Z_{chir}
(x; \{\bm{u},\bm{m}\},\{\bm{\mu},\bm{n}\},\{\bm{\tau}, \bm{p}\})\,,
\end{split}
\ee
where the contribution of the $\mathcal{N}=2$ vector multiplet is
\be
Z_{\text{vec}}(x; \{\bm{u},\bm{m}\})=\prod_{a,b=1}^{N-1}x^{-|m_a-m_b|}\left(1-(-1)^{m_a-m_b}x^{|m_a-m_b|}\left(\frac{z_a}{z_b}\right)^{\pm1}\right)\,,
\ee
while that of the chiral multiplets is
\be
\begin{split}
&Z_{\text{chir}}(x; \{\bm{u},\bm{m}\},\{\bm{\mu},\bm{n}\},\{\bm{\tau}, \bm{p}\})\\
&=\prod_{a=1}^{N-1}\prod_{i=1}^{2N}\left(u_a\mu_i^{-1}x^{1/2}\right)^{\frac{|n_i-m_a|}{2}}\frac{\xfac{(-1)^{n_i-m_a}u_a\mu_i^{-1}x^{3/2+|n_i-m_a|}}}{\xfac{(-1)^{n_i-m_a}u_a^{-1}\mu_i x^{1/2+|n_i-m_a|}}}\times\nn\\
&\times \left(u_a^{-1}\tau_ix^{1/2}\right)^{\frac{|m_a-p_i|}{2}}\frac{\xfac{(-1)^{m_a-p_i}u_a^{-1}\mu_ix^{3/2+|m_a-p_i|}}}{\xfac{(-1)^{m_a-p_i}u_a\mu_i^{-1} x^{1/2+|m_a-p_i|}}}
\end{split}
\ee
In the above expressions we denoted by $\{\bm{u},\bm{m}\}$ the fugacities and the magnetic fluxes respectively for the gauge symmetry and with $\{\bm{\mu},\bm{n}\},\{\bm{\tau}, \bm{p}\}$ those of the two $SU(2N)$ global symmetries, which have to satisfy the constraints $\prod_{i=1}^{2N}\mu_i=\prod_{i=1}^{2N}\tau_i=1$ and $\sum_{i=1}^{2N}n_i=\sum_{i=1}^{2N}p_i=0$.

The index of the dual theory \eqref{dualUNm1w2N} is related to that of \eref{UNm1w2N} by the following relation:
\begin{eqnarray}\label{indexmainduality}
\mathcal{I}_{\eqref{dualUNm1w2N} }(x;\{\bm{\mu},\bm{n}\},\{\bm{\tau}, \bm{p}\})&=&\prod_{i,j=1}^{2N}\left(\mu_i\tau_j^{-1}\right)^{-\frac{|n_i-p_j|}{2}}\frac{\xfac{(-1)^{n_i-p_j}\mu_i^{-1}\tau_jx^{1+|n_i-p_j|}}}{\xfac{(-1)^{n_i-p_j}\mu_i\tau_j^{-1}x^{1+|n_i-p_j|}}}\times\nn\\
&\times&\mathcal{I}_{\eref{UNm1w2N}}(x; \{\bm{\mu}^{-1},-\bm{n}\},\{\bm{\tau}^{-1}, -\bm{p}\})\, ,
\end{eqnarray}
where the right hand side of the first line is the contribution of the $4N^2$ gauge singlets $M$.  We point out that an analogous identity for the partition functions on $\mathbb{S}^3_b$ was actually derived in \cite{Benini:2017dud} as a limit of the identity for the $4d$ supersymmetric indices associated to Intriligator--Pouliot duality \cite{Intriligator:1995ne}, where the latter was proven in \cite{2014arXiv1408.0305R}. Although we shall not provide an analytic proof\footnote{Relation \eref{indexmainduality} could, in principle, be derived in a similar way to the one for the $\mathbb{S}^3_b$ partition functions if a generalization of Rains' results for the lens space index \cite{Benini:2011nc}, which is the partition function on $\mathbb{S}^3/\mathbb{Z}_p\times\mathbb{S}^1$, were known.} of the relation \eqref{indexmainduality} in this paper, it can be checked perturbatively by expanding both sides as power series in $x$ and matching each order of the power expansion. Moreover, as a further support of \eref{indexmainduality}, one may take an appropriate 2d limit of the index of each side in \eref{indexmainduality} to obtain certain complex integrals \cite{Pasquetti:2019uop}, which are related to CFT free field correlators; the equality of such integrals was proposed in \cite{Baseilhac:1998eq,Fateev:2007qn}.

\subsection{Inclusion of the 4d fields}
As a theory realised on the wall, one of the $SU(2N)$ symmetries (say, the one associated with the left square node) can be decomposed into a subgroup $SU(N) \times SU(N) \times U(1)$, where we shall refer to the latter $U(1)$ as $U(1)_q$.   Each of these $SU(N)$ can then be coupled to the $SU(N)$ gauge symmetry of the 4d theory on each side of the wall.  Moreover, the 3d chiral fields of the theory on the wall also couple non-trivially to the chiral fields coming from the 4d theory.  The appropriate quiver description for the 3d $\CN=2$ theory on the wall is
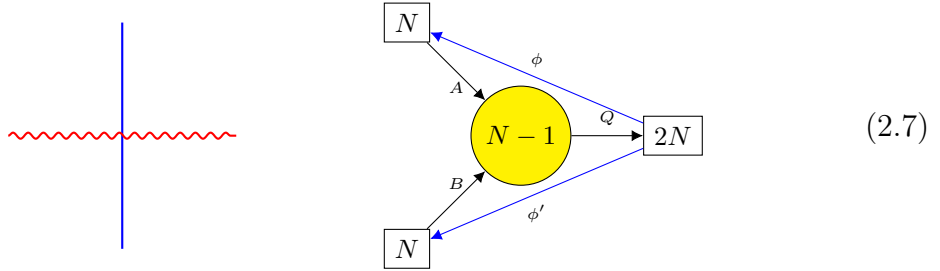
\be \label{basicblock}
\begin{tikzpicture}[baseline]
\tikzstyle{every node}=[font=\footnotesize, node distance=0.45cm]
\tikzset{decoration={snake,amplitude=.4mm,segment length=2mm,
                       post length=0mm,pre length=0mm}}
\draw[blue,thick] (0,-1.5)--(0,1.5) node[midway, right] {};  
\draw[decorate,red,thick] (-1.5,0) -- (1.5,0) node[right] {}; 
\end{tikzpicture}
\qquad \qquad 
\begin{tikzpicture}[baseline]
\tikzstyle{every node}=[font=\footnotesize]
\node[draw, circle, fill=yellow] (node1) at (0,0) {$N-1$};
\node[draw, rectangle] (sqnode1) at (2,0) {$2N$};
\node[draw, rectangle] (sqnode2) at (-1.5,1.5) {$N$};
\node[draw, rectangle] (sqnode3) at (-1.5,-1.5) {$N$};
\draw[draw=black,solid,->]  (sqnode2) to node[midway,below]{{\tiny $A$}} (node1) ; 
\draw[draw=black,solid,->]  (sqnode3) to node[midway,above]{{\tiny $B$}} (node1) ; 
\draw[draw=black,solid,->]  (node1) to node[midway,above]{{\tiny $Q$}} (sqnode1) ; 
\draw[draw=blue,solid,<-]  (sqnode2) to node[midway,above]{{\tiny $\phi$}} (sqnode1) ; 
\draw[draw=blue,solid,<-]  (sqnode3) to node[midway,below]{{\tiny $\phi'$}} (sqnode1) ; 
\end{tikzpicture}
\ee
where $\phi$ is one of the chiral fields contained in the hypermultiplets of the 4d $\CN=2$ $SU(N)$ gauge theory with $2N$ flavours on one side of the wall restricted to the interface.  The same is for $\phi'$ on the other side of the wall. The superpotential of \eref{basicblock} is
\be
W_{\eref{basicblock}} = V_+ + V_- + Q \phi A+ Q \phi' B~.
\ee
We shall, from now on, denote as blue arrows the chiral fields coming from the 4d theory.

The arrows in the right diagram are consistent with the decomposition rule of the fundamental representation of $SU(2N)$ to $SU(N) \times SU(N) \times U(1)_q$:
\be \label{decomp}
[1,0^{2N-2}]~\longrightarrow~q  [1,0^{N-2}; 0^{N-1}] + q^{-1} [0^{N-1}; 1, 0^{N-2}]~,
\ee
which correspond to chiral fields $A$ and $B$ respectively.  Note that $Q$ carries zero charge under $U(1)_q$, and so from the superpotential, $\phi$ and $\phi'$ carry $U(1)_q$ charges $-1$ and $+1$ respectively.

Let us now explain the ``skeleton'' diagram on the left of \eref{basicblock}. Each blue external leg (or each end of the blue line) denotes an $SU(N)$ global symmetry, and the wiggly red line denotes a duality wall, which brings about an $SU(2N) \times U(1)_q$ global symmetry.  Note that the latter is the symmetry of the 4d $\CN=2$ $SU(N)$ gauge theory with $2N$ flavours, where $U(1)_q$ plays a role as the baryonic symmetry.

One may, in fact, apply the duality \eref{mainduality} to the yellow node in \eref{basicblock}.  As a result, $\phi$ and $\phi'$ disappear, and the arrows of $A$, $B$ and $Q$ are reversed.  We denote the chiral fields in the dual theory as $\tilde{A}$, $\tilde{B}$ and $\tilde{Q}$; they carry opposite $U(1)_q$ charges with respect to $A$, $B$ and $Q$ respectively.  The dual theory is therefore
\be \label{dualbasicblock}
\begin{tikzpicture}[baseline]
\tikzstyle{every node}=[font=\footnotesize]
\node[draw, circle, fill=yellow] (node1) at (0,0) {$N-1$};
\node[draw, rectangle] (sqnode1) at (2,0) {$2N$};
\node[draw, rectangle] (sqnode2) at (-1.5,1.5) {$N$};
\node[draw, rectangle] (sqnode3) at (-1.5,-1.5) {$N$};
\draw[draw=black,solid,<-]  (sqnode2) to node[midway,below]{{\tiny $\tilde{A}$}} (node1) ; 
\draw[draw=black,solid,<-]  (sqnode3) to node[midway,above]{{\tiny $\tilde{B}$}} (node1) ; 
\draw[draw=black,solid,<-]  (node1) to node[midway,above]{{\tiny $\tilde{Q}$}} (sqnode1) ; 
\end{tikzpicture}
\ee
with the superpotential 
\be
W_{\eref{dualbasicblock}} = V_+ + V_- ~.
\ee

Theory \eref{basicblock} will be used as as a basic building block to construct other theories in the subsequent part of the paper.  For the sake of readability, we shall suppress the number $N-1$ in the yellow node from now on.

\subsection{Another representation of \eref{basicblock}}
There is another {\it equivalent way} to represent theory \eref{basicblock}.  We further decompose the $SU(2N)$ flavour node in quiver \eref{basicblock} into $SU(N) \times SU(N) \times U(1)_p$.  The resulting quiver is
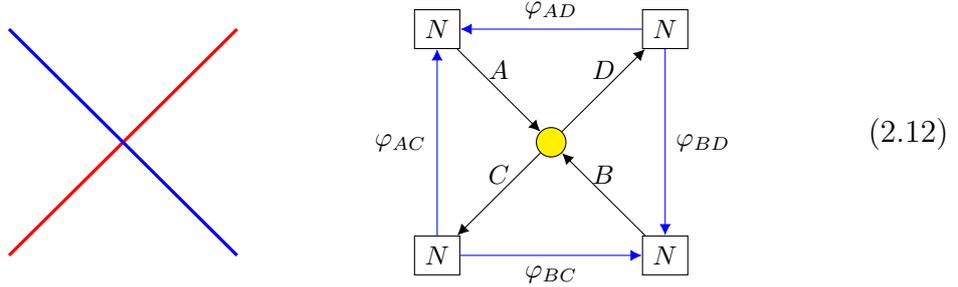
\be \label{basicblock1}
\begin{tikzpicture}[baseline]
\draw[draw=red,solid, very thick,-] (-1.5,-1.5) to (1.5,1.5);
\draw[draw=blue,solid, very thick,-] (-1.5,1.5) to (1.5,-1.5);
\end{tikzpicture}
\qquad \qquad
\begin{tikzpicture}[baseline]
\tikzstyle{every node}=[font=\footnotesize]
\node[draw, circle, fill=yellow] (node1) at (0,0) {}; 
\node[draw, rectangle] (sqnode1) at (1.5,1.5) {$N$};
\node[draw, rectangle] (sqnode4) at (1.5,-1.5) {$N$};
\node[draw, rectangle] (sqnode2) at (-1.5,1.5) {$N$};
\node[draw, rectangle] (sqnode3) at (-1.5,-1.5) {$N$};
\draw[draw=black,solid,->]  (sqnode2) to  node[midway,above]{$A$} (node1) ; 
\draw[draw=black,solid,<-]  (sqnode3) to node[midway,above]{$C$} (node1) ; 
\draw[draw=black,solid,->]  (node1) to node[midway,above]{$D$} (sqnode1) ;
\draw[draw=black,solid,<-]  (node1) to node[midway,above]{$B$} (sqnode4);  
\draw[draw=blue,solid,<-]  (sqnode2) to node[midway,above]{$\varphi_{AD}$}  (sqnode1) ; 
\draw[draw=blue,solid,<-]  (sqnode2) to node[midway,left]{$\varphi_{AC}$} (sqnode3) ; 
\draw[draw=blue,solid,->]  (sqnode3) to node[midway,below]{$\varphi_{BC}$} (sqnode4) ;
\draw[draw=blue,solid,<-]  (sqnode4) to node[midway,right]{$\varphi_{BD}$} (sqnode1) ; 
\end{tikzpicture}
\ee
Here $C$ and $D$ are the chiral fields that come from the decomposition of $Q$ in \eref{basicblock}, and $\varphi_{AD}$, $\varphi_{AC}$, $\varphi_{BC}$ and $\varphi_{BD}$ are the fields that come from the 4d theory.  The superpotential is
\be
W_{\eref{basicblock1}} = V_+ +V_- + A \varphi_{AD} D + A \varphi_{AC} C+ B \varphi_{BC} C +  B \varphi_{BD} D~. 
\ee
The $U(1)_p$ and $U(1)_q$ charge assignment is depicted as follows. 
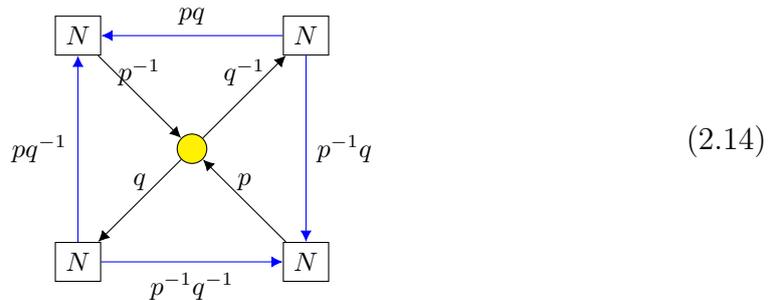
\be \label{basicblock1charges}
\begin{tikzpicture}[baseline]
\tikzstyle{every node}=[font=\footnotesize]
\node[draw, circle, fill=yellow] (node1) at (0,0) {}; 
\node[draw, rectangle] (sqnode1) at (1.5,1.5) {$N$};
\node[draw, rectangle] (sqnode4) at (1.5,-1.5) {$N$};
\node[draw, rectangle] (sqnode2) at (-1.5,1.5) {$N$};
\node[draw, rectangle] (sqnode3) at (-1.5,-1.5) {$N$};
\draw[draw=black,solid,->]  (sqnode2) to  node[midway,above]{$p^{-1}$} (node1) ; 
\draw[draw=black,solid,<-]  (sqnode3) to node[midway,above]{$q$} (node1) ; 
\draw[draw=black,solid,->]  (node1) to node[midway,above]{$q^{-1}$} (sqnode1) ;
\draw[draw=black,solid,<-]  (node1) to node[midway,above]{$p$} (sqnode4);  
\draw[draw=blue,solid,<-]  (sqnode2) to node[midway,above]{$p q$}  (sqnode1) ; 
\draw[draw=blue,solid,<-]  (sqnode2) to node[midway,left]{$p q^{-1}$} (sqnode3) ; 
\draw[draw=blue,solid,->]  (sqnode3) to node[midway,below]{$p^{-1} q^{-1}$} (sqnode4) ;
\draw[draw=blue,solid,<-]  (sqnode4) to node[midway,right]{$p^{-1}q$} (sqnode1) ; 
\end{tikzpicture}
\ee

We use the ``skeleton'' diagram on the left of \eref{basicblock1} to represent such a building block. Each red and blue external leg (or each end of the red and blue lines) corresponds to a flavour symmetry $SU(N)$.  The red colour indicates that the two $SU(N)$ symmetries come from the group decomposition $SU(2N)$ due to the duality wall.  The blue colour is the same as that used in \eref{basicblock}.  Observe the directions of the arrows of the chiral fields $A$, $B$, $C$, $D$ that are transformed under each $SU(N)$ flavour symmetry associated with each external legs: it is ingoing for blue and outcoming for red.

Similar to the discussion around \eref{dualbasicblock}, we may get rid of the 4d chiral fields $\varphi_{AD}$, $\varphi_{AC}$, $\varphi_{BC}$ and $\varphi_{BD}$ using the duality \eref{mainduality}.  This results in
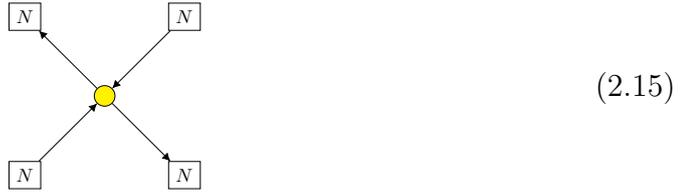
\be
\scalebox{0.7}{
\begin{tikzpicture}[baseline]
\tikzstyle{every node}=[font=\footnotesize]
\node[draw, circle, fill=yellow] (node1) at (0,0) {}; 
\node[draw, rectangle] (sqnode1) at (1.5,1.5) {$N$};
\node[draw, rectangle] (sqnode4) at (1.5,-1.5) {$N$};
\node[draw, rectangle] (sqnode2) at (-1.5,1.5) {$N$};
\node[draw, rectangle] (sqnode3) at (-1.5,-1.5) {$N$};
\draw[draw=black,solid,<-]  (sqnode2) to  node[midway,above]{} (node1) ; 
\draw[draw=black,solid,->]  (sqnode3) to node[midway,above]{} (node1) ; 
\draw[draw=black,solid,<-]  (node1) to node[midway,above]{} (sqnode1) ;
\draw[draw=black,solid,->]  (node1) to node[midway,above]{} (sqnode4);  
\end{tikzpicture}}
\ee
with the monopole superpotential $W= V_+ + V_-$.

\section{Gluing basic building blocks} \label{sec:glue}
Having discussed the basic building block, we now consider construction involving multiple duality walls.  The corresponding 3d theory can be obtained by gluing together the same number of basic building blocks in certain ways along the 4d fields (denoted by blue arrows in the quiver).  In the following, we discuss the prescription for the gluing in detail.  In fact, such a prescription is heavily motivated by that adopted in \cite{Kim:2017toz,Kim:2018lfo,Pasquetti:2019hxf} in the context of compactifications of 6d theories on a Riemann surface with fluxes for the global symmetries. We discuss the motivation and the similarity of our set-up and that of \cite{Kim:2017toz} in the last subsection of this section.


\subsection{Using basic building block \eqref{basicblock}}
We start by considering the cases in which we glue a number of copies of the basic block \eqref{basicblock}.  This corresponds to set-up involving the same number of duality walls. 

\subsubsection{Prescription}

Let us consider two copies of the basic building blocks \eref{basicblock}.  For the first copy, we assign the $U(1)_q \times SU(2N)$ fugacities $a_i = q\, u_i$ to $\phi$ (and hence $a'_i = q^{-1}\, u_i$ to $\phi'$), where $i=1, 2, \ldots, N$ and $u_i$ are the parameters that have to satisfy $\prod_{i=1}^{2N}u_i=1$ being $SU(2N)$ fugacities.  For the second copy, let us call the 4d fields $\tilde{\phi}$ and $\tilde{\phi}'$ and assign the $U(1)_{\tilde{q}} \times SU(2N)$ fugacities $\tilde{a}_i = \tilde{q} \, \tilde{u}_i$ to $\tilde{\phi}$ (and hence $\tilde{a}'_i = \tilde{q}^{-1}\, \tilde{u}_i$ to $\tilde{\phi}'$), again with the constraint $\prod_{i=1}^{2N}\tilde{u}_i=1$.



The prescription is that two building blocks can be glued along $\phi$ and $\tilde{\phi}'$ (or along $\phi'$ and $\tilde{\phi}$) if and only if one of the following conditions is satisfied:
\begin{eqnarray} \label{condglue}
&&\Phi\text{-gluing:}\qquad a_i=\tilde{a}'_i,\nn\\
&&S\text{-gluing:}\qquad a_i=\frac{1}{\tilde{a}'_i},\qquad\qquad\forall i=1,\cdots,2N\, .
\label{fugacitymap}
\end{eqnarray}
Let us illustrate this using explicit examples.  We can perform an $S$-gluing, but not a $\Phi$-gluing, for these two models along $\phi$ and $\tilde{\phi}'$ (or along $\phi'$ and $\tilde{\phi}$):
\be
\scalebox{0.8}{
\begin{tikzpicture}[baseline]
\tikzstyle{every node}=[font=\footnotesize]
\node[draw, circle, fill=yellow] (node1) at (0,0) {};
\node[draw, rectangle] (sqnode1) at (2,0) {$2N$};
\node[draw, rectangle] (sqnode2) at (-1.5,1.5) {$N$};
\node[draw, rectangle] (sqnode3) at (-1.5,-1.5) {$N$};
\draw[draw=black,solid,->]  (sqnode2) to node[midway,left]{\red $q^{-1}$} (node1) ; 
\draw[draw=black,solid,->]  (sqnode3) to node[midway,left]{\red $q$} (node1) ; 
\draw[draw=black,solid,->]  (node1) to node[midway,above]{} (sqnode1) ; 
\draw[draw=blue,solid,<-]  (sqnode2) to node[near end,above]{{$\phi$}}  node[midway,above]{{\red $q u_i$}} (sqnode1) ; 
\draw[draw=blue,solid,<-]  (sqnode3) to node[near end,below]{{$\phi'$}}  node[midway,below]{{\red $q^{-1} u_i$}} (sqnode1) ; 
\end{tikzpicture}}
\qquad \qquad
\scalebox{0.8}{
\begin{tikzpicture}[baseline]
\tikzstyle{every node}=[font=\footnotesize]
\node[draw, circle, fill=yellow] (node1) at (0,0) {};
\node[draw, rectangle] (sqnode1) at (2,0) {$2N$};
\node[draw, rectangle] (sqnode2) at (-1.5,1.5) {$N$};
\node[draw, rectangle] (sqnode3) at (-1.5,-1.5) {$N$};
\draw[draw=black,solid,<-]  (sqnode2) to node[midway,left]{\red $q$} (node1) ; 
\draw[draw=black,solid,<-]  (sqnode3) to node[midway,left]{\red $q^{-1}$} (node1) ; 
\draw[draw=black,solid,<-]  (node1) to node[midway,above]{} (sqnode1) ; 
\draw[draw=blue,solid,->]  (sqnode2) to node[near end,above]{{$\tilde{\phi}'$}}  node[midway,above]{{\red $q^{-1} u^{-1}_i$}} (sqnode1) ; 
\draw[draw=blue,solid,->]  (sqnode3) to node[near end,below]{{$\tilde{\phi}$}} node[midway,below]{{\red $q u^{-1}_i$}} (sqnode1) ; 
\end{tikzpicture}}
\ee
On the other hand, it is possible to perform a $\Phi$-gluing, but not an $S$-gluing, for these two model along $\phi$ and $\tilde{\phi}'$ (or along $\phi'$ and $\tilde{\phi}$):
\be
\scalebox{0.8}{
\begin{tikzpicture}[baseline]
\tikzstyle{every node}=[font=\footnotesize]
\node[draw, circle, fill=yellow] (node1) at (0,0) {};
\node[draw, rectangle] (sqnode1) at (2,0) {$2N$};
\node[draw, rectangle] (sqnode2) at (-1.5,1.5) {$N$};
\node[draw, rectangle] (sqnode3) at (-1.5,-1.5) {$N$};
\draw[draw=black,solid,->]  (sqnode2) to node[midway,left]{\red $q^{-1}$} (node1) ; 
\draw[draw=black,solid,->]  (sqnode3) to node[midway,left]{\red $q$} (node1) ; 
\draw[draw=black,solid,->]  (node1) to node[midway,above]{} (sqnode1) ; 
\draw[draw=blue,solid,<-]  (sqnode2) to node[near end,above]{{$\phi$}}    node[midway,above]{{\red $q u_i$}} (sqnode1) ; 
\draw[draw=blue,solid,<-]  (sqnode3) to node[near end,below]{{$\phi'$}}  node[midway,below]{{\red $q^{-1} u_i$}} (sqnode1) ; 
\end{tikzpicture}}
\qquad \qquad
\scalebox{0.8}{
\begin{tikzpicture}[baseline]
\tikzstyle{every node}=[font=\footnotesize]
\node[draw, circle, fill=yellow] (node1) at (0,0) {};
\node[draw, rectangle] (sqnode1) at (2,0) {$2N$};
\node[draw, rectangle] (sqnode2) at (-1.5,1.5) {$N$};
\node[draw, rectangle] (sqnode3) at (-1.5,-1.5) {$N$};
\draw[draw=black,solid,->]  (sqnode2) to node[midway,left]{\red $q^{-1}$} (node1) ; 
\draw[draw=black,solid,->]  (sqnode3) to node[midway,left]{\red $q$} (node1) ; 
\draw[draw=black,solid,->]  (node1) to node[midway,above]{} (sqnode1) ; 
\draw[draw=blue,solid,<-]  (sqnode2) to node[near end,above]{{$\tilde{\phi}'$}} node[midway,above]{{\red $q u_i$}} (sqnode1) ; 
\draw[draw=blue,solid,<-]  (sqnode3) to node[near end,below]{{$\tilde{\phi}$}}   node[midway,below]{{\red $q^{-1} u_i$}} (sqnode1) ; 
\end{tikzpicture}}
\ee

The next step is to turn on some superpotential terms to identify the 4d fields along which we glue.
\paragraph{The $\Phi$-gluing.} To identify $\phi$ with $\tilde{\phi}'$, we introduce an additional set of chiral fields $\Phi$ that are coupled to the 4d fields via the superpotential term
\be
\gd W=\Phi(\phi-\tilde{\phi}')\, ,
\ee
where the contraction of indices is understood. This is a mass term for the fields $\Phi$, $\phi$ and $\tilde{\phi}'$ and integrating them out we are left with only one combination of $\phi$ and $\tilde{\phi}'$. In the process, the equations of motion of $\Phi$ precisely identify $\phi=\tilde{\phi}'$ as desired. Moreover, this superpotential breaks the two $SU(N)$ symmetries from each copy of the building blocks to a diagonal combination, which we gauge with CS level $k$. 
Similarly, the two copies of the $SU(2N)$ symmetry are also broken to a diagonal subgroup, which remains as a flavour symmetry in the resulting theory.  In the quiver description, the $\Phi$-gluing and the resulting model are
\be \label{Phiglue}
\begin{split}
&
\scalebox{0.7}{
\begin{tikzpicture}[baseline]
\tikzstyle{every node}=[font=\footnotesize]
\node[draw, circle, fill=yellow] (node1) at (0,0) {};
\node[draw, rectangle] (sqnode1) at (0,-2) {$2N$};
\node[draw, rectangle] (sqnode2) at (2,1) {$N$};
\node[draw, rectangle] (sqnode3) at (-2,1) {$N$};
\draw[draw=black,solid,->]  (sqnode2) to node[midway,above]{\red $q^{-1}$} (node1) ; 
\draw[draw=black,solid,->]  (sqnode3) to node[midway,above]{\red $q$} (node1) ; 
\draw[draw=black,solid,->]  (node1) to node[midway,above]{} (sqnode1) ; 
\draw[draw=blue,solid,<-]  (sqnode2) to node[near end,right]{{$\phi$}}    node[midway,right]{{\red $q u_i$}} (sqnode1) ; 
\draw[draw=blue,solid,<-]  (sqnode3) to node[near end,left]{{$\phi'$}}  node[midway,left]{{\red $q^{-1} u_i$}} (sqnode1) ; 
\end{tikzpicture}}
\qquad 
\scalebox{0.7}{
\begin{tikzpicture}[baseline]
\tikzstyle{every node}=[font=\footnotesize]
\node[draw, rectangle] (sqnode1) at (0,-2) {$2N$};
\node[draw, rectangle] (sqnode2) at (0,1) {$N$};
\draw[draw=blue,solid,->]  (sqnode2) to node[midway,right]{$\Phi$}  node[midway,left]{\red $q^{-1} u_i^{-1}$} (sqnode1) ;  
\end{tikzpicture}}
\qquad
\scalebox{0.7}{
\begin{tikzpicture}[baseline]
\tikzstyle{every node}=[font=\footnotesize]
\node[draw, circle, fill=yellow] (node1) at (0,0) {};
\node[draw, rectangle] (sqnode1) at (0,-2) {$2N$};
\node[draw, rectangle] (sqnode3) at (2,1) {$N$};
\node[draw, rectangle] (sqnode2) at (-2,1) {$N$};
\draw[draw=black,solid,->]  (sqnode2) to node[midway,above]{\red $q^{-1}$} (node1) ; 
\draw[draw=black,solid,->]  (sqnode3) to node[midway,above]{\red $q$} (node1) ; 
\draw[draw=black,solid,->]  (node1) to node[midway,above]{} (sqnode1) ; 
\draw[draw=blue,solid,<-]  (sqnode2) to node[near end,left]{{$\tilde{\phi}'$}}    node[midway,left]{{\red $q u_i$}} (sqnode1) ; 
\draw[draw=blue,solid,<-]  (sqnode3) to node[near end,right]{{$\tilde{\phi}$}}  node[midway,right]{{\red $q^{-1} u_i$}} (sqnode1) ; 
\end{tikzpicture}}
\\ 
&= \qquad \quad
\scalebox{0.7}{
\begin{tikzpicture}[baseline]
\tikzstyle{every node}=[font=\footnotesize]
\node[draw, rectangle] (sqnode) at (0,0) {$2N$};
\node[draw, circle, fill=yellow] (yell1) at (-2,0) {};
\node[draw, circle, dashed] (circ1) at (0,2) {$N_{k}$};
\node[draw, rectangle] (sqnode1) at (-4,0) {$N$};
\node[draw, rectangle] (sqnode2) at (4,0) {$N$};
\node[draw, circle, fill=yellow] (yell2) at (2,0) {};
\draw[draw=black,solid,<-]  (sqnode) to node[midway,above] {$P$}   (yell1) ; 
\draw[draw=black,solid,<-]  (yell1) to node[midway,above] {$A$} node[near end,below] {\red $q^{-1}$}    (circ1) ; 
\draw[draw=black,solid,<-]  (yell1) to node[midway,below] {$B$} node[midway,above] {\red $q$}   (sqnode1) ; 
\draw[draw=black,solid,->]  (circ1) to node[midway,above] {$C$} node[near start,below] {\red $q^{-1}$}     (yell2) ; 
\draw[draw=black,solid,->]  (sqnode2) to node[midway,below] {$D$} node[midway,above] {\red $q$}   (yell2) ; 
\draw[draw=black,solid,->]  (yell2) to node[midway,above] {$Q$}   (sqnode) ; 
\draw[draw=blue,solid,<-]  (circ1) to node[midway,left] {\blue $\phi$} node[near end,right] {\red $q$}   (sqnode) ; 
\draw[draw=blue,solid,<-]  (sqnode1) to[bend right=80] node[midway,above] {\blue $\phi'$} node[midway,below] {\red $q^{-1}$} (sqnode);
\draw[draw=blue,solid,<-]  (sqnode2) to[bend left=80] node[midway,above] {\blue $\tilde{\phi}$} node[midway,below] {\red $q^{-1}$}  (sqnode);
\end{tikzpicture}}
\end{split} 
\ee
where the superpotential is
\be
W= V^{(1)}_++ V^{(1)}_- + V^{(2)}_++ V^{(2)}_- + A\phi P +C \phi Q + B \phi' P + D \tilde{\phi}Q~,
\ee
and we drop the fugacity $u_i$ in the lower diagram (the transformation rule of each chiral field under $SU(2N)$ is clear from the arrow).  Throughout the paper, we denote by $N_k$ in a dashed circle the $SU(N)$ gauge group with CS level $k$. Notice that the charges and the representations of all the chiral fields under the global symmetries implied by the $\Phi$-gluing condition are compatible with the cubic superpotential terms corresponding to each loop in the quiver.  We use the following skeleton diagram to denote the $\Phi$-gluing \eref{Phiglue}:
\be
\scalebox{0.8}{
\begin{tikzpicture}[baseline]
\tikzstyle{every node}=[font=\footnotesize, node distance=0.45cm]
\tikzset{decoration={snake,amplitude=.4mm,segment length=2mm,
                       post length=0mm,pre length=0mm}}
\draw[blue,thick] (0,-1.5)--(0,1.5) node[midway, right] {$\Phi$};  
\draw[decorate,red,thick] (-1,0.5) -- (1,0.5) node[right] {}; 
\draw[decorate,red,thick] (-1,-0.5) -- (1,-0.5) node[right] {}; 
\end{tikzpicture}}
\ee
The two blue external legs correspond to the two $SU(N)$ flavour nodes in the lower diagram of \eref{Phiglue}.  As discussed before, the two $SU(2N)$ symmetries coming from each duality wall (red wiggle line) are broken to a diagonal subgroup by the aforementioned superpotential, and this is denoted by the square node labelled by $2N$ in the bottom quiver in \eref{Phiglue}.

\paragraph{The $S$-gluing.} The $S$-gluing can be implemented by introducing the superpotential term
\be
\gd W=\phi\tilde{\phi}'\, .
\ee
This implies that both $\phi$ and $\tilde{\phi}'$ are integrated out and we are left with no field. Again the two $SU(N)$ symmetries are broken to a diagonal combination, which we gauge with a CS level $k$.  The two $SU(2N)$ symmetries are also broken to a diagonal subgroup.  In the quiver description, the $S$-gluing and the resulting model are
\be \label{Sglue}
\begin{split}
&
\scalebox{0.7}{
\begin{tikzpicture}[baseline]
\tikzstyle{every node}=[font=\footnotesize]
\node[draw, circle, fill=yellow] (node1) at (0,0) {};
\node[draw, rectangle] (sqnode1) at (0,-2) {$2N$};
\node[draw, rectangle] (sqnode2) at (2,1) {$N$};
\node[draw, rectangle] (sqnode3) at (-2,1) {$N$};
\draw[draw=black,solid,->]  (sqnode2) to node[midway,above]{\red $q^{-1}$} (node1) ; 
\draw[draw=black,solid,->]  (sqnode3) to node[midway,above]{\red $q$} (node1) ; 
\draw[draw=black,solid,->]  (node1) to node[midway,above]{} (sqnode1) ; 
\draw[draw=blue,solid,<-]  (sqnode2) to node[near end,right]{{$\phi$}}    node[midway,right]{{\red $q u_i$}} (sqnode1) ; 
\draw[draw=blue,solid,<-]  (sqnode3) to node[near end,left]{{$\phi'$}}  node[midway,left]{{\red $q^{-1} u_i$}} (sqnode1) ; 
\end{tikzpicture}}
\qquad 
\qquad
\scalebox{0.7}{
\begin{tikzpicture}[baseline]
\tikzstyle{every node}=[font=\footnotesize]
\node[draw, circle, fill=yellow] (node1) at (0,0) {};
\node[draw, rectangle] (sqnode1) at (0,-2) {$2N$};
\node[draw, rectangle] (sqnode3) at (2,1) {$N$};
\node[draw, rectangle] (sqnode2) at (-2,1) {$N$};
\draw[draw=black,solid,<-]  (sqnode2) to node[midway,above]{\red $q$} (node1) ; 
\draw[draw=black,solid,<-]  (sqnode3) to node[midway,above]{\red $q^{-1}$} (node1) ; 
\draw[draw=black,solid,<-]  (node1) to node[midway,above]{} (sqnode1) ; 
\draw[draw=blue,solid,->]  (sqnode2) to node[near end,left]{{$\tilde{\phi}'$}}    node[midway,left]{{\red $q^{-1} u_i^{-1}$}} (sqnode1) ; 
\draw[draw=blue,solid,->]  (sqnode3) to node[near end,right]{{$\tilde{\phi}$}}  node[midway,right]{{\red $q u_i^{-1}$}} (sqnode1) ; 
\end{tikzpicture}}
\\ 
&= \qquad
\scalebox{0.7}{
\begin{tikzpicture}[baseline]
\tikzstyle{every node}=[font=\footnotesize]
\node[draw, rectangle] (sqnode) at (0,0) {$2N$};
\node[draw, circle, fill=yellow] (yell1) at (-2,0) {};
\node[draw, circle, dashed] (circ1) at (0,2) {$N_{k}$};
\node[draw, rectangle] (sqnode1) at (-4,0) {$N$};
\node[draw, rectangle] (sqnode2) at (4,0) {$N$};
\node[draw, circle, fill=yellow] (yell2) at (2,0) {};
\draw[draw=black,solid,<-]  (sqnode) to node[midway,above] {$P$}   (yell1) ; 
\draw[draw=black,solid,<-]  (yell1) to node[midway,above] {$A$} node[near end,below] {\red $q^{-1}$}    (circ1) ; 
\draw[draw=black,solid,<-]  (yell1) to node[midway,below] {$B$} node[midway,above] {\red $q$}   (sqnode1) ; 
\draw[draw=black,solid,<-]  (circ1) to node[midway,above] {$C$} node[near start,below] {\red $q$}     (yell2) ; 
\draw[draw=black,solid,<-]  (sqnode2) to node[midway,below] {$D$} node[midway,above] {\red $q^{-1}$}   (yell2) ; 
\draw[draw=black,solid,<-]  (yell2) to node[midway,above] {$Q$}   (sqnode) ; 
\draw[draw=blue,solid,<-]  (sqnode1) to[bend right=80] node[midway,above] {\blue $\phi'$} node[midway,below] {\red $q^{-1}$} (sqnode);
\draw[draw=blue,solid,->]  (sqnode2) to[bend left=80] node[midway,above] {\blue $\tilde{\phi}$} node[midway,below] {\red $q$}  (sqnode);
\end{tikzpicture}}
\end{split} 
\ee
\noindent The superpotential of the resulting theory is
\be
W= V^{(1)}_++ V^{(1)}_- + V^{(2)}_++ V^{(2)}_- + CAPQ + B \phi' P + D \tilde{\phi} Q~.
\ee
Notice again that the charges and the representations of all the chiral fields under the global symmetries implied by the $S$-gluing condition are compatible with the cubic and quartic superpotential terms corresponding to each loop in the quiver. We use the following skeleton diagram to denote the $S$-gluing \eref{Sglue}:
\be
\scalebox{0.8}{
\begin{tikzpicture}[baseline]
\tikzstyle{every node}=[font=\footnotesize, node distance=0.45cm]
\tikzset{decoration={snake,amplitude=.4mm,segment length=2mm,
                       post length=0mm,pre length=0mm}}
\draw[blue,thick] (0,-1.5)--(0,1.5) node[midway, right] {$S$};  
\draw[decorate,red,thick] (-1,0.5) -- (1,0.5) node[right] {}; 
\draw[decorate,red,thick] (-1,-0.5) -- (1,-0.5) node[right] {}; 
\end{tikzpicture}}
\ee

Note that we can also treat odd number of duality walls in a similar way as described above.  For example, in the case of three duality walls, we can perform $S$-gluing in the following way:
\be
\scalebox{0.8}{
\begin{tikzpicture}[baseline]
\tikzstyle{every node}=[font=\footnotesize, node distance=0.45cm]
\tikzset{decoration={snake,amplitude=.4mm,segment length=2mm,
                       post length=0mm,pre length=0mm}}
\draw[blue,thick] (0,-1.5)--(0,1.5) node[midway, right] {};  
\draw[decorate,red,thick] (-1,1) -- (1,1) node[right] {}; 
\draw[decorate,red,thick] (-1,0) -- (1,0) node[right] {}; 
\draw[decorate,red,thick] (-1,-1) -- (1,-1) node[right] {}; 
\node[draw=none] at (0.5,0.5) {\blue $S$};
\node[draw=none] at (0.5,-0.5) {\blue $S$};
\end{tikzpicture}}
\ee
The corresponding theory is
\be \label{Sglue3walls}
\begin{split}
&
\scalebox{0.7}{
\begin{tikzpicture}[baseline]
\tikzstyle{every node}=[font=\footnotesize]
\node[draw, circle, fill=yellow] (node1) at (0,0) {};
\node[draw, rectangle] (sqnode1) at (0,-2) {$2N$};
\node[draw, rectangle] (sqnode2) at (2,1) {$N$};
\node[draw, rectangle] (sqnode3) at (-2,1) {$N$};
\draw[draw=black,solid,->]  (sqnode2) to node[midway,above]{\red $q^{-1}$} (node1) ; 
\draw[draw=black,solid,->]  (sqnode3) to node[midway,above]{\red $q$} (node1) ; 
\draw[draw=black,solid,->]  (node1) to node[midway,above]{} (sqnode1) ; 
\draw[draw=blue,solid,<-]  (sqnode2) to node[near end,right]{{$\hat{\phi}$}}    node[midway,right]{{\red $q u_i$}} (sqnode1) ; 
\draw[draw=blue,solid,<-]  (sqnode3) to node[near end,left]{{$\hat{\phi}'$}}  node[midway,left]{{\red $q^{-1} u_i$}} (sqnode1) ; 
\end{tikzpicture}}
\qquad 
\scalebox{0.7}{
\begin{tikzpicture}[baseline]
\tikzstyle{every node}=[font=\footnotesize]
\node[draw, circle, fill=yellow] (node1) at (0,0) {};
\node[draw, rectangle] (sqnode1) at (0,-2) {$2N$};
\node[draw, rectangle] (sqnode3) at (2,1) {$N$};
\node[draw, rectangle] (sqnode2) at (-2,1) {$N$};
\draw[draw=black,solid,<-]  (sqnode2) to node[midway,above]{\red $q$} (node1) ; 
\draw[draw=black,solid,<-]  (sqnode3) to node[midway,above]{\red $q^{-1}$} (node1) ; 
\draw[draw=black,solid,<-]  (node1) to node[midway,above]{} (sqnode1) ; 
\draw[draw=blue,solid,->]  (sqnode2) to node[near end,left]{{$\tilde{\phi}'$}}    node[midway,left]{{\red $q^{-1} u_i^{-1}$}} (sqnode1) ; 
\draw[draw=blue,solid,->]  (sqnode3) to node[near end,right]{{$\tilde{\phi}$}}  node[midway,right]{{\red $q u_i^{-1}$}} (sqnode1) ; 
\end{tikzpicture}}
\qquad 
\scalebox{0.7}{
\begin{tikzpicture}[baseline]
\tikzstyle{every node}=[font=\footnotesize]
\node[draw, circle, fill=yellow] (node1) at (0,0) {};
\node[draw, rectangle] (sqnode1) at (0,-2) {$2N$};
\node[draw, rectangle] (sqnode2) at (2,1) {$N$};
\node[draw, rectangle] (sqnode3) at (-2,1) {$N$};
\draw[draw=black,solid,->]  (sqnode2) to node[midway,above]{\red $q^{-1}$} (node1) ; 
\draw[draw=black,solid,->]  (sqnode3) to node[midway,above]{\red $q$} (node1) ; 
\draw[draw=black,solid,->]  (node1) to node[midway,above]{} (sqnode1) ; 
\draw[draw=blue,solid,<-]  (sqnode2) to node[near end,right]{{$\phi$}}    node[midway,right]{{\red $q u_i$}} (sqnode1) ; 
\draw[draw=blue,solid,<-]  (sqnode3) to node[near end,left]{{$\phi'$}}  node[midway,left]{{\red $q^{-1} u_i$}} (sqnode1) ; 
\end{tikzpicture}}\\
&= \qquad
\scalebox{0.7}{
\begin{tikzpicture}[baseline]
\tikzstyle{every node}=[font=\footnotesize]
\node[draw, rectangle] (sqnode) at (-1.5,-2) {$2N$};
\node[draw, circle, fill=yellow] (yell1) at (-1.5,0) {};
\node[draw, circle, dashed] (circ1) at (0,2) {$N_{k_1}$};
\node[draw, circle, dashed] (circ2) at (-3,2) {$N_{k_2}$};
\node[draw, rectangle] (sqnode2) at (4,0) {$N$};
\node[draw, rectangle] (sqnode3) at (-7,0) {$N$};
\node[draw, circle, fill=yellow] (yell2) at (1.5,0) {};
\node[draw, circle, fill=yellow] (yell3) at (-4.5,0) {};
\draw[draw=black,solid,->]  (sqnode) to node[midway,above] {}   (yell1) ; 
\draw[draw=black,solid,->]  (yell1) to node[midway,above] {}   (circ1) ; 
\draw[draw=black,solid,->]  (yell1) to node[midway,below] {}   (circ2) ; 
\draw[draw=black,solid,->]  (circ1) to node[midway,above] {}   (yell2) ; 
\draw[draw=black,solid,->]  (sqnode2) to node[midway,below] {}   (yell2) ; 
\draw[draw=black,solid,->]  (yell2) to node[midway,above] {}   (sqnode) ; 
\draw[draw=blue,solid,->]  (sqnode) to [bend left=0] node[midway,below]  {\blue $\hat{\phi}'$}   (sqnode3) ; \draw[draw=black,solid,->]  (circ2) to node[midway,left] {}   (yell3) ; 
\draw[draw=black,solid,->]  (sqnode3) to node[midway,left] {}   (yell3) ; 
\draw[draw=black,solid,->]  (yell3) to node[midway,right] {}   (sqnode) ; 
\draw[draw=blue,solid,<-]  (sqnode2) to[bend left=0] node[midway,below] {\blue $\phi$}  (sqnode);
\end{tikzpicture}}
\end{split}
\ee

\subsubsection{Self-gluing: closing external legs}
With the prescription for the $\Phi$- and $S$-gluing one can construct several other models, either adding more basic building blocks or gauging the remaining flavour symmetries. The latter corresponds to closing external legs of the skeleton diagram.  For example, in \eref{Phiglue} and \eref{Sglue} we can ``self-glue'' the theory along $\phi'$ and $\tilde{\phi}$ which results in gauging together the two remaining $SU(N)$ flavour symmetries. 

In the model \eref{Phiglue}, obtained from the $\Phi$-gluing of two basic building blocks, we can only perform a further $\Phi$-gluing along $\phi'$ and $\tilde{\phi}$.  The latter is because both $\phi'$ and $\tilde{\phi}$ carry the same $U(1)_q$ charge and transform the same way under $SU(N) \times SU(2N)$.  This leads to the model
\be  \label{twowallsphi}
\begin{tikzpicture}[baseline]
\tikzstyle{every node}=[font=\footnotesize, node distance=0.45cm]
\tikzset{decoration={snake,amplitude=.4mm,segment length=2mm,
                       post length=0mm,pre length=0mm}}
\draw[blue,thick] (0,0) circle (1cm) node[midway] {$\Phi$};  
\draw[decorate,red,thick] (0.5,0) -- (1.5,0) node[right] {}; 
\draw[decorate,red,thick] (-0.5,0) -- (-1.5,0) node[right] {}; 
\end{tikzpicture}
\qquad \qquad
\scalebox{0.8}{
\begin{tikzpicture}[baseline]
\tikzstyle{every node}=[font=\footnotesize]
\node[draw, rectangle] (sqnode) at (0,0) {$2N$};
\node[draw, circle, fill=yellow] (yell1) at (-2,0) {};
\node[draw, circle, dashed] (circ1) at (0,2) {$N_{k_1}$};
\node[draw, circle, dashed] (circ2) at (0,-2) {$N_{k_2}$};
\node[draw, circle, fill=yellow] (yell2) at (2,0) {};
\draw[draw=black,solid,<-]  (sqnode) to node[midway,above] {$P$}   (yell1) ; 
\draw[draw=black,solid,<-]  (yell1) to node[midway,above] {$A$}   (circ1) ; 
\draw[draw=black,solid,<-]  (yell1) to node[midway,below] {$B$}   (circ2) ; 
\draw[draw=black,solid,->]  (circ1) to node[midway,above] {$C$}   (yell2) ; 
\draw[draw=black,solid,->]  (circ2) to node[midway,below] {$D$}   (yell2) ; 
\draw[draw=black,solid,->]  (yell2) to node[midway,above] {$Q$}   (sqnode) ; 
\draw[draw=blue,solid,<-]  (circ1) to node[midway,left] {\blue $\phi$}  (sqnode) ; 
\draw[draw=blue,solid,<-]  (circ2) to node[midway,left] {\blue $\phi'$}  (sqnode) ; 
\end{tikzpicture}}
\ee
with superpotential
\be
W=V^{(1)}_++ V^{(1)}_- + V^{(2)}_++ V^{(2)}_- + A\phi P +B \phi' P + C \phi Q + D \phi' Q~.
\ee

On the other hand, in the model \eref{Sglue}, obtained from the $S$-gluing, we can only perform a further $S$-gluing along $\phi'$ and $\tilde{\phi}'$.  This is because $\phi'$ and $\tilde{\phi}$ carry opposite $U(1)_q$ charges and transform the opposite way under $SU(N) \times SU(2N)$.  We thus arrive at the following model
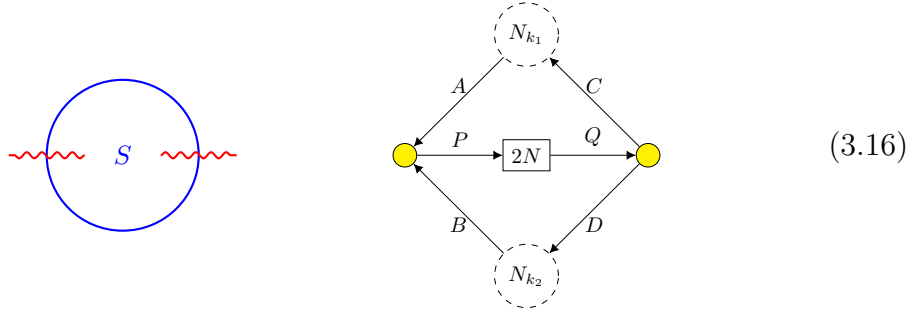
\be  \label{twowallswophi}
\begin{tikzpicture}[baseline]
\tikzstyle{every node}=[font=\footnotesize, node distance=0.45cm]
\tikzset{decoration={snake,amplitude=.4mm,segment length=2mm,
                       post length=0mm,pre length=0mm}}
\draw[blue,thick] (0,0) circle (1cm) node[midway] {$S$};  
\draw[decorate,red,thick] (0.5,0) -- (1.5,0) node[right] {}; 
\draw[decorate,red,thick] (-0.5,0) -- (-1.5,0) node[right] {}; 
\end{tikzpicture}
\qquad \qquad
\scalebox{0.8}{
\begin{tikzpicture}[baseline]
\tikzstyle{every node}=[font=\footnotesize]
\node[draw, rectangle] (sqnode) at (0,0) {$2N$};
\node[draw, circle, fill=yellow] (yell1) at (-2,0) {};
\node[draw, circle, dashed] (circ1) at (0,2) {$N_{k_1}$};
\node[draw, circle, dashed] (circ2) at (0,-2) {$N_{k_2}$};
\node[draw, circle, fill=yellow] (yell2) at (2,0) {};
\draw[draw=black,solid,<-]  (sqnode) to node[midway,above] {$P$}   (yell1) ; 
\draw[draw=black,solid,<-]  (yell1) to node[midway,above] {$A$}   (circ1) ; 
\draw[draw=black,solid,<-]  (yell1) to node[midway,below] {$B$}   (circ2) ; 
\draw[draw=black,solid,<-]  (circ1) to node[midway,above] {$C$}   (yell2) ; 
\draw[draw=black,solid,<-]  (circ2) to node[midway,below] {$D$}   (yell2) ; 
\draw[draw=black,solid,<-]  (yell2) to node[midway,above] {$Q$}   (sqnode) ; 
\end{tikzpicture}}
\ee
with superpotential
\be \label{suptwowallswophi}
W=V^{(1)}_++ V^{(1)}_- + V^{(2)}_++ V^{(2)}_- + PQCA +PQDB~.
\ee

Notice that the two previous models have similar structures, apart from the fact that in \eref{twowallswophi} the 4d fields $\phi$ and $\phi'$ are absent, and the $U(1)$ charges as well as the arrows of the right half of the quiver are inverted with respect to \eref{twowallsphi}. This is very similar to the difference between the models \eqref{basicblock} and \eqref{dualbasicblock}. Indeed, by applying duality \eqref{mainduality} locally on the right yellow node of \eref{twowallswophi}, one obtains \eref{twowallsphi}.  Models \eqref{twowallsphi} and \eqref{twowallswophi} are actually dual to each other for any $N \geq 2$:
\be
\eqref{twowallsphi} ~ \overset{\eref{mainduality}}{\longleftrightarrow} ~ \eqref{twowallswophi}~.
\ee
As a result there is no need to specify $\Phi$ or $S$ when we draw the skeleton diagram with all external legs being closed.

This result can be generalised for any {\it even} number of duality walls.  We state a general result as follows.
\begin{claim}
For given $N$ and the Chern--Simons levels as well as a topology of the skeleton diagram, if all external legs of the latter are closed, the theories associated with the $\Phi$-gluing and/or $S$-gluing of an even number of walls are {\it dual} to each other.
\end{claim}

Let us provide an example for theories associated with four duality walls such that all external legs are closed.
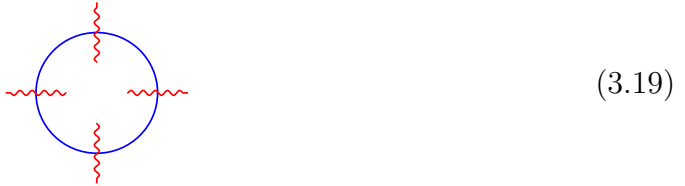
\be
\scalebox{0.8}{
\begin{tikzpicture}[baseline]
\tikzstyle{every node}=[font=\footnotesize, node distance=0.45cm]
\tikzset{decoration={snake,amplitude=.4mm,segment length=2mm,
                       post length=0mm,pre length=0mm}}
\draw[blue,thick] (0,0) circle (1cm) node[midway, right] {};  
\draw[decorate,red,thick] (0.5,0) -- (1.5,0) node[right] {}; 
\draw[decorate,red,thick] (-0.5,0) -- (-1.5,0) node[right] {}; 
\draw[decorate,red,thick] (0,0.5) -- (0,1.5) node[right] {}; 
\draw[decorate,red,thick] (0,-0.5) -- (0,-1.5) node[right] {}; 
\end{tikzpicture}}
\ee
We have eight duality frames with an ``octality'' that relates them to each other.
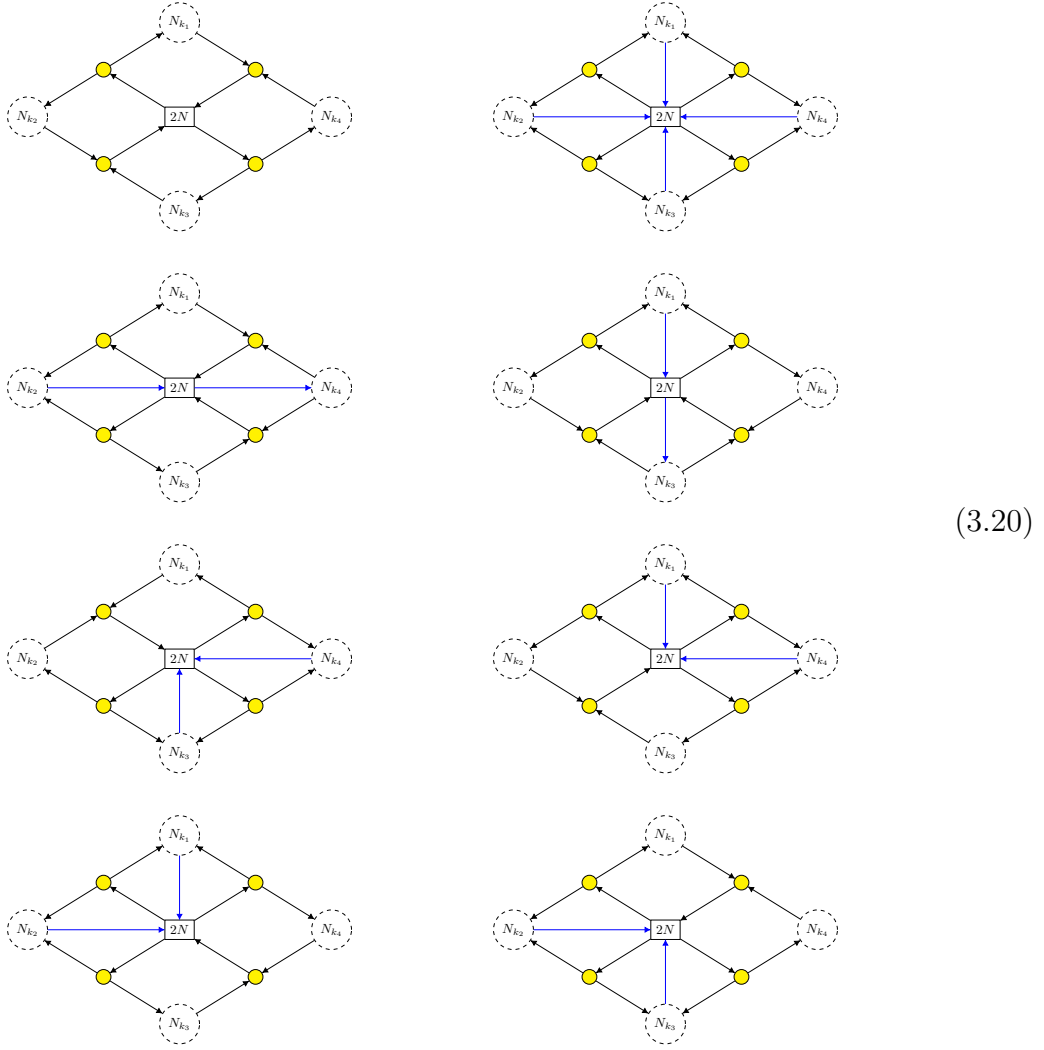
\be  \label{fourwallsquad}
\begin{array}{ll}
\scalebox{0.5}{
\begin{tikzpicture}[baseline]
\tikzstyle{every node}=[font=\footnotesize]
\node[draw, rectangle] (sqnode) at (0,0) {$2N$};
\node[draw, circle, fill=yellow] (yell1) at (-2,1.25) {};
\node[draw, circle, fill=yellow] (yell2) at (2,1.25) {};
\node[draw, circle, fill=yellow] (yell4) at (2,-1.25) {};
\node[draw, circle, fill=yellow] (yell3) at (-2,-1.25) {};
\node[draw, circle, dashed] (circ1) at (0,2.5) {$N_{k_1}$};
\node[draw, circle, dashed] (circ2) at (-4,0) {$N_{k_2}$};
\node[draw, circle, dashed] (circ4) at (4,0) {$N_{k_4}$};
\node[draw, circle, dashed] (circ3) at (0,-2.5) {$N_{k_3}$};
\draw[draw=black,solid,->]  (yell1) to node[midway,above] {}   (circ1) ; 
\draw[draw=black,solid,->]  (yell1) to node[midway,above] {}   (circ2) ; 
\draw[draw=black,solid,<-]  (yell3) to node[midway,above] {}   (circ2) ; 
\draw[draw=black,solid,<-]  (yell3) to node[midway,above] {}   (circ3) ; 
\draw[draw=black,solid,->]  (yell4) to node[midway,above] {}   (circ3) ; 
\draw[draw=black,solid,->]  (yell4) to node[midway,above] {}   (circ4) ;
\draw[draw=black,solid,<-]  (yell2) to node[midway,above] {}   (circ4) ; 
\draw[draw=black,solid,<-]  (yell2) to node[midway,above] {}   (circ1) ; 

\draw[draw=black,solid,<-]  (yell1) to node[midway,above] {}   (sqnode) ; 
\draw[draw=black,solid,->]  (yell2) to node[midway,above] {}   (sqnode) ;
\draw[draw=black,solid,->]  (yell3) to node[midway,above] {}   (sqnode) ; 
\draw[draw=black,solid,<-]  (yell4) to node[midway,above] {}   (sqnode) ;
\end{tikzpicture}} 
& \qquad \qquad 
\scalebox{0.5}{
\begin{tikzpicture}[baseline]
\tikzstyle{every node}=[font=\footnotesize]
\node[draw, rectangle] (sqnode) at (0,0) {$2N$};
\node[draw, circle, fill=yellow] (yell1) at (-2,1.25) {};
\node[draw, circle, fill=yellow] (yell2) at (2,1.25) {};
\node[draw, circle, fill=yellow] (yell4) at (2,-1.25) {};
\node[draw, circle, fill=yellow] (yell3) at (-2,-1.25) {};
\node[draw, circle, dashed] (circ1) at (0,2.5) {$N_{k_1}$};
\node[draw, circle, dashed] (circ2) at (-4,0) {$N_{k_2}$};
\node[draw, circle, dashed] (circ4) at (4,0) {$N_{k_4}$};
\node[draw, circle, dashed] (circ3) at (0,-2.5) {$N_{k_3}$};
\draw[draw=black,solid,->]  (yell1) to node[midway,above] {}   (circ1) ; 
\draw[draw=black,solid,->]  (yell1) to node[midway,above] {}   (circ2) ; 
\draw[draw=black,solid,->]  (yell3) to node[midway,above] {}   (circ2) ; 
\draw[draw=black,solid,->]  (yell3) to node[midway,above] {}   (circ3) ; 
\draw[draw=black,solid,->]  (yell4) to node[midway,above] {}   (circ3) ; 
\draw[draw=black,solid,->]  (yell4) to node[midway,above] {}   (circ4) ;
\draw[draw=black,solid,->]  (yell2) to node[midway,above] {}   (circ4) ; 
\draw[draw=black,solid,->]  (yell2) to node[midway,above] {}   (circ1) ; 

\draw[draw=black,solid,<-]  (yell1) to node[midway,above] {}   (sqnode) ; 
\draw[draw=black,solid,<-]  (yell2) to node[midway,above] {}   (sqnode) ;
\draw[draw=black,solid,<-]  (yell3) to node[midway,above] {}   (sqnode) ; 
\draw[draw=black,solid,<-]  (yell4) to node[midway,above] {}   (sqnode) ;

\draw[draw=blue,solid,->]  (circ1) to node[midway,above] {}   (sqnode) ; 
\draw[draw=blue,solid,->]  (circ3) to node[midway,above] {}   (sqnode) ;
\draw[draw=blue,solid,->]  (circ2) to node[midway,above] {}   (sqnode) ; 
\draw[draw=blue,solid,->]  (circ4) to node[midway,above] {}   (sqnode) ;
\end{tikzpicture}} \\& \\
\scalebox{0.5}{
\begin{tikzpicture}[baseline]
\tikzstyle{every node}=[font=\footnotesize]
\node[draw, rectangle] (sqnode) at (0,0) {$2N$};
\node[draw, circle, fill=yellow] (yell1) at (-2,1.25) {};
\node[draw, circle, fill=yellow] (yell2) at (2,1.25) {};
\node[draw, circle, fill=yellow] (yell4) at (2,-1.25) {};
\node[draw, circle, fill=yellow] (yell3) at (-2,-1.25) {};
\node[draw, circle, dashed] (circ1) at (0,2.5) {$N_{k_1}$};
\node[draw, circle, dashed] (circ2) at (-4,0) {$N_{k_2}$};
\node[draw, circle, dashed] (circ4) at (4,0) {$N_{k_4}$};
\node[draw, circle, dashed] (circ3) at (0,-2.5) {$N_{k_3}$};
\draw[draw=black,solid,->]  (yell1) to node[midway,above] {}   (circ1) ; 
\draw[draw=black,solid,->]  (yell1) to node[midway,above] {}   (circ2) ; 
\draw[draw=black,solid,->]  (yell3) to node[midway,above] {}   (circ2) ; 
\draw[draw=black,solid,->]  (yell3) to node[midway,above] {}   (circ3) ; 
\draw[draw=black,solid,<-]  (yell4) to node[midway,above] {}   (circ3) ; 
\draw[draw=black,solid,<-]  (yell4) to node[midway,above] {}   (circ4) ;
\draw[draw=black,solid,<-]  (yell2) to node[midway,above] {}   (circ4) ; 
\draw[draw=black,solid,<-]  (yell2) to node[midway,above] {}   (circ1) ; 

\draw[draw=black,solid,<-]  (yell1) to node[midway,above] {}   (sqnode) ; 
\draw[draw=black,solid,->]  (yell2) to node[midway,above] {}   (sqnode) ;
\draw[draw=black,solid,<-]  (yell3) to node[midway,above] {}   (sqnode) ; 
\draw[draw=black,solid,->]  (yell4) to node[midway,above] {}   (sqnode) ;

\draw[draw=blue,solid,->]  (circ2) to node[midway,above] {}   (sqnode) ; 
\draw[draw=blue,solid,<-]  (circ4) to node[midway,above] {}   (sqnode) ;
\end{tikzpicture}}
& \qquad \qquad
\scalebox{0.5}{
\begin{tikzpicture}[baseline]
\tikzstyle{every node}=[font=\footnotesize]
\node[draw, rectangle] (sqnode) at (0,0) {$2N$};
\node[draw, circle, fill=yellow] (yell1) at (-2,1.25) {};
\node[draw, circle, fill=yellow] (yell2) at (2,1.25) {};
\node[draw, circle, fill=yellow] (yell4) at (2,-1.25) {};
\node[draw, circle, fill=yellow] (yell3) at (-2,-1.25) {};
\node[draw, circle, dashed] (circ1) at (0,2.5) {$N_{k_1}$};
\node[draw, circle, dashed] (circ2) at (-4,0) {$N_{k_2}$};
\node[draw, circle, dashed] (circ4) at (4,0) {$N_{k_4}$};
\node[draw, circle, dashed] (circ3) at (0,-2.5) {$N_{k_3}$};
\draw[draw=black,solid,->]  (yell1) to node[midway,above] {}   (circ1) ; 
\draw[draw=black,solid,->]  (yell1) to node[midway,above] {}   (circ2) ; 
\draw[draw=black,solid,<-]  (yell3) to node[midway,above] {}   (circ2) ; 
\draw[draw=black,solid,<-]  (yell3) to node[midway,above] {}   (circ3) ; 
\draw[draw=black,solid,<-]  (yell4) to node[midway,above] {}   (circ3) ; 
\draw[draw=black,solid,<-]  (yell4) to node[midway,above] {}   (circ4) ;
\draw[draw=black,solid,->]  (yell2) to node[midway,above] {}   (circ4) ; 
\draw[draw=black,solid,->]  (yell2) to node[midway,above] {}   (circ1) ; 

\draw[draw=black,solid,<-]  (yell1) to node[midway,above] {}   (sqnode) ; 
\draw[draw=black,solid,<-]  (yell2) to node[midway,above] {}   (sqnode) ;
\draw[draw=black,solid,->]  (yell3) to node[midway,above] {}   (sqnode) ; 
\draw[draw=black,solid,->]  (yell4) to node[midway,above] {}   (sqnode) ;

\draw[draw=blue,solid,->]  (circ1) to node[midway,above] {}   (sqnode) ; 
\draw[draw=blue,solid,<-]  (circ3) to node[midway,above] {}   (sqnode) ;
\end{tikzpicture}} \\ & \\
\scalebox{0.5}{
\begin{tikzpicture}[baseline]
\tikzstyle{every node}=[font=\footnotesize]
\node[draw, rectangle] (sqnode) at (0,0) {$2N$};
\node[draw, circle, fill=yellow] (yell1) at (-2,1.25) {};
\node[draw, circle, fill=yellow] (yell2) at (2,1.25) {};
\node[draw, circle, fill=yellow] (yell4) at (2,-1.25) {};
\node[draw, circle, fill=yellow] (yell3) at (-2,-1.25) {};
\node[draw, circle, dashed] (circ1) at (0,2.5) {$N_{k_1}$};
\node[draw, circle, dashed] (circ2) at (-4,0) {$N_{k_2}$};
\node[draw, circle, dashed] (circ4) at (4,0) {$N_{k_4}$};
\node[draw, circle, dashed] (circ3) at (0,-2.5) {$N_{k_3}$};
\draw[draw=black,solid,<-]  (yell1) to node[midway,above] {}   (circ1) ; 
\draw[draw=black,solid,<-]  (yell1) to node[midway,above] {}   (circ2) ; 
\draw[draw=black,solid,->]  (yell3) to node[midway,above] {}   (circ2) ; 
\draw[draw=black,solid,->]  (yell3) to node[midway,above] {}   (circ3) ; 
\draw[draw=black,solid,->]  (yell4) to node[midway,above] {}   (circ3) ; 
\draw[draw=black,solid,->]  (yell4) to node[midway,above] {}   (circ4) ;
\draw[draw=black,solid,->]  (yell2) to node[midway,above] {}   (circ4) ; 
\draw[draw=black,solid,->]  (yell2) to node[midway,above] {}   (circ1) ; 

\draw[draw=black,solid,->]  (yell1) to node[midway,above] {}   (sqnode) ; 
\draw[draw=black,solid,<-]  (yell2) to node[midway,above] {}   (sqnode) ;
\draw[draw=black,solid,<-]  (yell3) to node[midway,above] {}   (sqnode) ; 
\draw[draw=black,solid,<-]  (yell4) to node[midway,above] {}   (sqnode) ;

\draw[draw=blue,solid,->]  (circ3) to node[midway,above] {}   (sqnode) ;
\draw[draw=blue,solid,->]  (circ4) to node[midway,above] {}   (sqnode) ;
\end{tikzpicture}} 
& \qquad \qquad
\scalebox{0.5}{
\begin{tikzpicture}[baseline]
\tikzstyle{every node}=[font=\footnotesize]
\node[draw, rectangle] (sqnode) at (0,0) {$2N$};
\node[draw, circle, fill=yellow] (yell1) at (-2,1.25) {};
\node[draw, circle, fill=yellow] (yell2) at (2,1.25) {};
\node[draw, circle, fill=yellow] (yell4) at (2,-1.25) {};
\node[draw, circle, fill=yellow] (yell3) at (-2,-1.25) {};
\node[draw, circle, dashed] (circ1) at (0,2.5) {$N_{k_1}$};
\node[draw, circle, dashed] (circ2) at (-4,0) {$N_{k_2}$};
\node[draw, circle, dashed] (circ4) at (4,0) {$N_{k_4}$};
\node[draw, circle, dashed] (circ3) at (0,-2.5) {$N_{k_3}$};
\draw[draw=black,solid,->]  (yell1) to node[midway,above] {}   (circ1) ; 
\draw[draw=black,solid,->]  (yell1) to node[midway,above] {}   (circ2) ; 
\draw[draw=black,solid,<-]  (yell3) to node[midway,above] {}   (circ2) ; 
\draw[draw=black,solid,<-]  (yell3) to node[midway,above] {}   (circ3) ; 
\draw[draw=black,solid,->]  (yell4) to node[midway,above] {}   (circ3) ; 
\draw[draw=black,solid,->]  (yell4) to node[midway,above] {}   (circ4) ;
\draw[draw=black,solid,->]  (yell2) to node[midway,above] {}   (circ4) ; 
\draw[draw=black,solid,->]  (yell2) to node[midway,above] {}   (circ1) ; 

\draw[draw=black,solid,<-]  (yell1) to node[midway,above] {}   (sqnode) ; 
\draw[draw=black,solid,<-]  (yell2) to node[midway,above] {}   (sqnode) ;
\draw[draw=black,solid,->]  (yell3) to node[midway,above] {}   (sqnode) ; 
\draw[draw=black,solid,<-]  (yell4) to node[midway,above] {}   (sqnode) ;

\draw[draw=blue,solid,->]  (circ1) to node[midway,above] {}   (sqnode) ; 
\draw[draw=blue,solid,->]  (circ4) to node[midway,above] {}   (sqnode) ;
\end{tikzpicture}} \\ & \\
\scalebox{0.5}{
\begin{tikzpicture}[baseline]
\tikzstyle{every node}=[font=\footnotesize]
\node[draw, rectangle] (sqnode) at (0,0) {$2N$};
\node[draw, circle, fill=yellow] (yell1) at (-2,1.25) {};
\node[draw, circle, fill=yellow] (yell2) at (2,1.25) {};
\node[draw, circle, fill=yellow] (yell4) at (2,-1.25) {};
\node[draw, circle, fill=yellow] (yell3) at (-2,-1.25) {};
\node[draw, circle, dashed] (circ1) at (0,2.5) {$N_{k_1}$};
\node[draw, circle, dashed] (circ2) at (-4,0) {$N_{k_2}$};
\node[draw, circle, dashed] (circ4) at (4,0) {$N_{k_4}$};
\node[draw, circle, dashed] (circ3) at (0,-2.5) {$N_{k_3}$};
\draw[draw=black,solid,->]  (yell1) to node[midway,above] {}   (circ1) ; 
\draw[draw=black,solid,->]  (yell1) to node[midway,above] {}   (circ2) ; 
\draw[draw=black,solid,->]  (yell3) to node[midway,above] {}   (circ2) ; 
\draw[draw=black,solid,->]  (yell3) to node[midway,above] {}   (circ3) ; 
\draw[draw=black,solid,<-]  (yell4) to node[midway,above] {}   (circ3) ; 
\draw[draw=black,solid,<-]  (yell4) to node[midway,above] {}   (circ4) ;
\draw[draw=black,solid,->]  (yell2) to node[midway,above] {}   (circ4) ; 
\draw[draw=black,solid,->]  (yell2) to node[midway,above] {}   (circ1) ; 

\draw[draw=black,solid,<-]  (yell1) to node[midway,above] {}   (sqnode) ; 
\draw[draw=black,solid,<-]  (yell2) to node[midway,above] {}   (sqnode) ;
\draw[draw=black,solid,<-]  (yell3) to node[midway,above] {}   (sqnode) ; 
\draw[draw=black,solid,->]  (yell4) to node[midway,above] {}   (sqnode) ;

\draw[draw=blue,solid,->]  (circ1) to node[midway,above] {}   (sqnode) ; 
\draw[draw=blue,solid,->]  (circ2) to node[midway,above] {}   (sqnode) ;
\end{tikzpicture}}
& \qquad \qquad 
\scalebox{0.5}{
\begin{tikzpicture}[baseline]
\tikzstyle{every node}=[font=\footnotesize]
\node[draw, rectangle] (sqnode) at (0,0) {$2N$};
\node[draw, circle, fill=yellow] (yell1) at (-2,1.25) {};
\node[draw, circle, fill=yellow] (yell2) at (2,1.25) {};
\node[draw, circle, fill=yellow] (yell4) at (2,-1.25) {};
\node[draw, circle, fill=yellow] (yell3) at (-2,-1.25) {};
\node[draw, circle, dashed] (circ1) at (0,2.5) {$N_{k_1}$};
\node[draw, circle, dashed] (circ2) at (-4,0) {$N_{k_2}$};
\node[draw, circle, dashed] (circ4) at (4,0) {$N_{k_4}$};
\node[draw, circle, dashed] (circ3) at (0,-2.5) {$N_{k_3}$};
\draw[draw=black,solid,->]  (yell1) to node[midway,above] {}   (circ1) ; 
\draw[draw=black,solid,->]  (yell1) to node[midway,above] {}   (circ2) ; 
\draw[draw=black,solid,->]  (yell3) to node[midway,above] {}   (circ2) ; 
\draw[draw=black,solid,->]  (yell3) to node[midway,above] {}   (circ3) ; 
\draw[draw=black,solid,->]  (yell4) to node[midway,above] {}   (circ3) ; 
\draw[draw=black,solid,->]  (yell4) to node[midway,above] {}   (circ4) ;
\draw[draw=black,solid,<-]  (yell2) to node[midway,above] {}   (circ4) ; 
\draw[draw=black,solid,<-]  (yell2) to node[midway,above] {}   (circ1) ; 

\draw[draw=black,solid,<-]  (yell1) to node[midway,above] {}   (sqnode) ; 
\draw[draw=black,solid,->]  (yell2) to node[midway,above] {}   (sqnode) ;
\draw[draw=black,solid,<-]  (yell3) to node[midway,above] {}   (sqnode) ; 
\draw[draw=black,solid,<-]  (yell4) to node[midway,above] {}   (sqnode) ;

\draw[draw=blue,solid,->]  (circ2) to node[midway,above] {}   (sqnode) ; 
\draw[draw=blue,solid,->]  (circ3) to node[midway,above] {}   (sqnode) ;
\end{tikzpicture}}
\end{array} 
\ee
The superpotential of each theory contains the basic monopole operators from each yellow node; the cubic terms coming from every closed triangular loop that contains one blue line as an edge; and the quartic terms coming from every closed rectangular loop that does not contain a blue line. 

As a final remark, we point out that in the case of odd number of duality walls, it is not possible to close all external legs in the skeleton diagram.  For example, in \eref{Sglue3walls}, $\hat{\phi}'$ and $\phi$ carry the fugacities $q^{-1} u_i$ and $q u_i$ respectively.  These do not satisfy the gluing condition \eref{condglue} and so we cannot glue the theory along $\hat{\phi}'$ and $\phi$ and hence the external legs cannot be closed.  A way to evade this problem is to use \eref{basicblock1} as a basic building block instead of \eref{basicblock}.  We discuss this in further detail in section \ref{sec:oddblocks}.

\subsection{Using basic building block \eref{basicblock1}}
\subsubsection{Rectangular gluing}
Instead of using \eref{basicblock}, we can perform a $\Phi$ gluing or an $S$-gluing for multiple copies of the building block \eref{basicblock1}.   For example, if we take two copies of \eref{basicblock1} and perform a {\bf $\Phi$-gluing} along $\varphi_{BD}$ in both copies, the resulting theory is 
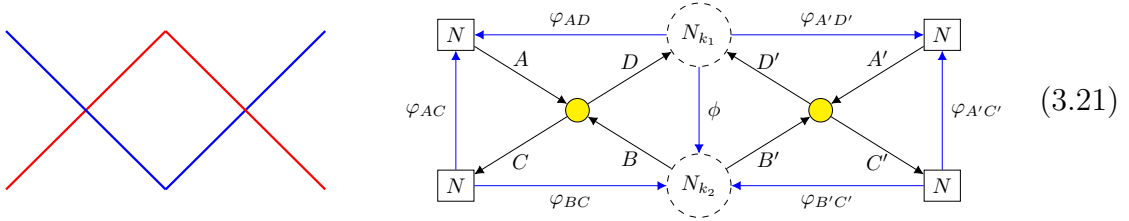
\be \label{ex1Phi}
\scalebox{0.7}{
\begin{tikzpicture}[baseline]
\draw[draw=red,solid, very thick,-] (-1.5,-1.5) to (1.5,1.5);
\draw[draw=blue,solid, very thick,-] (-1.5,1.5) to (1.5,-1.5);
\draw[draw=blue,solid, very thick,-] (-1.5+3,-1.5) to (1.5+3,1.5);
\draw[draw=red,solid, very thick,-] (-1.5+3,1.5) to (1.5+3,-1.5);
\end{tikzpicture}}
\qquad
\scalebox{0.8}{
\begin{tikzpicture}[baseline]
\tikzstyle{every node}=[font=\footnotesize]
\node[draw, rectangle] (sqnode1a) at (-4,1.25) {$N$};
\node[draw, rectangle] (sqnode1b) at (-4,-1.25) {$N$};
\node[draw, circle, fill=yellow] (yell1) at (-2,0) {};
\node[draw, circle, dashed] (circ1) at (0,1.25) {$N_{k_1}$};
\node[draw, circle, dashed] (circ2) at (0,-1.25) {$N_{k_2}$};
\node[draw, circle, fill=yellow] (yell2) at (2,0) {};
\node[draw, rectangle] (sqnode2a) at (4,1.25) {$N$};
\node[draw, rectangle] (sqnode2b) at (4,-1.25) {$N$};
\draw[draw=black,solid,->]  (sqnode1a) to node[midway,above] {$A$}   (yell1) ; 
\draw[draw=black,solid,<-]  (sqnode1b) to node[midway,below] {$C$}   (yell1) ; 
\draw[draw=black,solid,->]  (yell1) to node[midway,above] {$D$}   (circ1) ; 
\draw[draw=black,solid,<-]  (yell1) to node[midway,below] {$B$}   (circ2) ; 
\draw[draw=black,solid,<-]  (circ1) to node[midway,above] {$D'$}   (yell2) ; 
\draw[draw=black,solid,->]  (circ2) to node[midway,below] {$B'$}   (yell2) ; 
\draw[draw=black,solid,<-]  (yell2) to node[midway,above] {$A'$}   (sqnode2a) ; 
\draw[draw=black,solid,->]  (yell2) to node[midway,below] {$C'$}   (sqnode2b) ; 
\draw[draw=blue,solid,<-]  (sqnode1a) to node[midway,above] {$\varphi_{AD}$}   (circ1) ; 
\draw[draw=blue,solid,<-]  (sqnode1a) to node[midway,left] {$\varphi_{AC}$}   (sqnode1b) ; 
\draw[draw=blue,solid,->]  (sqnode1b) to node[midway,below] {$\varphi_{BC}$}   (circ2) ;
\draw[draw=blue,solid,<-]  (sqnode2a) to node[midway,above] {$\varphi_{A'D'}$}   (circ1) ;
\draw[draw=blue,solid,->]  (sqnode2b) to node[midway,below] {$\varphi_{B'C'}$}   (circ2) ;   
\draw[draw=blue,solid,<-]  (sqnode2a) to node[midway,right] {$\varphi_{A'C'}$}   (sqnode2b) ; 
\draw[draw=blue,solid,->]  (circ1) to node[midway,right] {$\phi$}   (circ2) ; 
\end{tikzpicture}}
\ee
with superpotential containing the basic monopole operators from both yellow nodes and the cubic terms coming from every closed triangular loop in the quiver that contains one blue line. 
Upon gluing, we have gauged the upper and lower $SU(N)$ symmetries with CS levels $k_1$ and $k_2$ respectively.   In the skeleton diagram, for the $\Phi$-gluing,  a blue ({\it resp.} red) line joins with another blue ({\it resp.} red) line.  Topologically, the skeleton diagram has genus 1, as well as 2 red and 2 blue external legs.

Let us now consider the {\bf $S$-gluing}.  We take two copies of \eref{basicblock1} and glue them along $\varphi_{BD}$ of one copy and $\varphi_{AC}$ of the other copy.  As a result we obtain 
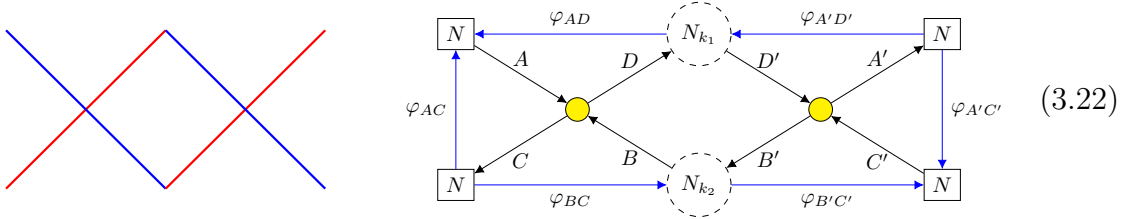
\be \label{ex1S}
\scalebox{0.7}{
\begin{tikzpicture}[baseline]
\draw[draw=red,solid, very thick,-] (-1.5,-1.5) to (1.5,1.5);
\draw[draw=blue,solid, very thick,-] (-1.5,1.5) to (1.5,-1.5);
\draw[draw=red,solid, very thick,-] (-1.5+3,-1.5) to (1.5+3,1.5);
\draw[draw=blue,solid, very thick,-] (-1.5+3,1.5) to (1.5+3,-1.5);
\end{tikzpicture}}
\qquad
\scalebox{0.8}{
\begin{tikzpicture}[baseline]
\tikzstyle{every node}=[font=\footnotesize]
\node[draw, rectangle] (sqnode1a) at (-4,1.25) {$N$};
\node[draw, rectangle] (sqnode1b) at (-4,-1.25) {$N$};
\node[draw, circle, fill=yellow] (yell1) at (-2,0) {};
\node[draw, circle, dashed] (circ1) at (0,1.25) {$N_{k_1}$};
\node[draw, circle, dashed] (circ2) at (0,-1.25) {$N_{k_2}$};
\node[draw, circle, fill=yellow] (yell2) at (2,0) {};
\node[draw, rectangle] (sqnode2a) at (4,1.25) {$N$};
\node[draw, rectangle] (sqnode2b) at (4,-1.25) {$N$};
\draw[draw=black,solid,->]  (sqnode1a) to node[midway,above] {$A$}   (yell1) ; 
\draw[draw=black,solid,<-]  (sqnode1b) to node[midway,below] {$C$}   (yell1) ; 
\draw[draw=black,solid,->]  (yell1) to node[midway,above] {$D$}   (circ1) ; 
\draw[draw=black,solid,<-]  (yell1) to node[midway,below] {$B$}   (circ2) ; 
\draw[draw=black,solid,->]  (circ1) to node[midway,above] {$D'$}   (yell2) ; 
\draw[draw=black,solid,<-]  (circ2) to node[midway,below] {$B'$}   (yell2) ; 
\draw[draw=black,solid,->]  (yell2) to node[midway,above] {$A'$}   (sqnode2a) ; 
\draw[draw=black,solid,<-]  (yell2) to node[midway,below] {$C'$}   (sqnode2b) ; 
\draw[draw=blue,solid,<-]  (sqnode1a) to node[midway,above] {$\varphi_{AD}$}   (circ1) ; 
\draw[draw=blue,solid,<-]  (sqnode1a) to node[midway,left] {$\varphi_{AC}$}   (sqnode1b) ; 
\draw[draw=blue,solid,->]  (sqnode1b) to node[midway,below] {$\varphi_{BC}$}   (circ2) ;
\draw[draw=blue,solid,->]  (sqnode2a) to node[midway,above] {$\varphi_{A'D'}$}   (circ1) ;
\draw[draw=blue,solid,<-]  (sqnode2b) to node[midway,below] {$\varphi_{B'C'}$}   (circ2) ;   
\draw[draw=blue,solid,->]  (sqnode2a) to node[midway,right] {$\varphi_{A'C'}$}   (sqnode2b) ; 
\end{tikzpicture}}
\ee
The superpotential of the resulting theory contains the basic monopole operators from each yellow node; the cubic terms coming from every closed triangular loop in the quiver that contains one blue line; and the quartic term $D D' B' B$ coming from the middle rectangular loop.  In the skeleton diagram, for the $S$-gluing,  a blue ({\it resp.} red) line joins with another red ({\it resp.} blue) line -- this is opposite to the $\Phi$-gluing.  

Observe that as a result of such gluing, which involves {\bf two pairs} of external legs at the same time, we end up with a rectangle in the skeleton diagram.  We will refer to these types of gluing as {\bf rectangular $\Phi$-gluing} and {\bf rectangular $S$-gluing} respectively.  There is also another type of gluing which is not a rectangular gluing.  For example, one may self-glue the left part of the skeleton diagram of \eref{basicblock1} to obtain
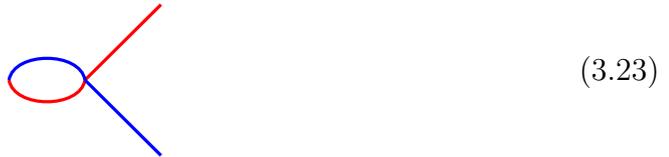
\be
\begin{tikzpicture}[baseline]
\draw[draw=red,solid, very thick,-] (-1,0) to[bend right=80] (0,0);
\draw[draw=red,solid, very thick,-] (0,0) to (1,1);
\draw[draw=blue,solid, very thick,-] (-1,0) to[bend left=80]  (0,0);
\draw[draw=blue,solid, very thick,-] (0,0) to (1,-1);
\end{tikzpicture}
\ee
First of all, the ``loop'' on the left is not rectangular.   Secondly, this type of gluing involves only one pair of external legs, not two pairs as for the rectangular gluing.  We postpone the discussion of the non-rectangular gluing until later.

Theories \eref{ex1Phi} and \eref{ex1S} will be analysed in detail in section \ref{sec:twowallsalternative}.

\paragraph{Gluing amusement.} As a final remark, we can further perform a rectangular self-$\Phi$-gluing on \eref{ex1Phi} such that the blue ({\it resp.} red) external leg on the left is joined with the blue ({\it resp.} red) external leg on the right.  As a result, we obtain the skeleton diagram (as well as the quiver diagram) whose topology is an ``strip'', whose face containing two rectangles.  Similarly, we can further perform a rectangular self-$S$-gluing on \eref{ex1S} such that the blue external legs are joined with the red ones.  The topology of the diagram is also a strip, but with half of the face ``flipped'' with respect to the former.

\subsubsection{Odd number of basic building blocks} \label{sec:oddblocks}
As we have discussed in the paragraph below \eref{fourwallsquad}, it is not possible to close all external legs for odd number of duality walls, provided that we use \eref{basicblock} as a basic building block.  This can also be seen in the case of one duality wall.  In particular, it is not possible to perform the following self-gluing:
\be
\scalebox{0.8}{
\begin{tikzpicture}[baseline]
\tikzstyle{every node}=[font=\footnotesize, node distance=0.45cm]
\tikzset{decoration={snake,amplitude=.4mm,segment length=2mm,
                       post length=0mm,pre length=0mm}}
\draw[blue,thick] (0,0) circle (1cm) node[midway, right] {};  
\draw[decorate,red,thick] (0.5,0) -- (1.5,0) node[right] {}; 
\end{tikzpicture}}
\ee
This is because none of the conditions in \eref{condglue} is satisfied, since $\phi$ carries a fugacity $q\,u_i$, whereas $\phi'$ carries a fugacity $q^{-1}\,u_i$).  However, if we instead use \eqref{basicblock1} as a basic building block, we can perform a {\bf rectangular (self-)$S$-gluing} along the opposite blue edges, namely along $(\varphi_{AD}, \varphi_{BC})$ or along $(\varphi_{AC}, \varphi_{BD})$.  For definiteness, let us consider the former option. In terms of the skeleton diagram, we can identify the left ({\it resp.} right) blue external leg with the left ({\it resp.} right) red external leg.  As a result, we obtain
\be \label{twoloops}
\begin{tikzpicture}[baseline]
\draw[draw=red,solid, very thick,-] (-1,0) to[bend right=80] (0,0);
\draw[draw=red,solid, very thick,-] (0,0) to[bend left=80]  (1,0);
\draw[draw=blue,solid, very thick,-] (-1,0) to[bend left=80]  (0,0);
\draw[draw=blue,solid, very thick,-] (0,0) to[bend right=80]  (1,0);
\end{tikzpicture}
\qquad \qquad
\scalebox{0.7}{
\begin{tikzpicture}[baseline]
\tikzstyle{every node}=[font=\footnotesize]
\node[draw, circle, fill=yellow] (node1) at (0,0) {};
\node[draw, circle, dashed] (sqnode1) at (2,0) {$N_{k_2}$};
\node[draw, circle, dashed] (circ) at (-2,0) {$N_{k_1}$};
\draw[draw=blue,solid, <-] (circ) edge [out=135,in=-135,loop,looseness=5] node[midway,left]{$\varphi_{AC}$}  (circ);
\draw[draw=black,solid,->]  (circ) to[bend left=60]  node[midway,above] {$A$}   (node1) ; 
\draw[draw=black,solid,<-]  (circ) to[bend right=60] node[midway,below] {$C$}  (node1) ; 
\draw[draw=black,solid,->]  (node1)to[bend left=60]  node[midway,above] {$B$} (sqnode1) ; 
\draw[draw=black,solid,<-]  (node1) to[bend right=60]  node[midway,below] {$D$} (sqnode1) ;  
\draw[draw=blue,solid,->]  (sqnode1) edge [out=45,in=-45,loop,looseness=5] node[midway,right]{$\varphi_{BD}$} (sqnode1) ;  
\node[draw=none] at (0,-2) {$W=V_+ + V_- + C\varphi_{AC} A+ B \varphi_{BD} D$};
\end{tikzpicture}}
\ee
Note that the skeleton diagram is rectangular in the sense that it has four sides.  Also, since this gluing involves two pairs of external legs at the same time, it is qualified as a rectangular gluing. Moreover, the blue lines that connect $SU(N)_{k_1}$ and $SU(N)_{k_2}$ disappear because we have performed an $S$-gluing.  We can further apply duality \eref{mainduality} to the yellow node of the quiver \eref{twoloops} and obtain the following dual theory:
\be \label{dualtwoloops}
\eref{twoloops} \quad \overset{\eref{mainduality}}{\longleftrightarrow} \quad
\scalebox{0.7}{
\begin{tikzpicture}[baseline]
\tikzstyle{every node}=[font=\footnotesize]
\node[draw, circle, fill=yellow] (node1) at (0,0) {};
\node[draw, circle, dashed] (sqnode1) at (2,0) {$N_{k_2}$};
\node[draw, circle, dashed] (circ) at (-2,0) {$N_{k_1}$};
\draw[draw=black,solid,->]  (circ) to[bend left=60]  node[midway,below] {}   (node1) ; 
\draw[draw=black,solid,<-]  (circ) to[bend right=60] node[midway,above] {}  (node1) ; 
\draw[draw=black,solid,->]  (node1)to[bend left=60]  node[midway,above] {} (sqnode1) ; 
\draw[draw=black,solid,<-]  (node1) to[bend right=60]  node[midway,above] {} (sqnode1) ;  
\draw[draw=blue,solid,<-]  (circ) to [bend left=70] node[midway,above] {}  (sqnode1) ;
\draw[draw=blue,solid,->]  (circ) to [bend right=70] node[midway,below] {}  (sqnode1) ;
\end{tikzpicture}}
\ee
with the monopole superpotential and the two cubic terms coming from the upper and lower triangular loops.
We further explore these theories in section \ref{sec:genus2noleg}, where we find two more dual theories.  These four theories are then related to each other by a quadrality as shown in \eref{quadralitygenus2noleg}.

\subsubsection{Non-rectangular gluing}
Let us now consider a closure of one pair of external legs.  We propose the following prescription:
\be \label{fish} 
\begin{tikzpicture}[baseline]
\draw[draw=red,solid, very thick,-] (-1,0) to[bend right=80] (0,0);
\draw[draw=red,solid, very thick,-] (0,0) to (1,1);
\draw[draw=blue,solid, very thick,-] (-1,0) to[bend left=80]  (0,0);
\draw[draw=blue,solid, very thick,-] (0,0) to (1,-1);
\end{tikzpicture}
\qquad \qquad 
\scalebox{0.8}{
\begin{tikzpicture}[baseline]
\tikzstyle{every node}=[font=\footnotesize]
\node[draw, circle, fill=yellow] (node1) at (0,0) {};
\node[draw, rectangle] (sqnode1) at (2,1) {$N$};
\node[draw, rectangle] (sqnode2) at (2,-1) {$N$};
\node[draw, circle, dashed] (circ) at (-2,0) {$N_k$};
\draw[draw=blue,solid, <-] (circ) edge [out=135,in=-135,loop,looseness=5] node[midway,left]{$\varphi_{AC}$}  (circ);
\draw[draw=black,solid,->]  (circ) to[bend left=60]  node[midway,below] {$A$}   (node1) ; 
\draw[draw=black,solid,<-]  (circ) to[bend right=60] node[midway,above] {$C$}  (node1) ; 
\draw[draw=black,solid,->]  (node1) to node[midway,above] {$D$} (sqnode1) ; 
\draw[draw=black,solid,<-]  (node1) to node[midway,above] {$B$} (sqnode2) ;  
\draw[draw=blue,solid,->]  (sqnode1) to node[midway,right] {$\varphi_{BD}$} (sqnode2) ;  
\node[draw=none] at (0,-2) {$W=V_+ + V_- + C \varphi_{AC} A + D \varphi_{BD} B$};
\end{tikzpicture}}
\ee
When a pair of external legs is glued together, the corresponding $SU(N)$ flavour symmetries associated to those legs are commonly gauged with a certain CS level $k$.  The 4d fields that was connecting the two $SU(N)$ flavour symmetries becomes an adjoint field and a singlet under the gauge group $SU(N)_k$ (this is $\varphi_{AC}$ in the above example).  We also remove the 4d fields connecting the $SU(N)_k$ gauge groups to other $SU(N)$ flavour symmetries (hence $\varphi_{AD}$ and $\varphi_{BC}$ are absent in the above example).

The reason we proposed such a prescription for the non-rectangular gluing is the consistency with \eref{twoloops}.  Observe that when we also close the right pair of external legs in \eref{fish}, we obtain precisely \eref{twoloops}.  

Notice also that the above prescription for closing a pair of external legs {\it commutes} with duality \eref{mainduality}.  In \eref{dualtwoloops}, we first closed all external legs and then applied duality \eref{mainduality} to the yellow node to obtain the right quiver diagram.  Now suppose that we first apply duality \eref{mainduality} to the yellow node in \eref{fish} to obtain\footnote{We emphasise that, upon applying duality \eref{mainduality}, all black arrows in \eref{dualfish} have to be reversed with respect to those in \eref{fish}. (The directions of the blue arrows are then fixed.) However, since the quiver has a horizontal symmetry, we draw the quiver as it is in \eref{dualfish}. One should keep in mind that the roles of the upper and lower nodes in \eref{dualfish} are reversed with respect to those of \eref{fish}.} 
\be \label{dualfish} 
\eref{fish} \quad \longleftrightarrow \quad
\scalebox{0.7}
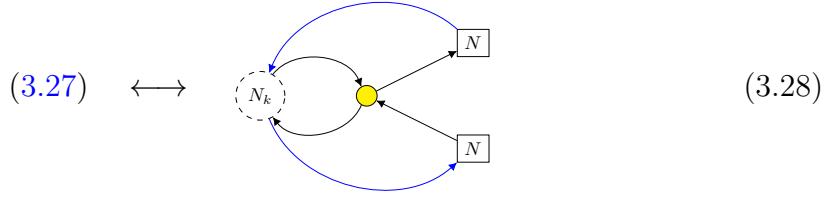
{
\begin{tikzpicture}[baseline]
\tikzstyle{every node}=[font=\footnotesize]
\node[draw, circle, fill=yellow] (node1) at (0,0) {};
\node[draw, rectangle] (sqnode1) at (2,1) {$N$};
\node[draw, rectangle] (sqnode2) at (2,-1) {$N$};
\node[draw, circle, dashed] (circ) at (-2,0) {$N_k$};
\draw[draw=black,solid,->]  (circ) to[bend left=60]  node[midway,below] {}   (node1) ; 
\draw[draw=black,solid,<-]  (circ) to[bend right=60] node[midway,above] {}  (node1) ; 
\draw[draw=black,solid,->]  (node1) to node[midway,above] {} (sqnode1) ; 
\draw[draw=black,solid,<-]  (node1) to node[midway,above] {} (sqnode2) ;  
\draw[draw=blue,solid,<-]  (circ) to [bend left=55] node[midway,above] {}  (sqnode1) ;
\draw[draw=blue,solid,->]  (circ) to [bend right=55] node[midway,below] {}  (sqnode2) ;
\end{tikzpicture}}
\ee
with the monopole superpotential and the two cubic terms coming from the upper and lower triangular loops.
Upon closing the right pair of external legs using the above prescription, one obtain precisely the quiver in \eref{dualtwoloops}.  In section \ref{sec:singlewallnoleg}, we analyse \eref{fish} and \eref{dualfish} in more detail.

This prescription can, of course, be applied to a more complicated theory.  For example, we have
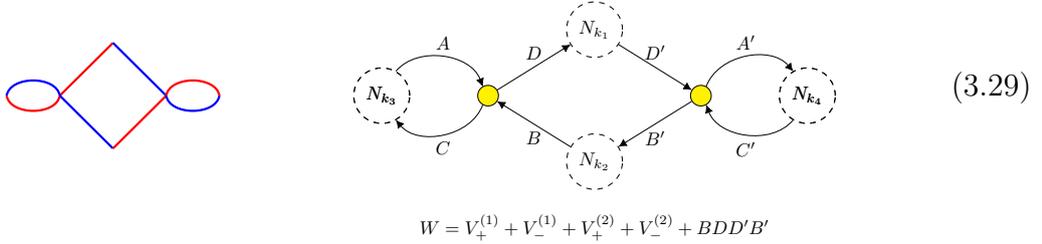
\be
\scalebox{0.7}{
\begin{tikzpicture}[baseline]
\draw[draw=red,solid, very thick,-] (-1,0) to[bend right=80] (0,0);
\draw[draw=red,solid, very thick,-] (0,0) to (1,1);
\draw[draw=blue,solid, very thick,-] (-1,0) to[bend left=80]  (0,0);
\draw[draw=blue,solid, very thick,-] (0,0) to (1,-1);
\draw[draw=red,solid, very thick,-] (-1+2,-1) to (2,0);
\draw[draw=blue,solid, very thick,-] (-1+2,1) to (2,0);
\draw[draw=blue,solid, very thick,-] (2,0) to[bend right=80] (3,0);
\draw[draw=red,solid, very thick,-] (2,0) to[bend left=80]  (3,0);
\end{tikzpicture}}
\qquad \qquad
\scalebox{0.7}{
\begin{tikzpicture}[baseline]
\tikzstyle{every node}=[font=\footnotesize]
\node[draw, circle, dashed] (sqnode1a) at (-4,0) {$N_{k_3}$};
\node[draw, circle, dashed] (sqnode1b) at (-4,0) {$N_{k_3}$};
\node[draw, circle, fill=yellow] (yell1) at (-2,0) {};
\node[draw, circle, dashed] (circ1) at (0,1.25) {$N_{k_1}$};
\node[draw, circle, dashed] (circ2) at (0,-1.25) {$N_{k_2}$};
\node[draw, circle, fill=yellow] (yell2) at (2,0) {};
\node[draw, circle, dashed] (sqnode2a) at (4,0) {$N_{k_4}$};
\node[draw, circle, dashed] (sqnode2b) at (4,0) {$N_{k_4}$};
\draw[draw=black,solid,->]  (sqnode1a) to[bend left=60] node[midway,above] {$A$}   (yell1) ; 
\draw[draw=black,solid,<-]  (sqnode1b) to[bend right=60] node[midway,below] {$C$}   (yell1) ; 
\draw[draw=black,solid,->]  (yell1) to node[midway,above] {$D$}   (circ1) ; 
\draw[draw=black,solid,<-]  (yell1) to node[midway,below] {$B$}   (circ2) ; 
\draw[draw=black,solid,->]  (circ1) to node[midway,above] {$D'$}   (yell2) ; 
\draw[draw=black,solid,<-]  (circ2) to node[midway,below] {$B'$}   (yell2) ; 
\draw[draw=black,solid,->]  (yell2) to[bend left=60] node[midway,above] {$A'$}   (sqnode2a) ; 
\draw[draw=black,solid,<-]  (yell2) to[bend right=60] node[midway,below] {$C'$}   (sqnode2b) ; 
\node[draw=none] at (0,-2.5) {$W= V^{(1)}_+ +  V^{(1)}_- + V^{(2)}_+ +  V^{(2)}_- + B D D' B'$};
\end{tikzpicture}}
\ee

\subsection{Comparison with the gluing prescription in \cite{Kim:2017toz}} \label{sec:motivationglue}
As mentioned earlier, the gluing prescription adopted in this paper is heavily motivated by that used in \cite{Kim:2017toz}.  The reason why we adopted the latter is due to the similarity of our construction and \cite{Kim:2017toz}.  

Let us first briefly summarise the construction of \cite{Kim:2017toz}. In that reference, the basic building block arises from the 6d $E$-string theory compactified on a sphere with two punctures (a tube) with a particular choice of flux that breaks the $E_8$ symmetry of the $E$-string theory to $E_7 \times U(1)_F$.  The latter was then realised from the 5d $E$-string theory with a duality domain wall \cite{Gaiotto:2015una}, which gives rise to a subgroup $SU(8) \times U(1)_F$ of the former symmetry.  The $U(1)_F$ charge on one side of the domain wall flips its sign as we cross to the other side.

We now turn to our construction in this paper.  We consider duality domain walls in 4d $\CN=2$ $SU(N)$ gauge theory with $2N$ flavours.  In this case, the duality wall gives rise to a symmetry $SU(2N) \times U(1)_q$, which is also the flavour symmetry of the 4d theory.  The analog of $U(1)_F$ in \cite{Kim:2017toz} is indeed $U(1)_q$ in this paper.  As we explained around \eref{basicblock}, each of the two $SU(N)$ flavour symmetries are coupled to the $SU(N)$ gauge symmetry of the 4d theory on each side of the wall.  Since $A$ and $B$ as well as $\phi$ and $\phi'$ carry opposite charges under $U(1)_q$, we see that, indeed, the $U(1)_q$ charge on the left flips its sign on the right of the duality wall.  The cubic superpotential terms also appear in the same way as described in \cite{Kim:2017toz}.

Although we do not have a realisation of our theory as coming from a 5d theory on a Riemann surface (analog of 6d $E$-string theory on a Riemann surface in \cite{Kim:2017toz}), we have a very similar geometric analog of the Riemann surface, namely the skeleton diagram. The genus and the external legs of the latter play the same roles as the genus and the puncture of the Riemann surface in \cite{Kim:2017toz}.  It would be nice to understand the theory studied in this paper as coming from compactification of a higher dimensional theory.  We postpone this to future work.

\section{A single duality wall}  \label{sec:singlewall}
In this section we consider the case of a single duality wall, whose skeleton diagram has genus one.  We first discuss theories with two external legs and then move on to those with zero external legs and genus two.  We study indices of such theories and discuss various dualities among them. 

\subsection{Two external legs} \label{sec:singlewallnoleg}
We have already introduced two dual theories associated with the skeleton diagram with genus one and two external legs, namely \eref{fish} and \eref{dualfish}.  In this subsection, we introduce two more theories that are closely related to the former.  The first one is
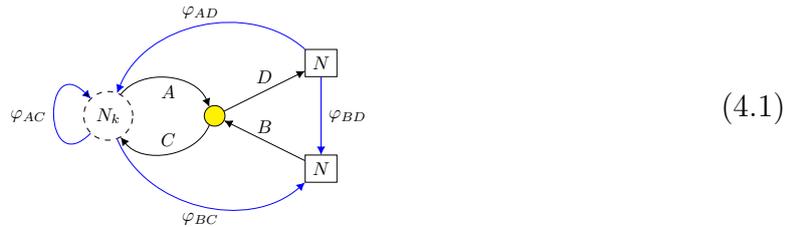
\be \label{modelI}
\scalebox{0.7}{
\begin{tikzpicture}[baseline]
\tikzstyle{every node}=[font=\footnotesize]
\node[draw, circle, fill=yellow] (node1) at (0,0) {};
\node[draw, rectangle] (sqnode1) at (2,1) {$N$};
\node[draw, rectangle] (sqnode2) at (2,-1) {$N$};
\node[draw, circle, dashed] (circ) at (-2,0) {$N_k$};
\draw[draw=blue,solid, <-] (circ) edge [out=135,in=-135,loop,looseness=5] node[midway,left]{$\varphi_{AC}$}  (circ);
\draw[draw=black,solid,->]  (circ) to[bend left=60]  node[midway,below] {$A$}   (node1) ; 
\draw[draw=black,solid,<-]  (circ) to[bend right=60] node[midway,above] {$C$}  (node1) ; 
\draw[draw=black,solid,->]  (node1) to node[midway,above] {$D$} (sqnode1) ; 
\draw[draw=black,solid,<-]  (node1) to node[midway,above] {$B$} (sqnode2) ;  
\draw[draw=blue,solid,->]  (sqnode1) to node[midway,right] {$\varphi_{BD}$} (sqnode2) ;  
\draw[draw=blue,solid,<-]  (circ) to [bend left=55] node[midway,above] {\color{black} $\varphi_{AD}$}  (sqnode1) ;
\draw[draw=blue,solid,->]  (circ) to [bend right=55] node[midway,below] {\color{black} $\varphi_{BC}$}  (sqnode2) ;
\node[draw=none] at (0,-3) {\large $W=V_+ + V_- + C \varphi_{AC} A + D \varphi_{BD} B + D \varphi_{AD} A +C \varphi_{BC} B$};
\end{tikzpicture}}
\ee

To obtain the second theory, we apply duality \eref{mainduality} to the yellow node.  We get rid of $\varphi_{AD}$,  $\varphi_{BC}$, $\varphi_{BD}$ and $\varphi_{AC}$, and reverse all the black arrows.  However, since the quiver has a horizontal symmetry, we can draw the quiver for the dual theory as follows:
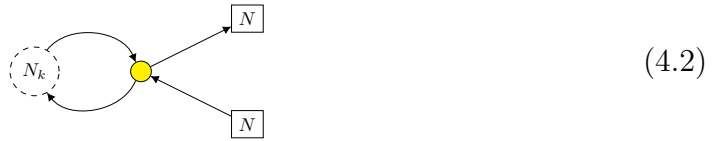
\be \label{modelII}
\scalebox{0.7}{
\begin{tikzpicture}[baseline]
\tikzstyle{every node}=[font=\footnotesize]
\node[draw, circle, fill=yellow] (node1) at (0,0) {};
\node[draw, rectangle] (sqnode1) at (2,1) {$N$};
\node[draw, rectangle] (sqnode2) at (2,-1) {$N$};
\node[draw, circle, dashed] (circ) at (-2,0) {$N_k$};
\draw[draw=black,solid,->]  (circ) to[bend left=60]  node[midway,below] {}   (node1) ; 
\draw[draw=black,solid,<-]  (circ) to[bend right=60] node[midway,above] {}  (node1) ; 
\draw[draw=black,solid,->]  (node1) to node[midway,above] {} (sqnode1) ; 
\draw[draw=black,solid,<-]  (node1) to node[midway,above] {} (sqnode2) ;  
\node[draw=none] at (0,-2) {\large $W=V_+ + V_-$};
\end{tikzpicture}}
\ee
where we emphasise that the roles of the upper and lower flavour nodes are reversed with respect to that of \eref{modelI}.

Let us summarise the four closely related theories:
\be \label{SS}
\begin{split}
\eref{modelI}~ &\overset{\eref{mainduality}}{\longleftrightarrow}~ \eref{modelII} \\
\eref{fish}~ &\overset{\eref{mainduality}}{\longleftrightarrow}~ \eref{dualfish}
\end{split}
\ee
where each pair is related by the duality \eref{mainduality}.  We shall discuss in the next subsection that, for $N=2$, the four theories are, in fact, dual to each other.  However, for $N>2$, the theories in the first line are not dual to those in the second line.

\subsubsection{Quadrality for the case of $N=2$} 
In the special case of $N=2$, as we shall discuss below, the indices of the four models in \eref{SS} are equal.  We thus conjecture that the four models are related by a quadrality:
\be \label{quadrality}
\begin{array}{ccc}
\eref{modelI}~ &\overset{\eref{mainduality}}{\longleftrightarrow}~ \eref{modelII}  ~\overset{\text{for $N=2$}}{\longleftrightarrow}~ \eref{dualfish}~ \overset{\eref{mainduality}}{\longleftrightarrow}~ \eref{fish}
\end{array}
\ee

\subsubsection*{The indices for the theories in \eref{quadrality} with $N=2$}
Let us first fix the convention in drawing the quivers in \eref{quadrality}.  All black lines in every model in \eref{quadrality} are drawn in the following way and carry the following $U(1)_p \times U(1)_q$ fugacities:
\be
\scalebox{0.7}{
\begin{tikzpicture}[baseline]
\tikzstyle{every node}=[font=\footnotesize]
\node[draw, circle, fill=yellow] (node1) at (0,0) {};
\node[draw, rectangle] (sqnode1) at (2,1) {$N$};
\node[draw, rectangle] (sqnode2) at (2,-1) {$N$};
\node[draw, circle, dashed] (circ) at (-2,0) {$N_k$};
\draw[draw=black,solid,->]  (circ) to[bend left=60]  node[midway,above] {$p^{-1}$}  node[midway,below] {}   (node1) ; 
\draw[draw=black,solid,<-]  (circ) to[bend right=60] node[midway,below] {$q$} node[midway,above] {}  (node1) ; 
\draw[draw=black,solid,->]  (node1) to node[midway,above] {$q^{-1}$} node[midway,below] {} (sqnode1) ; 
\draw[draw=black,solid,<-]  (node1) to node[midway,below] {$p$} node[midway,above] {} (sqnode2) ;  
\end{tikzpicture}}
\ee
The $U(1)_p \times U(1)_q$ charges of the chiral fields corresponding to the blue line then follow from the superpotential.  For example, $\varphi_{BD}$ in \eref{modelI} carries the $U(1)_p \times U(1)_q$ fugacity $p^{-1} q$.

We first examine theory \eref{modelI}. For $N=2$ and the CS level $k \geq 2$\footnote{Throughout this paper, we take  the CS level to be generic; unless specified otherwise, we take its absolute value to be larger than or equal to $2$.  When the CS level is taken to be $0$ or $1$, for example, the index may diverge depending on the cases we are considering.}, the index reads
\be \label{indexmodel1}
\begin{split}
&\CI^{N=2}_{\eref{modelI}}(x; \vec y, \vec z, p, q) \\
&= 1+ C_1(\vec y, \vec z, p, q) x + C_2(\vec y, \vec z, p, q) x^2 + C_3(\vec y, \vec z, p, q) x^3 + \ldots~.
\end{split}
\ee
where the coefficients $C_1$, $C_2$ and $C_3$ are as follows:
\be \label{coeffonewallN2}
\begin{split}
C_1(\vec y, \vec z, p, q) &= p^{-1} q [1;1] + p q^{-1} \\
C_2(\vec y, \vec z, p, q) &= p^{-2} q^2 [2;2] +[1;1] + p^2 q^{-2} -([2;0]+[0;2]+2[0;0])\\
C_3(\vec y, \vec z, p, q) &=  p^{-3} q^3 [3;3]+p^{-1} q [2;2]+2 p q^{-1} [1;1] + p^{-1} q + p^3 q^{-3} \\
& \quad  - p^{-1} q ([1;3]+[3;1]) - p q^{-1} ([2;0]+ [0;2]) - 2 p^{-1} q [1;1] \\
& \quad - p^{-3} q^{-1}[2;0] - 2p q^{-1}~.
\end{split}
\ee
Here we use the shorthand notation $[a;b]$ to denote the characters $\chi^{SU(2)}_{[a]}(\vec y) \chi^{SU(2)}_{[b]}(\vec z)$ of the representation $[a;b]$ of the global symmetry $SU(2) \times SU(2)$, with the first slot $a$ corresponding to the upper node and second slot to the lower node.

We find that the indices of the other theories in \eref{quadrality} are related to that of \eref{indexmodel1} by the following relation:
\be
\begin{split}
\CI^{N=2}_{\eref{modelI}}(x; \vec y, \vec z, p, q) &= \CI^{N=2}_{\eref{modelII}}(x; \vec y, \vec z, p^{-1}, q^{-1})\\
= \CI^{N=2}_{\eref{fish}}(x; \vec z, \vec y, p,q)  &=  \CI^{N=2}_{\eref{dualfish}} (x; \vec z, \vec y, p^{-1},q^{-1})~. 
\end{split}
\ee
This serves as a non-trivial test for the quadrality proposed in \eref{quadrality}.

Let us label the chiral fields in \eref{modelII} and their $U(1)$ charges as follows.
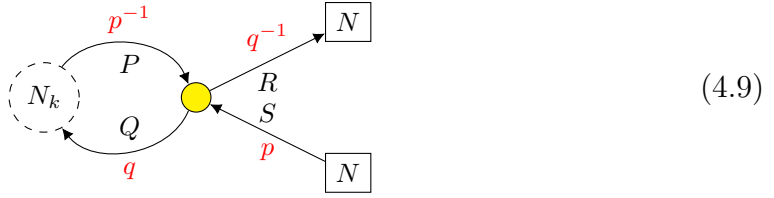
\be \label{quivlabelonewall}
\begin{tikzpicture}[baseline]
\tikzstyle{every node}=[font=\footnotesize]
\node[draw, circle, fill=yellow] (node1) at (0,0) {};
\node[draw, rectangle] (sqnode1) at (2,1) {$N$};
\node[draw, rectangle] (sqnode2) at (2,-1) {$N$};
\node[draw, circle, dashed] (circ) at (-2,0) {$N_k$};
\draw[draw=black,solid,->]  (circ) to[bend left=60]  node[midway,above] {\red $p^{-1}$}  node[midway,below] {$P$}   (node1) ; 
\draw[draw=black,solid,<-]  (circ) to[bend right=60] node[midway,below] {\red $q$} node[midway,above] {$Q$}  (node1) ; 
\draw[draw=black,solid,->]  (node1) to node[midway,above] {\red $q^{-1}$} node[midway,below] {$R$} (sqnode1) ; 
\draw[draw=black,solid,<-]  (node1) to node[midway,below] {\red $p$} node[midway,above] {$S$} (sqnode2) ;  
\end{tikzpicture}
\ee
Recall that for this theory we need to invert $p$ and $q$ in \eref{coeffonewallN2}.  The terms in the coefficient coefficient $C_1$ correspond to the following gauge invariant quantities:
\be
\begin{split}
p q^{-1} [1;1]:& \qquad X_i^{i'} := R_i S^{i'}~, \\  
p^{-1}q: & \qquad Y:= P^\alpha Q_\alpha~.
\end{split}
\ee
where $i, j=1, 2$ and $i',j'=1,2$ are the flavour indices for the upper and lower square nodes respectively, and $\alpha,\beta=1,2$ are the $SU(2)_k$ gauge indices.  These are relevant operators.  The positive terms in $C_2$ correspond to 
\be
\begin{split}
p^2 q^{-2} [2;2]: &\quad X_i^{i'} X_j^{j'}~, \\ 
[1;1]: & \quad X_i^{i'} Y~, \\
p^2 q^{-2}: & \quad Y^2~.
\end{split}
\ee
These are marginal operators.  The negative terms in $C_2$ indicate that the global symmetry is $SU(2)^2 \times U(1)^2$, as is manifest in the quiver diagram.

\subsubsection*{The indices for the case of $N=3$} 
Let us take $N=3$ and $k\geq 2$.  The indices of \eref{modelI} and \eref{modelII} are given by
\be
\begin{split}
 \CI_{ \eref{modelII}} (\vec y, \vec z, p, q)  &= \CI_{\eref{modelI}} (\vec y, \vec z, p^{-1}, q^{-1}) \\
&= 1+ C_1(\vec y, \vec z, p, q) x + C_2(\vec y, \vec z, p, q) x^2  + \ldots~,
\end{split}
\ee
where
\be \label{coeffonewallN3a}
\begin{split}
C_1(\vec y, \vec z, p, q) &=   p q^{-1} [1,0 ; 0,1] + p^{-1} q\\
C_2(\vec y, \vec z, p, q) &= p^2 q^{-2} [2,0; 0,2]+ 2 p^{-2} q^2  +p^2 q^{-2} [0,1; 1,0] +2 [1,0; 0,1]  \\
& \quad  - [1,1;0,0] - [0,0;1,1] -2 - p^2 q^{-2} ~.
\end{split}
\ee
Here the notation $[\mathbf{R}_1; \mathbf{R}_2]$ denote a representation of the $SU(3) \times SU(3)$ flavour symmetry, where the first slot corresponds to the lower node and the second slot corresponds to the upper node of  \eref{modelII} (which becomes the upper and lower nodes of the dual theory \eref{modelI}).
Let us use the notation as in \eref{quivlabelonewall} and take $N=3$.  Now the yellow node is $U(2)$, whose indices will be denoted by $a,b=1,2$.  The terms in the coefficient $C_1$ correspond to
\be \label{rel1}
\begin{split}
p q^{-1} [1,0 ; 0,1]: &\qquad X_i^{i'} := R^a_i S^{i'}_a~,  \\
p^{-1} q: &\qquad Y:= P^\alpha_a Q_\alpha^a~.
\end{split}
\ee
These are the relevant operators.  The positive terms in $C_2$ correspond to 
\be \label{mar1}
\begin{array}{ll}
p^2 q^{-2} [2,0; 0,2]: &\qquad X_i^{i'} X_j^{j'}~, \\  
2 p^{-2} q^2: & \qquad Y^2~, \quad P^\alpha_a Q^a_\beta P^\beta_b Q^b_\alpha~, \\ 
p^2 q^{-2} [0,1; 1,0]: & \qquad \epsilon_{ab} \epsilon^{cd} R_i^a  R_j^b S^{i'}_c S^{j'}_d ~, \\
2 [1,0; 0,1]: &\qquad X_i^{i'} Y~, \quad  S^{i'}_a Q_\alpha^a P^\alpha_b  R^b_i~.
\end{array}
\ee
These are the marginal operators. 

On the other hand, the indices of \eref{fish} and \eref{dualfish} are given by
\be
\begin{split}
 \CI_{\eref{dualfish}} (\vec y, \vec z, p, q)&= \CI_{\eref{fish}} (\vec y, \vec z, p^{-1}, q^{-1})\\
&= 1+ c_1(\vec y, \vec z, p, q) x + c_2(\vec y, \vec z, p, q) x^2  + \ldots~,
\end{split}
\ee
where
\be   \label{coeffonewallN3b}
\begin{split}
c_1(\vec y, \vec z, p, q) &=  p q^{-1} [1,0 ; 0,1]+p^{-1} q  \\
c_2(\vec y, \vec z, p, q) &= p^{2} q^{-2} [2,0; 0,2] + 2 p^{-2} q^{2} + [1,0; 0,1] +(1+p^{2} q^{-2}) [0,1; 1,0]  \\
& \quad  - [1,1;0,0] - [0,0;1,1] -2 - p^{2} q^{-2} ~.
\end{split}
\ee
Let us analyse theory \eref{dualfish}.  The relevant operators, corresponding to the terms in $c_1$, are
\be \label{rel2}
\begin{split}
p q^{-1} [1,0 ; 0,1]: &\qquad X^{i'}_{i} := B^{i'}_a D^a_i ~, \\  
p^{-1} q: &\qquad Y:= A^\alpha_a C_\alpha^a~.
\end{split}
\ee
The marginal operators, corresponding to the terms in $c_2$, are
\be \label{mar2}
\begin{array}{ll}
~p^{2} q^{-2} [2,0; 0,2]: & \qquad X^{i'}_{i} X^{j'}_{j} \\
~  2 p^{-2} q^{2}: &\qquad Y^2~, \quad  A^\alpha_a C^a_\beta A^\beta_b C^b_\alpha \\
~[1,0; 0,1]: &\qquad  X^{i'}_{i} Y \\
~[0,1; 1,0]: &\qquad  (\varphi_{AD})^i_a (\varphi_{BC})^a_{i'} \\
~p^{2} q^{-2}[0,1; 1,0]: & \qquad \epsilon_{ab} \epsilon^{cd} D_i^a  D_j^b B^{i'}_c B^{j'}_d
\end{array} 
\ee

Let us now compare the two sets of results.  Observe that the operators in \eref{rel1} are in correspondence with \eref{rel2}, and so as the first two lines of \eref{mar1} and \eref{mar2}.  However, the last two lines of \eref{mar1} do not agree with \eref{mar2}. In particular, in the former the representation $[1,0;0,1]$ appears with multiplicity $2$, whereas it appears with multiplicity $1$ in the latter.  For this reason we conclude that for $N>2$, the two sets of theories stated in \eref{SS} are not dual to each other.  

\subsubsection*{Superconformal fixed points}
Let us focus on $N=3$, and {\it assume} that theories \eref{modelI}, \eref{modelII}, \eref{fish} and \eref{dualfish} flow to superconformal fixed points. Due to the dualities \eref{SS}, theory \eref{modelI} flows to the same fixed point as theory \eref{modelII}, and theory \eref{fish} flows to the same fixed point as theory \eref{dualfish}.  Due to the previous discussion, we expect that the two fixed points are different for $N=3$.  

Under the assumption of the existence of the superconformal fixed point, the negative terms in $C_2$ of \eref{coeffonewallN3a} and those in $c_2$ of \eref{coeffonewallN3b} correspond to the conserved current of each set of theories. Both contain a term $-p^2 q^{-2}$, which should correspond to a $U(1)$ conserved current and should appear in the index as $1$ (since its character is $1$). Therefore our assumption on the conformality forces us to set $p=q$.  The terms $2 p^{-2} q^{2}  -2 - p^{2} q^{-2}$ thus combine into $-1$, and we are left with the negative terms $- [1,1;0,0] - [0,0;1,1]-1$ in both $C_2$ and $c_2$.  These indicate that the global symmetries of both superconformal fixed points are indeed $SU(3) \times SU(3) \times U(1)$, where the fugacity of such a $U(1)$ symmetry is identified with $p=q$.  Another possible interpretation of this phenomenon is as follows: if we deform theory \eref{coeffonewallN3a} or theory \eref{coeffonewallN3b} by two real mass deformations, one associated with $U(1)_p$ and the other with $U(1)_q$, then we reach the aftermentioned fixed point only when the two real masses are set to be equal.

\subsection{Zero external leg and quadrality} \label{sec:genus2noleg}
In this subsection, the two $SU(N)$ global symmetry in each of the theories in  \eref{SS} are commonly gauged with CS level $k_2$, and let us denote the CS level $k$ for the former $SU(N)$ gauge group by $k_1$.  We have introduced actually two of the resulting theories in \eref{twoloops} and \eref{dualtwoloops}, whose skeleton diagram has genus two and zero external leg.  In this subsection we discuss the relation between the four theories after such gauging.

We find that the indices of the following four theories are equal for {\bf any $N \geq 2$} and for $k_1, k_2 \geq 2$:
\be\label{quadralitygenus2noleg}
\begin{array}{ll}
\scalebox{0.6}{
\begin{tikzpicture}[baseline]
\tikzstyle{every node}=[font=\footnotesize]
\node[draw, circle, fill=yellow] (node1) at (0,0) {};
\node[draw, circle, dashed] (sqnode1) at (2,0) {$N_{k_2}$};
\node[draw, circle, dashed] (circ) at (-2,0) {$N_{k_1}$};
\draw[draw=blue,solid, <-] (circ) edge [out=135,in=-135,loop,looseness=5] node[midway,left]{}  (circ);
\draw[draw=black,solid,->]  (circ) to[bend left=60]  node[midway,below] {}   (node1) ; 
\draw[draw=black,solid,<-]  (circ) to[bend right=60] node[midway,above] {}  (node1) ; 
\draw[draw=black,solid,->]  (node1)to[bend left=60]  node[midway,above] {} (sqnode1) ; 
\draw[draw=black,solid,<-]  (node1) to[bend right=60]  node[midway,above] {} (sqnode1) ;  
\draw[draw=blue,solid,->]  (sqnode1) edge [out=45,in=-45,loop,looseness=5] node[midway,right]{} (sqnode1) ;  
\draw[draw=blue,solid,<-]  (circ) to [bend left=70] node[midway,above] {}  (sqnode1) ;
\draw[draw=blue,solid,->]  (circ) to [bend right=70] node[midway,below] {}  (sqnode1) ;
\end{tikzpicture}}
\qquad \qquad \qquad& \qquad
\scalebox{0.6}{
\begin{tikzpicture}[baseline]
\tikzstyle{every node}=[font=\footnotesize]
\node[draw, circle, fill=yellow] (node1) at (0,0) {};
\node[draw, circle, dashed] (circ2) at (2,0) {$N_{k_2}$};
\node[draw, circle, dashed] (circ) at (-2,0) {$N_{k_1}$};
\draw[draw=black,solid,->]  (circ) to[bend left=60]  node[midway,below] {}   (node1) ; 
\draw[draw=black,solid,<-]  (circ) to[bend right=60] node[midway,above] {}  (node1) ; 
\draw[draw=black,solid,->]  (node1) to[bend left=60] node[midway,below] {} (circ2) ; 
\draw[draw=black,solid,<-]  (node1) to[bend right=60] node[midway,above] {} (circ2) ;  
\end{tikzpicture}}
\\ \hspace{0.5cm}
\scalebox{0.6}{
\begin{tikzpicture}[baseline]
\tikzstyle{every node}=[font=\footnotesize]
\node[draw, circle, fill=yellow] (node1) at (0,0) {};
\node[draw, circle, dashed] (sqnode1) at (2,0) {$N_{k_2}$};
\node[draw, circle, dashed] (circ) at (-2,0) {$N_{k_1}$};
\draw[draw=black,solid,->]  (circ) to[bend left=60]  node[midway,below] {}   (node1) ; 
\draw[draw=black,solid,<-]  (circ) to[bend right=60] node[midway,above] {}  (node1) ; 
\draw[draw=black,solid,->]  (node1)to[bend left=60]  node[midway,above] {} (sqnode1) ; 
\draw[draw=black,solid,<-]  (node1) to[bend right=60]  node[midway,above] {} (sqnode1) ;  
\draw[draw=blue,solid,<-]  (circ) to [bend left=70] node[midway,above] {}  (sqnode1) ;
\draw[draw=blue,solid,->]  (circ) to [bend right=70] node[midway,below] {}  (sqnode1) ;
\end{tikzpicture}}
\qquad \qquad \qquad& \hspace{0.3cm}
\scalebox{0.6}{
\begin{tikzpicture}[baseline]
\tikzstyle{every node}=[font=\footnotesize]
\node[draw, circle, fill=yellow] (node1) at (0,0) {};
\node[draw, circle, dashed] (sqnode1) at (2,0) {$N_{k_2}$};
\node[draw, circle, dashed] (circ) at (-2,0) {$N_{k_1}$};
\draw[draw=blue,solid, <-] (circ) edge [out=135,in=-135,loop,looseness=5] node[midway,left]{}  (circ);
\draw[draw=black,solid,->]  (circ) to[bend left=60]  node[midway,below] {}   (node1) ; 
\draw[draw=black,solid,<-]  (circ) to[bend right=60] node[midway,above] {}  (node1) ; 
\draw[draw=black,solid,->]  (node1)to[bend left=60]  node[midway,above] {} (sqnode1) ; 
\draw[draw=black,solid,<-]  (node1) to[bend right=60]  node[midway,above] {} (sqnode1) ;  
\draw[draw=blue,solid,->]  (sqnode1) edge [out=45,in=-45,loop,looseness=5] node[midway,right]{} (sqnode1) ;  
\end{tikzpicture}}
\end{array}
\ee
where, for each quiver, there is a monopole superpotential due to the yellow node and the cubic superpotential terms coming from every closed triangular loop that contains one blue line as an edge.  We thus claim that these four theories are related to each other by a {\it quadrality}.  Note that for the special case of $N=2$, such a quadrality is an immediate consequence of that discussed in \eref{quadrality}.

Let us analyse such theories in more detail.  For definiteness, we choose one of the theories from the above list, say
\be \label{doublegaugeonewall}
\begin{tikzpicture}[baseline]
\tikzstyle{every node}=[font=\footnotesize]
\node[draw, circle, fill=yellow] (node1) at (0,0) {};
\node[draw, circle, dashed] (circ2) at (2,0) {$N_{k_2}$};
\node[draw, circle, dashed] (circ) at (-2,0) {$N_{k_1}$};
\draw[draw=black,solid,->]  (circ) to[bend left=60]  node[midway,below] {$A_1$}   (node1) ; 
\draw[draw=black,solid,<-]  (circ) to[bend right=60] node[midway,above] {$\tilde{A}_1$}  (node1) ; 
\draw[draw=black,solid,->]  (node1) to[bend left=60] node[midway,below] {$A_2$} (circ2) ; 
\draw[draw=black,solid,<-]  (node1) to[bend right=60] node[midway,above] {$\tilde{A}_2$} (circ2) ;  
\end{tikzpicture}
\ee
with the superpotential
\be
W = V_+ + V_- ~.
\ee

\subsubsection*{The case of $N=2$}
For $k_1 \geq 1$ and $k_2 \geq 2$ (or $k_2 \geq 1$ and $k_1 \geq 2$), the first few orders of the power expansion of the index are\footnote{We find that for $k_1=k_2=1$, the index is equal to unity, and if either $k_1$ or $k_2$ is zero, the index diverges.}
\be \label{indexexpgenus2}
\CI_{\eref{doublegaugeonewall}}(x; u) = 1 + C_1(p,q) x + C_2(p,q) x^2 + C_3(p,q) x^3 + C_4(p,q) x^4+ \ldots~,
\ee
where
\be \label{C1C2N2}
\begin{split}
C_1(p,q)  &= p q^{-1}+p^{-1} q \\
C_2(p,q)  &= (p^2 q^{-2}+1+p^{-2} q^2) -2 \\
C_3(p,q)  &= p^3 q^{-3} + p^{-3}  q^3\\
C_4(p,q)  &= p^4q^{-4}+p^{-4} q^4+p^2 q^2+p^{-2} q^{-2}+c_{k_1,k_2}~,
\end{split}
\ee
with $c_{k_1,k_2}$ a positive interger that depends on the values of $k_1$ and $k_2$. For example, $c_{2,2}=1$, $c_{2,k} =2$ for $k \geq3$, and $c_{k_1, k_2} =3$ for $k_1, k_2 \geq 3$.

Using the assignment as in \eref{basicblock1charges}, we see that the relevant operators, corresponding to the terms $p q^{-1}$ and $p^{-1} q$ in $C_1(p,q)$, are 
\be \label{defX1X2}
X_1:=(A_1)^\alpha (\tilde{A}_1)_\alpha~, \qquad X_2:= (A_2)_{\alpha'} (\tilde{A}_2)^{\alpha'}~,
\ee
where $\alpha=1,2, \ldots, N$ and $\alpha'=1,2, \ldots, N$ are the gauge indices for $SU(N)_{k_1}$ and $SU(N)_{k_2}$ respectively.  The marginal operators, corresponding to the terms $p^2 q^{-2}$, $1$ and $p^{-2} q^2$ in $C_2(p,q)$, are $X^2_1, \, X_1 X_2, \, X_2^2$.  The term $-2$ in $C_2(p,q)$, corresponding to the conserved current, confirms that the global symmetry of the theory is $U(1)_p \times U(1)_q$.  Note that due to terms in $C_4(p,q)$, it is not possible to rewrite the fugacities $p$, $q$ in terms of characters of $SU(2)$ representations. 

\subsubsection*{The case of $N=3$}
For $N=3$ with $k_1 \geq 2$ and $k_2 \geq 2$, the first few coefficients of the index \eref{indexexpgenus2} are
\be \label{C1C2N3}
\begin{split}
C_1(p,q)  &= p q^{-1}+p^{-1} q \\
C_2(p,q)  &= (x_{k_1} p^2 q^{-2}+2+ x_{k_2} p^{-2} q^2) -2~,
\end{split}
\ee
where
\be
x_{k} = \begin{cases} 1 &\qquad \text{if $k=2$} \\  2 &\qquad \text{if $k\geq 3$} \end{cases}~.
\ee

The term $-2$ in $C_2(p,q)$ in \eref{C1C2N3} indicates that the global symmetry of the theory is $U(1) \times U(1)$, whose fugacities are denoted by $p$ and $q$.  Let us now explain the other terms in $C_2(p,q)$, as well as those in $C_1(p,q)$.

Let us first consider the case of $k_1=k_2=2$, so that $x_{k_1} = x_{k_2} = 1$. A crucial difference between the coefficient $C_2(p,q)$ for $N=3$ and that for $N=2$ in \eref{C1C2N2} is that there is an extra marginal operator that carry zero charges under both $U(1)_p$ and $U(1)_q$ in the former case.  For $N \geq 3$, the relevant operators are similar to \eref{defX1X2}:
\be \label{defX1X2N3}
X_1 = (A_1)^\alpha_a (\tilde{A}_1)^a_\alpha~, \qquad  X_2:= (A_2)^a_{\alpha'} (\tilde{A}_2)_a^{\alpha'}~,
\ee
where $a,b=1,2, \ldots, N-1$ are the indices for the $U(N-1)$ gauge group denoted by the yellow node.  These correspond to the terms in the coefficient $C_1(p,q)$.  In addition to $X_1^2,\,  X_1 X_2, \, X_2^2$, there is an extra marginal operator given by
\be
Q = (A_1)^\alpha_a  (A_2)^a_{\alpha'} (\tilde{A}_2)_b^{\alpha'}  (\tilde{A}_1)^b_\alpha~,
\ee
which is different from $X_1 X_2$, for $N\geq 3$, and is neutral under both $U(1)_p$ and $U(1)_q$.  These four marginal operators correspond to the terms in the bracket in the coefficient $C_2(p,q)$.

Now let us assume that one of $k_1$ and $k_2$ or both are strictly greater than $2$.  The left or right gauge nodes can be regarded as $SU(3)_{k}$ with $2$ flavours, where $k$ is either $k_1$ or $k_2$.  To analyse this, we find that it is convenient to apply the duality (3.23) of \cite{Aharony:2013dha}.  The dual theory is $U(k-1)_{-k, -1}$ with $2$ flavours $q$, $\tilde{q}$, the chiral field $M$ in the adjoint representation of the yellow $U(2)$ node in \eref{doublegaugeonewall}, and the superpotential $W= M q \tilde{q}$.  Let us denote by $M_1$ and $M_2$ the adjoint fields of the yellow $U(2)$ node that arise from dualising the $SU(3)_{k_1}$ and $SU(3)_{k_2}$ respectively.  The gauge invariant quantities $\tr(M_1)$ and $\tr(M_2)$ in this dual theory can be mapped to $X_1$ and $X_2$ in the original theory \eref{defX1X2N3}.

Let us consider the case of $k=2$. The dual theory has the $U(1)_{-1}$ gauge group. The $F$-terms with respect to $q$ and $\tilde{q}$ are $M^a_b q^b = 0$ and  $M^a_b \tilde{q}_a=0$.  Then, $M$ can be regarded as a two by two matrix of rank $1$, since $M$ maps a vector to zero and so the dimension of its kernel is one.  Since $M$ has rank $1$, it can be written as a product of two vectors and it follows that $\tr(M^2)= (\tr M)^2$.  Therefore, in the case of $k_1=k_2=1$, the operators $X_1^2,\,  X_1 X_2, \, X_2^2, \, Q$ can be mapped to $\tr(M_1^2), \, \tr(M_1) \tr(M_2), \, \tr(M_2^2), \, \tr (M_1 M_2)$, where $M_{1,2}$ satisfy $\tr(M_{1}^2)= (\tr M_{1})^2$ and $\tr(M_{2}^2)= (\tr M_{2})^2$.

However, when $k >2$, the dual gauge group $U(k-1)_{-k, 1}$ has a higher rank.  On the contrary to the case of $k=2$, $M$ has rank greater than $1$.  As a consequence, $\tr(M^2)$ and $(\tr M)^2$ are not identical and they correspond to two different operators.  This explains the presence of $x_{k_1}$ and $x_{k_2}$ in $C_2(p,q)$ in \eref{C1C2N3}.  In particular, if $k_1, k_2 >2$, the marginal operators corresponding to the terms in the brackets in $C_2(p,q)$ are $\tr(M_{1}^2), \, (\tr M_{1})^2, \, \tr(M_{2}^2), \, (\tr M_{2})^2$, corresponding to $2 p^2 q^{-2}+ 2 q^2 p^{-2}$, and $\tr(M_1) \tr(M_2), \tr (M_1 M_2)$, corresponding to 2.

\section{Two duality walls: using \eref{basicblock} as a building block} \label{sec:twowallsclassic}
Let us now consider the theories associated with two duality walls.  As discussed in section \ref{sec:glue}, if we use \eref{basicblock} as a basic building block, we obtain two theories \eref{Phiglue} and \eref{Sglue} from $\Phi$-gluing and $S$-gluing respectively. One can perform further gauging to close the external legs and obtain \eref{twowallsphi} and \eref{twowallswophi}.  We discuss in detail the four models in this section.  One the other hand, we discuss the case of two duality walls if we use \eref{basicblock1} as a basic building block in section \ref{sec:twowallsalternative}.

\subsection{Indices of models \eref{Phiglue} and \eref{Sglue} for $N=2$}
We first analyse model \eref{Sglue}.  
\be
\scalebox{0.7}
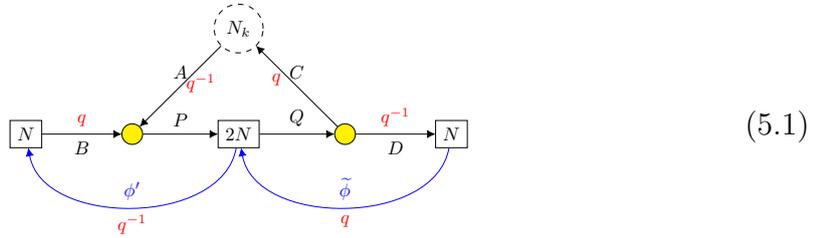
{
\begin{tikzpicture}[baseline]
\tikzstyle{every node}=[font=\footnotesize]
\node[draw, rectangle] (sqnode) at (0,0) {$2N$};
\node[draw, circle, fill=yellow] (yell1) at (-2,0) {};
\node[draw, circle, dashed] (circ1) at (0,2) {$N_{k}$};
\node[draw, rectangle] (sqnode1) at (-4,0) {$N$};
\node[draw, rectangle] (sqnode2) at (4,0) {$N$};
\node[draw, circle, fill=yellow] (yell2) at (2,0) {};
\draw[draw=black,solid,<-]  (sqnode) to node[midway,above] {$P$}   (yell1) ; 
\draw[draw=black,solid,<-]  (yell1) to node[midway,above] {$A$} node[near end,below] {\red $q^{-1}$}    (circ1) ; 
\draw[draw=black,solid,<-]  (yell1) to node[midway,below] {$B$} node[midway,above] {\red $q$}   (sqnode1) ; 
\draw[draw=black,solid,<-]  (circ1) to node[midway,above] {$C$} node[near start,below] {\red $q$}     (yell2) ; 
\draw[draw=black,solid,<-]  (sqnode2) to node[midway,below] {$D$} node[midway,above] {\red $q^{-1}$}   (yell2) ; 
\draw[draw=black,solid,<-]  (yell2) to node[midway,above] {$Q$}   (sqnode) ; 
\draw[draw=blue,solid,<-]  (sqnode1) to[bend right=80] node[midway,above] {\blue $\phi'$} node[midway,below] {\red $q^{-1}$} (sqnode);
\draw[draw=blue,solid,->]  (sqnode2) to[bend left=80] node[midway,above] {\blue $\tilde{\phi}$} node[midway,below] {\red $q$}  (sqnode);
\end{tikzpicture}}
\ee
For $k \geq 2$, the index for $N=2$ reads
\be
\CI^{N=2}_{\eref{Sglue}}(x; y_L, \vec y, y_R) = 1+ C_1 x + C_2 x^2 + \ldots
\ee
where
\be \label{coeffsSglue}
\begin{split}
C_1&= q^{-1} [1; \, 1,0,0; \, 0]+q [0; \, 0,0,1; \, 1]~, \\
C_2 &= q^{-2} [2; \, 2,0,0; \, 0] + q^{2} [0; \, 0,0,2; \, 2] + \left\{ [1; \, 1,0,1; \, 1]+[1; \, 0,0,0;\, 1]  \right\}\\
&\quad  - [2; \, 0,0,0; \, 0]  - [0; \, 1,0,1; \, 0]  - [0; \, 0,0,0; \, 2] -1~.
\end{split}
\ee
The unrefined index for this model is
\be \label{unrefSglue}
\CI^{N=2}_{\eref{Sglue}}(x; y_L=1, \vec y=1, y_R=1) = 1 + 16 x + 102 x^2 + 288 x^3 + 396 x^4+\ldots~.
\ee
The terms in $C_1$ in \eref{coeffsSglue} correspond to $\phi'$ and $\tilde{\phi}$ respectively.  The terms in the curly brackets in $C_2$ come from the tensor product of the two terms in $C_1$.  The second symmetric power of the representation $q^{-1} [1; \,1,0,0; \,0]$ in $C_1$ is $q^{-2} [2; \,2,0,0;\, 0]+ q^{-2} [0;\,0,1,0; \, 0]$.  However, the gauge invariant combinations $\epsilon^{\alpha \beta} M^{ij}_{\alpha \beta}$\footnote{Recall that we take $N=2$.  Here $i,j=1,\ldots, 4$ are the $SU(4)$ flavour indices, and $\alpha,\beta=1,2$ are the indices for the left $SU(2)$ flavour node.  Note also that, due to the definition of $M$, $M^{[ij]}_{\alpha \beta}$, with an antisymmetrisation on $i$ and $j$, can be written as $\epsilon^{\alpha \beta}  M^{ij}_{\alpha \beta}$.}, with $M^{ij}_{\alpha \beta}=(\phi')^{i}_\alpha (\phi')^{j}_\beta $ associated with the latter representation vanish in the chiral ring, due to the quantum rank condition (in the same way as in the Seiberg duality).  This can be seen by applying duality \eref{mainduality} to the left yellow node; we see that $(\phi')^i_\alpha$ is identified with $\tilde{B}_\alpha \tilde{P}^i$, where $\tilde{B}$ and $\tilde{P}$ are the chiral fields (whose arrows are in the opposite direction to $B$ and $P$) in the dual theory, and so $\epsilon^{\alpha \beta} M^{ij}_{\alpha \beta} = \epsilon^{\alpha \beta} \tilde{B}_\alpha \tilde{B}_\beta \tilde{P}^i  \tilde{P}^j=0$.  This is the reason why only $q^{-2} [2; \,2,0,0;\, 0]$ survives in the index. The similar argument can be applied to the second symmetric power of $q [0; \, 0,0,1; \, 1]$.  The negative terms in $C_2$ tell us that the global symmetry of the theory is $SU(2)^2 \times SU(4) \times U(1)$.

Now let us analyse model \eref{Phiglue}.
\be
\scalebox{0.7}
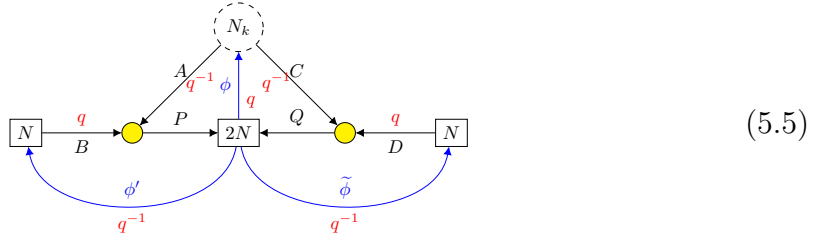
{
\begin{tikzpicture}[baseline]
\tikzstyle{every node}=[font=\footnotesize]
\node[draw, rectangle] (sqnode) at (0,0) {$2N$};
\node[draw, circle, fill=yellow] (yell1) at (-2,0) {};
\node[draw, circle, dashed] (circ1) at (0,2) {$N_{k}$};
\node[draw, rectangle] (sqnode1) at (-4,0) {$N$};
\node[draw, rectangle] (sqnode2) at (4,0) {$N$};
\node[draw, circle, fill=yellow] (yell2) at (2,0) {};
\draw[draw=black,solid,<-]  (sqnode) to node[midway,above] {$P$}   (yell1) ; 
\draw[draw=black,solid,<-]  (yell1) to node[midway,above] {$A$} node[near end,below] {\red $q^{-1}$}    (circ1) ; 
\draw[draw=black,solid,<-]  (yell1) to node[midway,below] {$B$} node[midway,above] {\red $q$}   (sqnode1) ; 
\draw[draw=black,solid,->]  (circ1) to node[midway,above] {$C$} node[near start,below] {\red $q^{-1}$}     (yell2) ; 
\draw[draw=black,solid,->]  (sqnode2) to node[midway,below] {$D$} node[midway,above] {\red $q$}   (yell2) ; 
\draw[draw=black,solid,->]  (yell2) to node[midway,above] {$Q$}   (sqnode) ; 
\draw[draw=blue,solid,<-]  (circ1) to node[midway,left] {\blue $\phi$} node[near end,right] {\red $q$}   (sqnode) ; 
\draw[draw=blue,solid,<-]  (sqnode1) to[bend right=80] node[midway,above] {\blue $\phi'$} node[midway,below] {\red $q^{-1}$} (sqnode);
\draw[draw=blue,solid,<-]  (sqnode2) to[bend left=80] node[midway,above] {\blue $\tilde{\phi}$} node[midway,below] {\red $q^{-1}$}  (sqnode);
\end{tikzpicture}}
\ee
The index for $N=2$ and for $k\geq 2$ reads
\be
\CI^{N=2}_{\eref{Phiglue}}(x; y_L, \vec y, y_R) = 1+ c_1 x + c_2 x^2 + \ldots
\ee
where
\be \label{coeffPhiglue}
\begin{split}
c_1&= q^{-1} [1; \, 1,0,0; \, 0]+q^{-1} [0; \, 1,0,0; \, 1]~, \\
c_2 &=  q^{-2}[2; \, 2,0,0; \, 0]+ q^{-2}[0; \, 2,0,0; \, 2] + \{ q^{-2}  [1; \, 0,1,0; \, 1] +q^{-2}[1; \, 2,0,0; \, 1] \} \\
& \quad {\blue +q^{-2} [0; \, 0,1,0;\; 0]  - q^{2} [0; \, 0,1,0;\; 0]} - [2; \, 0,0,0; \, 0]  - [0; \, 1,0,1; \, 0]  \\
& \quad - [0; \, 0,0,0; \, 2] -1~.
\end{split}
\ee
The unrefined index of this theory turns out to be equal to that of \eref{Phiglue}, which is given by \eref{unrefSglue}:
\be
\CI^{N=2}_{\eref{Phiglue}}(x; y_L=1, \vec y=1, y_R=1) =\CI^{N=2}_{\eref{Sglue}}(x; y_L=1, \vec y=1, y_R=1)~.
\ee
The interpretation for \eref{coeffPhiglue} is very similar to the above.  The terms in $c_1$ correspond to $\phi'$ and $\tilde{\phi}$.  The terms in the curly brackets in $c_2$ come from the tensor product of the two terms in $c_1$.  The term $+q^{-2} [0; \, 0,1,0;\; 0]$ can be conveniently explained using another duality frame.  If we dualise both left and right yellow nodes using duality \eref{mainduality}, the chiral fields $\phi'$ and $\tilde{\phi}$ disappear and we replace $\phi$ by a chiral field $\chi$, whose arrow is in the opposite direction of $\phi$ and carrying the $U(1)_q$ fugacity $q^{-1}$.  (The arrows for $A$, $B$, $P$, $C$, $Q$, $D$ also reverse their directions.)  We can construct the gauge invariant quantity $\epsilon_{ab} \chi^{a}_i \chi^b_j$, where $a,b=1,2$ is the $SU(2)_k$ gauge indices and $i,j=1,\ldots, 4$ are the $SU(4)$ flavour indices, in the representation $q^{-2} [0; \, 0,1,0;\; 0]$, as required.

It is interesting to point out that even though the unrefined indices of the two models are equal, their refined indices are different.  In particular, the representation $[1; \, 1,0,1; \, 1]+[1; \, 0,0,0;\, 1]$ that appears in the former but not the latter, and the representation [1; 0,1,0; 1] +[1; 0,0,2; 1] that appears in the latter but not in the former.  Although their dimensions are equal and they both come from the tensor products of the two terms of the coefficient of $x$, their characters are different.

Another important point is the negative term $- q^{-2} [0; \, 0,1,0;\; 0]$ that appears in $c_2$ in \eref{coeffPhiglue}.  Since this is not the adjoint representation, it cannot correspond to a conserved current.  If we {\it assume} that theory \eref{Phiglue} flows to a fixed point, this negative term cannot be there by itself. Indeed, if we set $q=1$, such a term cancels with another positive term (both are indicated in blue).  After the cancellation, the negative terms indicate that the global symmetry of the theory is $SU(2)^2 \times SU(4) \times U(1)$.  Since the fugacity $q$ has already been set to $1$, the index no longer has a manifest $U(1)$ fugacity, and we interpret such a $U(1)$ global symmetry as emergent in the infrared.

\subsection{Various dualities for any $N \geq 2$}
Given models \eref{Phiglue} and \eref{Sglue}, we can generate a number of dualities that hold for any $N\geq 2$ by applying duality \eref{mainduality} to each yellow node.   For \eref{Phiglue}, we have a {\it triality} between the following theories:
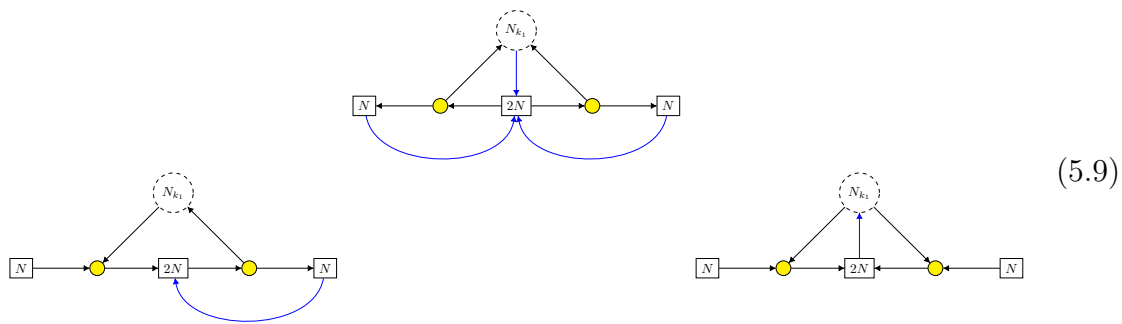
\be
\begin{array}{lll}
&
\scalebox{0.5}{
\begin{tikzpicture}[baseline]
\tikzstyle{every node}=[font=\footnotesize]
\node[draw, rectangle] (sqnode) at (0,0) {$2N$};
\node[draw, circle, fill=yellow] (yell1) at (-2,0) {};
\node[draw, circle, dashed] (circ1) at (0,2) {$N_{k_1}$};
\node[draw, rectangle] (sqnode1) at (-4,0) {$N$};
\node[draw, rectangle] (sqnode2) at (4,0) {$N$};
\node[draw, circle, fill=yellow] (yell2) at (2,0) {};
\draw[draw=black,solid,->]  (sqnode) to node[midway,above] {}   (yell1) ; 
\draw[draw=black,solid,->]  (yell1) to node[midway,above] {}   (circ1) ; 
\draw[draw=black,solid,->]  (yell1) to node[midway,below] {}   (sqnode1) ; 
\draw[draw=black,solid,<-]  (circ1) to node[midway,above] {}   (yell2) ; 
\draw[draw=black,solid,<-]  (sqnode2) to node[midway,below] {}   (yell2) ; 
\draw[draw=black,solid,<-]  (yell2) to node[midway,above] {}   (sqnode) ; 
\draw[draw=blue,solid,->]  (circ1) to node[midway,left] {}  (sqnode) ; 
\draw[draw=blue,solid,->]  (sqnode1) to[bend right=80] node[midway,above] {}  (sqnode);
\draw[draw=blue,solid,->]  (sqnode2) to[bend left=80] node[midway,above] {}  (sqnode);
\end{tikzpicture}}&
\\
\scalebox{0.5}{
\begin{tikzpicture}[baseline]
\tikzstyle{every node}=[font=\footnotesize]
\node[draw, rectangle] (sqnode) at (0,0) {$2N$};
\node[draw, circle, fill=yellow] (yell1) at (-2,0) {};
\node[draw, circle, dashed] (circ1) at (0,2) {$N_{k_1}$};
\node[draw, rectangle] (sqnode1) at (-4,0) {$N$};
\node[draw, rectangle] (sqnode2) at (4,0) {$N$};
\node[draw, circle, fill=yellow] (yell2) at (2,0) {};
\draw[draw=black,solid,<-]  (sqnode) to node[midway,above] {}   (yell1) ; 
\draw[draw=black,solid,<-]  (yell1) to node[midway,above] {}   (circ1) ; 
\draw[draw=black,solid,<-]  (yell1) to node[midway,below] {}   (sqnode1) ; 
\draw[draw=black,solid,<-]  (circ1) to node[midway,above] {}   (yell2) ; 
\draw[draw=black,solid,<-]  (sqnode2) to node[midway,below] {}   (yell2) ; 
\draw[draw=black,solid,<-]  (yell2) to node[midway,above] {}   (sqnode) ; 
\draw[draw=blue,solid,->]  (sqnode2) to[bend left=80] node[midway,above] {}  (sqnode);
\end{tikzpicture}}
&&
\scalebox{0.5}{
\begin{tikzpicture}[baseline]
\tikzstyle{every node}=[font=\footnotesize]
\node[draw, rectangle] (sqnode) at (0,0) {$2N$};
\node[draw, circle, fill=yellow] (yell1) at (-2,0) {};
\node[draw, circle, dashed] (circ1) at (0,2) {$N_{k_1}$};
\node[draw, rectangle] (sqnode1) at (-4,0) {$N$};
\node[draw, rectangle] (sqnode2) at (4,0) {$N$};
\node[draw, circle, fill=yellow] (yell2) at (2,0) {};
\draw[draw=black,solid,<-]  (sqnode) to node[midway,above] {}   (yell1) ; 
\draw[draw=black,solid,<-]  (yell1) to node[midway,above] {}   (circ1) ; 
\draw[draw=black,solid,<-]  (yell1) to node[midway,below] {}   (sqnode1) ; 
\draw[draw=black,solid,->]  (circ1) to node[midway,above] {}   (yell2) ; 
\draw[draw=black,solid,->]  (sqnode2) to node[midway,below] {}   (yell2) ; 
\draw[draw=black,solid,->]  (yell2) to node[midway,above] {}   (sqnode) ; 
\draw[draw=blue,solid,<-]  (circ1) to node[midway,left] {}  (sqnode) ; 
\end{tikzpicture}}
\end{array}
\ee
where the top and bottom left theories are related by dualising the left yellow node, and the bottom left and bottom right theories are related by dualising the right yellow node.  For each quiver, there is a monopole superpotential due to the yellow node, the cubic superpotential terms coming from every closed triangular loop that contains one blue line as an edge, and there is also a quartic term for the bottom left quiver coming from the middle triangle.

For \eref{Sglue}, we have a {\it triality} between the following theories:
\be
\begin{array}{lll}
&
\scalebox{0.5}{
\begin{tikzpicture}[baseline]
\tikzstyle{every node}=[font=\footnotesize]
\node[draw, rectangle] (sqnode) at (0,0) {$2N$};
\node[draw, circle, fill=yellow] (yell1) at (-2,0) {};
\node[draw, circle, dashed] (circ1) at (0,2) {$N_{k_1}$};
\node[draw, rectangle] (sqnode1) at (-4,0) {$N$};
\node[draw, rectangle] (sqnode2) at (4,0) {$N$};
\node[draw, circle, fill=yellow] (yell2) at (2,0) {};
\draw[draw=black,solid,->]  (sqnode) to node[midway,above] {}   (yell1) ; 
\draw[draw=black,solid,->]  (yell1) to node[midway,above] {}   (circ1) ; 
\draw[draw=black,solid,->]  (yell1) to node[midway,below] {}   (sqnode1) ; 
\draw[draw=black,solid,->]  (circ1) to node[midway,above] {}   (yell2) ; 
\draw[draw=black,solid,->]  (sqnode2) to node[midway,below] {}   (yell2) ; 
\draw[draw=black,solid,->]  (yell2) to node[midway,above] {}   (sqnode) ; 
\draw[draw=blue,solid,->]  (sqnode1) to[bend right=80] node[midway,above] {}  (sqnode);
\draw[draw=blue,solid,<-]  (sqnode2) to[bend left=80] node[midway,above] {}  (sqnode);
\end{tikzpicture}} &
\\
\scalebox{0.5}{
\begin{tikzpicture}[baseline]
\tikzstyle{every node}=[font=\footnotesize]
\node[draw, rectangle] (sqnode) at (0,0) {$2N$};
\node[draw, circle, fill=yellow] (yell1) at (-2,0) {};
\node[draw, circle, dashed] (circ1) at (0,2) {$N_{k_1}$};
\node[draw, rectangle] (sqnode1) at (-4,0) {$N$};
\node[draw, rectangle] (sqnode2) at (4,0) {$N$};
\node[draw, circle, fill=yellow] (yell2) at (2,0) {};
\draw[draw=black,solid,<-]  (sqnode) to node[midway,above] {}   (yell1) ; 
\draw[draw=black,solid,<-]  (yell1) to node[midway,above] {}   (circ1) ; 
\draw[draw=black,solid,<-]  (yell1) to node[midway,below] {}   (sqnode1) ; 
\draw[draw=black,solid,->]  (circ1) to node[midway,above] {}   (yell2) ; 
\draw[draw=black,solid,->]  (sqnode2) to node[midway,below] {}   (yell2) ; 
\draw[draw=black,solid,->]  (yell2) to node[midway,above] {}   (sqnode) ; 
\draw[draw=blue,solid,<-]  (circ1) to node[midway,left] {}  (sqnode) ; 
\draw[draw=blue,solid,<-]  (sqnode2) to[bend left=80] node[midway,above] {}  (sqnode);
\end{tikzpicture}}
&&
\scalebox{0.5}{
\begin{tikzpicture}[baseline]
\tikzstyle{every node}=[font=\footnotesize]
\node[draw, rectangle] (sqnode) at (0,0) {$2N$};
\node[draw, circle, fill=yellow] (yell1) at (-2,0) {};
\node[draw, circle, dashed] (circ1) at (0,2) {$N_{k_1}$};
\node[draw, rectangle] (sqnode1) at (-4,0) {$N$};
\node[draw, rectangle] (sqnode2) at (4,0) {$N$};
\node[draw, circle, fill=yellow] (yell2) at (2,0) {};
\draw[draw=black,solid,<-]  (sqnode) to node[midway,above] {}   (yell1) ; 
\draw[draw=black,solid,<-]  (yell1) to node[midway,above] {}   (circ1) ; 
\draw[draw=black,solid,<-]  (yell1) to node[midway,below] {}   (sqnode1) ; 
\draw[draw=black,solid,<-]  (circ1) to node[midway,above] {}   (yell2) ; 
\draw[draw=black,solid,<-]  (sqnode2) to node[midway,below] {}   (yell2) ; 
\draw[draw=black,solid,<-]  (yell2) to node[midway,above] {}   (sqnode) ; 
\end{tikzpicture}}
\end{array}
\ee
The superpotential for each quiver is in the same way as described above.

If we commonly gauge the two $SU(N)$ flavour symmetry corresponding to the left and right square nodes, we obtain models \eref{twowallsphi} and \eref{twowallswophi} and their {\it duality}.  We discuss this in detail below.  

\subsubsection{Duality between models \eref{twowallsphi} and \eref{twowallswophi}}
Applying duality \eref{mainduality} to either of the yellow nodes, we find that models \eref{twowallsphi} and \eref{twowallswophi} are dual to each other for {\bf any $N\geq 2$}.
Indeed, we find that the indices for \eref{twowallsphi} and \eref{twowallswophi} are equal.

In particular, for $N=2$ and $k_1, k_2 \geq 2$, their indices are
\be \label{indextwowallsN2}
\CI^{N=2}_{\eref{twowallsphi}}(x; q, \vec y) = \CI^{N=2}_{\eref{twowallswophi}}(x; q, \vec y) = 1+ 0 x + 0 x^2 +0 x^3 + C_4(q, \vec y) x^4+\ldots~,
\ee
where the coefficients of $x, \, x^2, \, x^3$ vanish, and 
\be
C_4(q) = \chi^{SU(4)}_{[1,0,1]}(\vec y)+\chi^{SU(4)}_{[0,2,0]}(\vec y) + 2(q^2+q^{-2}) \chi^{SU(4)}_{[0,1,0]}(\vec y)+ q^4 +1 +q^{-4}~,
\ee
where $\vec y = (y_1, y_2, y_3)$ are fugacities of the $SU(4)$ flavour symmetry and $q$ is a fugacity of the $U(1)$ global symmetry.
 
The vanishing coefficient of $x^2$ in \eref{indextwowallsN2} deserves some explanations.  Models \eref{twowallsphi} and \eref{twowallswophi} in fact have the global symmetry $SU(4) \times U(1)$.  The contribution $-\chi^{SU(4)}_{[1,0,1]}(\vec y)-1$ at order $x^2$ of the conserved current is cancelled by the contribution $\chi^{SU(4)}_{[1,0,1]}(\vec y)+1$ of the marginal operators.  For model \eref{twowallswophi}, such marginal operators are  $A^\alpha C_\alpha Q^i P_j$, corresponding to the close path in the upper triangle.  Note that these are equal to $-B^{\alpha'} D_{\alpha'} Q^i P_j$, corresponding to the close path in the lower triangle, due to the $F$-terms that are the derivatives with respect to $P_j$ of the superpotential \eref{suptwowallswophi}.

\section{Two duality walls: using \eref{basicblock1} as a building block}  \label{sec:twowallsalternative}
In this section, we consider the theories associated with two duality walls, using \eref{basicblock1} as a basic building block.  We consider the theories arising from $\Phi$-gluing and $S$-gluing and their dual theories.  We finally compute their indices and discuss the duality for the case of $N=2$.
\subsection*{$\Phi$-gluing}
The theory associated with the $\Phi$-gluing of two building blocks has already been introduced in \eref{ex1Phi}. We present such a theory, with the fugacities for $U(1)_p \times U(1)_q \times U(1)_{p'} \times U(1)_{q'}$ for each chiral fields, along with its duals below.
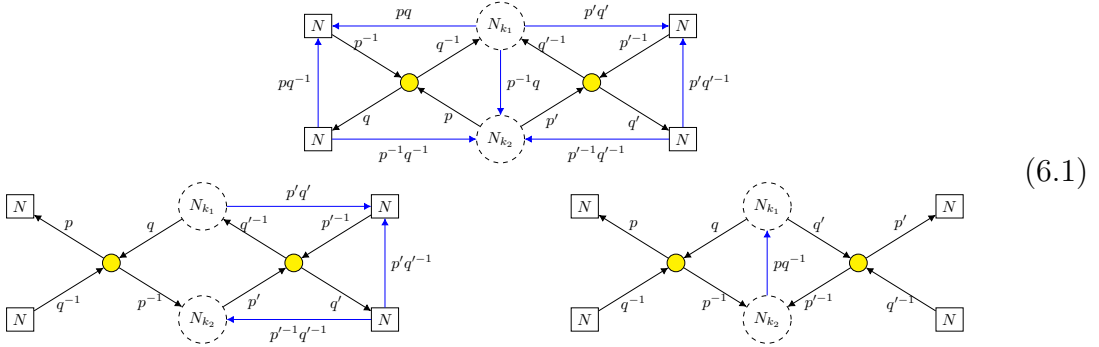
\be \label{finalPhiglue}
\begin{split}
& \hspace{3.5cm}
\scalebox{0.6}{
\begin{tikzpicture}[baseline]
\tikzstyle{every node}=[font=\footnotesize]
\node[draw, rectangle] (sqnode1a) at (-4,1.25) {$N$};
\node[draw, rectangle] (sqnode1b) at (-4,-1.25) {$N$};
\node[draw, circle, fill=yellow] (yell1) at (-2,0) {};
\node[draw, circle, dashed] (circ1) at (0,1.25) {$N_{k_1}$};
\node[draw, circle, dashed] (circ2) at (0,-1.25) {$N_{k_2}$};
\node[draw, circle, fill=yellow] (yell2) at (2,0) {};
\node[draw, rectangle] (sqnode2a) at (4,1.25) {$N$};
\node[draw, rectangle] (sqnode2b) at (4,-1.25) {$N$};
\draw[draw=black,solid,->]  (sqnode1a) to node[midway,above] {$p^{-1}$}   (yell1) ; 
\draw[draw=black,solid,<-]  (sqnode1b) to node[midway,below] {$q$}   (yell1) ; 
\draw[draw=black,solid,->]  (yell1) to node[midway,above] {$q^{-1}$}   (circ1) ; 
\draw[draw=black,solid,<-]  (yell1) to node[midway,below] {$p$}   (circ2) ; 
\draw[draw=black,solid,<-]  (circ1) to node[midway,above] {$q'^{-1}$}   (yell2) ; 
\draw[draw=black,solid,->]  (circ2) to node[midway,below] {$p'$}   (yell2) ; 
\draw[draw=black,solid,<-]  (yell2) to node[midway,above] {$p'^{-1}$}   (sqnode2a) ; 
\draw[draw=black,solid,->]  (yell2) to node[midway,below] {$q'$}   (sqnode2b) ; 
\draw[draw=blue,solid,<-]  (sqnode1a) to node[midway,above] {$pq$}   (circ1) ; 
\draw[draw=blue,solid,<-]  (sqnode1a) to node[midway,left] {$pq^{-1}$}   (sqnode1b) ; 
\draw[draw=blue,solid,->]  (sqnode1b) to node[midway,below] {$p^{-1}q^{-1}$}   (circ2) ;
\draw[draw=blue,solid,<-]  (sqnode2a) to node[midway,above] {$p'q'$}   (circ1) ;
\draw[draw=blue,solid,->]  (sqnode2b) to node[midway,below] {$p'^{-1}q'^{-1}$}   (circ2) ;   
\draw[draw=blue,solid,<-]  (sqnode2a) to node[midway,right] {$p'q'^{-1}$}   (sqnode2b) ; 
\draw[draw=blue,solid,->]  (circ1) to node[midway,right] {$p^{-1}q$}   (circ2) ; 
\end{tikzpicture}} \\
&
\scalebox{0.6}{
\begin{tikzpicture}[baseline]
\tikzstyle{every node}=[font=\footnotesize]
\node[draw, rectangle] (sqnode1a) at (-4,1.25) {$N$};
\node[draw, rectangle] (sqnode1b) at (-4,-1.25) {$N$};
\node[draw, circle, fill=yellow] (yell1) at (-2,0) {};
\node[draw, circle, dashed] (circ1) at (0,1.25) {$N_{k_1}$};
\node[draw, circle, dashed] (circ2) at (0,-1.25) {$N_{k_2}$};
\node[draw, circle, fill=yellow] (yell2) at (2,0) {};
\node[draw, rectangle] (sqnode2a) at (4,1.25) {$N$};
\node[draw, rectangle] (sqnode2b) at (4,-1.25) {$N$};
\draw[draw=black,solid,<-]  (sqnode1a) to node[midway,above] {$p$}   (yell1) ; 
\draw[draw=black,solid,->]  (sqnode1b) to node[midway,below] {$q^{-1}$}   (yell1) ; 
\draw[draw=black,solid,<-]  (yell1) to node[midway,above] {$q$}   (circ1) ; 
\draw[draw=black,solid,->]  (yell1) to node[midway,below] {$p^{-1}$}   (circ2) ; 
\draw[draw=black,solid,<-]  (circ1) to node[midway,above] {$q'^{-1}$}   (yell2) ; 
\draw[draw=black,solid,->]  (circ2) to node[midway,below] {$p'$}   (yell2) ; 
\draw[draw=black,solid,<-]  (yell2) to node[midway,above] {$p'^{-1}$}   (sqnode2a) ; 
\draw[draw=black,solid,->]  (yell2) to node[midway,below] {$q'$}   (sqnode2b) ; 
\draw[draw=blue,solid,<-]  (sqnode2a) to node[midway,above] {$p'q'$}   (circ1) ;
\draw[draw=blue,solid,->]  (sqnode2b) to node[midway,below] {$p'^{-1}q'^{-1}$}   (circ2) ;   
\draw[draw=blue,solid,<-]  (sqnode2a) to node[midway,right] {$p'q'^{-1}$}   (sqnode2b) ; 
\end{tikzpicture}}
\qquad \qquad
\scalebox{0.6}{
\begin{tikzpicture}[baseline]
\tikzstyle{every node}=[font=\footnotesize]
\node[draw, rectangle] (sqnode1a) at (-4,1.25) {$N$};
\node[draw, rectangle] (sqnode1b) at (-4,-1.25) {$N$};
\node[draw, circle, fill=yellow] (yell1) at (-2,0) {};
\node[draw, circle, dashed] (circ1) at (0,1.25) {$N_{k_1}$};
\node[draw, circle, dashed] (circ2) at (0,-1.25) {$N_{k_2}$};
\node[draw, circle, fill=yellow] (yell2) at (2,0) {};
\node[draw, rectangle] (sqnode2a) at (4,1.25) {$N$};
\node[draw, rectangle] (sqnode2b) at (4,-1.25) {$N$};
\draw[draw=black,solid,<-]  (sqnode1a) to node[midway,above] {$p$}   (yell1) ; 
\draw[draw=black,solid,->]  (sqnode1b) to node[midway,below] {$q^{-1}$}   (yell1) ; 
\draw[draw=black,solid,<-]  (yell1) to node[midway,above] {$q$}   (circ1) ; 
\draw[draw=black,solid,->]  (yell1) to node[midway,below] {$p^{-1}$}   (circ2) ; 
\draw[draw=black,solid,->]  (circ1) to node[midway,above] {$q'$}   (yell2) ; 
\draw[draw=black,solid,<-]  (circ2) to node[midway,below] {$p'^{-1}$}   (yell2) ; 
\draw[draw=black,solid,->]  (yell2) to node[midway,above] {$p'$}   (sqnode2a) ; 
\draw[draw=black,solid,<-]  (yell2) to node[midway,below] {$q'^{-1}$}   (sqnode2b) ; 
\draw[draw=blue,solid,<-]  (circ1) to node[midway,right] {$pq^{-1}$}   (circ2) ; 
\end{tikzpicture}}
\end{split}
\ee
where the bottom left and right quivers come from applying \eref{mainduality} to the left yellow node and to both yellow nodes of the top diagram, respectively.  There are monopole superpotential terms, the cubic superpotential terms coming from each triangular loop in the quiver that contains one blue line as an edge, and the quartic superpotential term for the bottom left quiver coming from rectangular loop in the middle.  Such a superpotential imposes the following condition on the $U(1)$ fugacities:
\be \label{cond1}
p^{-1}qp'q'^{-1}=1~.
\ee

\subsubsection*{$S$-gluing}
The theory associated with the $S$-gluing of two building blocks has already been introduced in \eref{ex1S}. We present such a theory, with the fugacities for $U(1)_p \times U(1)_q \times U(1)_{p'} \times U(1)_{q'}$ for each chiral fields, along with its duals below.
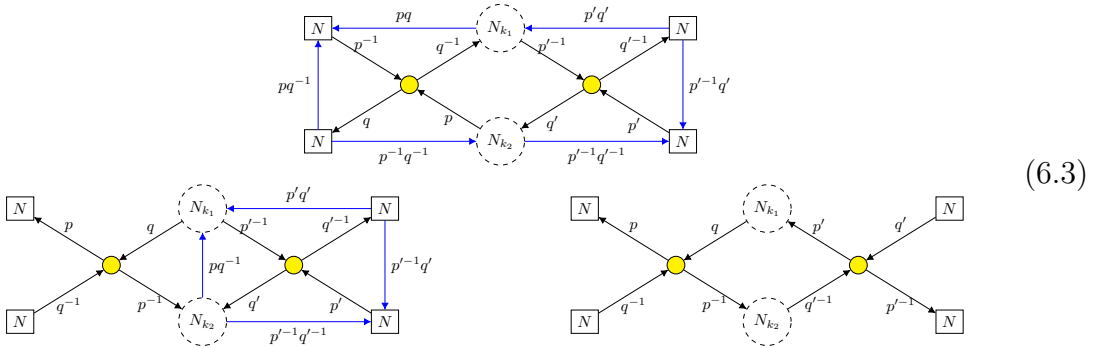
\be \label{finalSglue}
\begin{split}
&  \hspace{3.5cm}
\scalebox{0.6}{
\begin{tikzpicture}[baseline]
\tikzstyle{every node}=[font=\footnotesize]
\node[draw, rectangle] (sqnode1a) at (-4,1.25) {$N$};
\node[draw, rectangle] (sqnode1b) at (-4,-1.25) {$N$};
\node[draw, circle, fill=yellow] (yell1) at (-2,0) {};
\node[draw, circle, dashed] (circ1) at (0,1.25) {$N_{k_1}$};
\node[draw, circle, dashed] (circ2) at (0,-1.25) {$N_{k_2}$};
\node[draw, circle, fill=yellow] (yell2) at (2,0) {};
\node[draw, rectangle] (sqnode2a) at (4,1.25) {$N$};
\node[draw, rectangle] (sqnode2b) at (4,-1.25) {$N$};
\draw[draw=black,solid,->]  (sqnode1a) to node[midway,above] {$p^{-1}$}   (yell1) ; 
\draw[draw=black,solid,<-]  (sqnode1b) to node[midway,below] {$q$}   (yell1) ; 
\draw[draw=black,solid,->]  (yell1) to node[midway,above] {$q^{-1}$}   (circ1) ; 
\draw[draw=black,solid,<-]  (yell1) to node[midway,below] {$p$}   (circ2) ; 
\draw[draw=black,solid,->]  (circ1) to node[midway,above] {$p'^{-1}$}   (yell2) ; 
\draw[draw=black,solid,<-]  (circ2) to node[midway,below] {$q'$}   (yell2) ; 
\draw[draw=black,solid,->]  (yell2) to node[midway,above] {$q'^{-1}$}   (sqnode2a) ; 
\draw[draw=black,solid,<-]  (yell2) to node[midway,below] {$p'$}   (sqnode2b) ; 
\draw[draw=blue,solid,<-]  (sqnode1a) to node[midway,above] {$pq$}   (circ1) ; 
\draw[draw=blue,solid,<-]  (sqnode1a) to node[midway,left] {$pq^{-1}$}   (sqnode1b) ; 
\draw[draw=blue,solid,->]  (sqnode1b) to node[midway,below] {$p^{-1}q^{-1}$}   (circ2) ;
\draw[draw=blue,solid,->]  (sqnode2a) to node[midway,above] {$p'q'$}   (circ1) ;
\draw[draw=blue,solid,<-]  (sqnode2b) to node[midway,below] {$p'^{-1}q'^{-1}$}   (circ2) ;   
\draw[draw=blue,solid,->]  (sqnode2a) to node[midway,right] {$p'^{-1}q'$}   (sqnode2b) ; 
\end{tikzpicture}}
\\
&
\scalebox{0.6}{
\begin{tikzpicture}[baseline]
\tikzstyle{every node}=[font=\footnotesize]
\node[draw, rectangle] (sqnode1a) at (-4,1.25) {$N$};
\node[draw, rectangle] (sqnode1b) at (-4,-1.25) {$N$};
\node[draw, circle, fill=yellow] (yell1) at (-2,0) {};
\node[draw, circle, dashed] (circ1) at (0,1.25) {$N_{k_1}$};
\node[draw, circle, dashed] (circ2) at (0,-1.25) {$N_{k_2}$};
\node[draw, circle, fill=yellow] (yell2) at (2,0) {};
\node[draw, rectangle] (sqnode2a) at (4,1.25) {$N$};
\node[draw, rectangle] (sqnode2b) at (4,-1.25) {$N$};
\draw[draw=black,solid,<-]  (sqnode1a) to node[midway,above] {$p$}   (yell1) ; 
\draw[draw=black,solid,->]  (sqnode1b) to node[midway,below] {$q^{-1}$}   (yell1) ; 
\draw[draw=black,solid,<-]  (yell1) to node[midway,above] {$q$}   (circ1) ; 
\draw[draw=black,solid,->]  (yell1) to node[midway,below] {$p^{-1}$}   (circ2) ; 
\draw[draw=black,solid,->]  (circ1) to node[midway,above] {$p'^{-1}$}   (yell2) ; 
\draw[draw=black,solid,<-]  (circ2) to node[midway,below] {$q'$}   (yell2) ; 
\draw[draw=black,solid,->]  (yell2) to node[midway,above] {$q'^{-1}$}   (sqnode2a) ; 
\draw[draw=black,solid,<-]  (yell2) to node[midway,below] {$p'$}   (sqnode2b) ; 
\draw[draw=blue,solid,->]  (sqnode2a) to node[midway,above] {$p'q'$}   (circ1) ;
\draw[draw=blue,solid,<-]  (sqnode2b) to node[midway,below] {$p'^{-1}q'^{-1}$}   (circ2) ;   
\draw[draw=blue,solid,->]  (sqnode2a) to node[midway,right] {$p'^{-1}q'$}   (sqnode2b) ; 
\draw[draw=blue,solid,<-]  (circ1) to node[midway,right] {$pq^{-1}$}   (circ2) ; 
\end{tikzpicture}}
\qquad \qquad
\scalebox{0.6}{
\begin{tikzpicture}[baseline]
\tikzstyle{every node}=[font=\footnotesize]
\node[draw, rectangle] (sqnode1a) at (-4,1.25) {$N$};
\node[draw, rectangle] (sqnode1b) at (-4,-1.25) {$N$};
\node[draw, circle, fill=yellow] (yell1) at (-2,0) {};
\node[draw, circle, dashed] (circ1) at (0,1.25) {$N_{k_1}$};
\node[draw, circle, dashed] (circ2) at (0,-1.25) {$N_{k_2}$};
\node[draw, circle, fill=yellow] (yell2) at (2,0) {};
\node[draw, rectangle] (sqnode2a) at (4,1.25) {$N$};
\node[draw, rectangle] (sqnode2b) at (4,-1.25) {$N$};
\draw[draw=black,solid,<-]  (sqnode1a) to node[midway,above] {$p$}   (yell1) ; 
\draw[draw=black,solid,->]  (sqnode1b) to node[midway,below] {$q^{-1}$}   (yell1) ; 
\draw[draw=black,solid,<-]  (yell1) to node[midway,above] {$q$}   (circ1) ; 
\draw[draw=black,solid,->]  (yell1) to node[midway,below] {$p^{-1}$}   (circ2) ; 
\draw[draw=black,solid,<-]  (circ1) to node[midway,above] {$p'$}   (yell2) ; 
\draw[draw=black,solid,->]  (circ2) to node[midway,below] {$q'^{-1}$}   (yell2) ; 
\draw[draw=black,solid,<-]  (yell2) to node[midway,above] {$q'$}   (sqnode2a) ; 
\draw[draw=black,solid,->]  (yell2) to node[midway,below] {$p'^{-1}$}   (sqnode2b) ; 
\end{tikzpicture}}
\end{split}
\ee
where the bottom left and right quivers come from applying \eref{mainduality} to the left yellow node and to both yellow nodes of the top diagram, respectively.  There are monopole superpotential terms, the cubic superpotential terms coming from each triangular loop in the quiver that contains one blue line as an edge, and the quartic superpotential term for the top and bottom right quivers coming from the rectangular loop in the middle.  Such a superpotential imposes the following condition on the $U(1)$ fugacities:
\be \label{cond2}
pq^{-1} p'^{-1} q'=1~.
\ee

\subsection{The indices of \eref{finalPhiglue} and \eref{finalSglue} for $N=2$}
We focus only on the case of $N=2$ and $k_1, k_2 \geq 2$.
\subsubsection*{Theory  \eref{finalSglue}}
The index of this theory is
\be
\begin{split}
\CI^{N=2}_{\eref{finalSglue}}(x; p, q, p', q', y_1, \ldots, y_4) = 1+ C_1 x + C_2 x^2  +\ldots~,
\end{split}
\ee
where the coefficients $C_i$ are functions of  $p, q, p', q', y_1, \ldots, y_4$.   Here we report only the two coefficients $C_1$ and $C_2$ in full:
\be
\begin{split}
C_1&=\frac{p}{q} \begin{bmatrix} 1 & 0 \\ 1 & 0 \end{bmatrix} + \frac{q'}{p'} \begin{bmatrix} 0 & 1 \\ 0 & 1 \end{bmatrix} \overset{\eref{cond2}}= \frac{p}{q} \begin{bmatrix} 1 & 0 \\ 1 & 0 \end{bmatrix} + \frac{q}{p} \begin{bmatrix} 0 & 1 \\ 0 & 1 \end{bmatrix}  ~, \\
C_2&=\frac{p^2}{q^2} \begin{bmatrix} 2 & 0 \\ 2 & 0 \end{bmatrix} + \frac{q'^2}{p'^2} \begin{bmatrix} 0 & 2 \\ 0 & 2 \end{bmatrix} + \frac{p q'}{q p' }  \begin{bmatrix} 1 & 1 \\ 1 & 1 \end{bmatrix} + p q p' q'  \begin{bmatrix} 1 & 1 \\ 0 & 0 \end{bmatrix} + \frac{1}{p q p' q' } \begin{bmatrix} 0 & 0 \\ 1 & 1 \end{bmatrix}\\
& \quad  -   \begin{bmatrix} 2 & 0 \\ 0 & 0 \end{bmatrix}  -   \begin{bmatrix} 0 & 2 \\ 0 & 0 \end{bmatrix} -   \begin{bmatrix} 0 & 0 \\ 2 & 0 \end{bmatrix} -   \begin{bmatrix} 0 & 0\\ 0 & 2 \end{bmatrix} -4  + \underbrace{\frac{q p'}{p q'}}_{\overset{\eref{cond2}}{=}1}~.
\end{split}
\ee
We have used the notation $\begin{bmatrix} a & b \\ c & d \end{bmatrix}$ to denote the representation $[a; b; c; d]$ of the flavour symmetry $SU(2)^4$ associated with each corner of the quiver.  Upon setting $p, q, p', q', y_1, \ldots, y_4$ to $1$, the unrefined index for $(k_1,k_2) = (2,2)$ is
\be \label{unrefS}
1 + 8 x + 27 x^2 + 24 x^3 - 14 x^4 +\ldots~.
\ee

We now focus on gauge invariant combinations of chiral fields corresponding to various terms in the index.  For convenience, we consider the bottom right quiver in \eref{finalSglue} and label the chiral fields as follows:
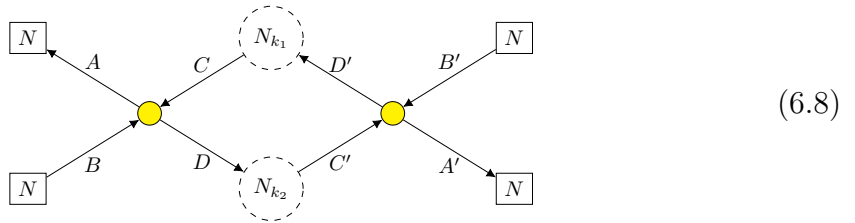
\be 
\scalebox{0.8}{
\begin{tikzpicture}[baseline]
\tikzstyle{every node}=[font=\footnotesize]
\node[draw, rectangle] (sqnode1a) at (-4,1.25) {$N$};
\node[draw, rectangle] (sqnode1b) at (-4,-1.25) {$N$};
\node[draw, circle, fill=yellow] (yell1) at (-2,0) {};
\node[draw, circle, dashed] (circ1) at (0,1.25) {$N_{k_1}$};
\node[draw, circle, dashed] (circ2) at (0,-1.25) {$N_{k_2}$};
\node[draw, circle, fill=yellow] (yell2) at (2,0) {};
\node[draw, rectangle] (sqnode2a) at (4,1.25) {$N$};
\node[draw, rectangle] (sqnode2b) at (4,-1.25) {$N$};
\draw[draw=black,solid,<-]  (sqnode1a) to node[midway,above] {$A$}   (yell1) ; 
\draw[draw=black,solid,->]  (sqnode1b) to node[midway,below] {$B$}   (yell1) ; 
\draw[draw=black,solid,<-]  (yell1) to node[midway,above] {$C$}   (circ1) ; 
\draw[draw=black,solid,->]  (yell1) to node[midway,below] {$D$}   (circ2) ; 
\draw[draw=black,solid,<-]  (circ1) to node[midway,above] {$D'$}   (yell2) ; 
\draw[draw=black,solid,->]  (circ2) to node[midway,below] {$C'$}   (yell2) ; 
\draw[draw=black,solid,<-]  (yell2) to node[midway,above] {$B'$}   (sqnode2a) ; 
\draw[draw=black,solid,->]  (yell2) to node[midway,below] {$A'$}   (sqnode2b) ; 
\end{tikzpicture}}
\ee
Explicitly, the superpotential of the above quiver is $W= V_+ +V_- +V'_+ +V'_- + C D C' D'$.  Let us use the indices $\begin{bmatrix} i,j &~~  m, n \\ i', j' &~~  m',n'\end{bmatrix}$, each of which takes values $1, \, 2$, for the flavour symmetry $SU(2)^4$ associated with each corner of the quiver.  We use $a,b=1,2$ and $a',b'=1,2$ to denote the $SU(2)_{k_1}$ and $SU(2)_{k_2}$ gauge indices respectively.

The terms in the coefficient $C_1$ corresponds to the following gauge invariant combinations:
\be
X^{i'}_{i} = A_i B^{i'}~, \qquad (X')^m_{m'} = A'_{m'} B'^{m}~.
\ee
Indeed, $X$ and $X'$ are the relevant operators.
The positive terms of the coefficient $C_2$ correspond to the following gauge invariant combinations:
\be
\begin{split}
&X^{i'}_{i} X^{j'}_{j}  ~, \quad  (X')^m_{m'}  (X')^n_{n'}~, \quad X^{i'}_{i} (X')^m_{m'} ~,\\
&Y^m_i := A_i C^a D'_a B'^m~, ~ Y'^{i'}_{m'} := B^{i'} D_a C'^a B'^{i'}
\end{split}
\ee
These are the marginal operators.  From the negative terms in the coefficient $C_2$, we see that the global symmetry of the theory is $SU(2)^4 \times U(1)^3$. Indeed, the $SU(2)^4$ symmetry is manifest as the four square nodes in the quiver, and the three copies of $U(1)$ correspond to the fugacities $p, \, q, \, p', \, q'$ subject to \eref{cond2}.

\subsubsection*{Theory  \eref{finalPhiglue}}
The index of this theory is
\be
\begin{split}
\CI^{N=2}_{\eref{finalPhiglue}}(x; p, q, p', q', y_1, \ldots, y_4) = 1+ c_1 x + c_2 x^2+\ldots~,
\end{split}
\ee
where the coefficients $c_i$ are functions of  $p, q, p', q', y_1, \ldots, y_4$.  We report only $c_1$ and $c_2$ in full:
\be \label{coeffindexPhiglue}
\begin{split}
c_1&=\frac{p}{q} \begin{bmatrix} 1 & 0 \\ 1 & 0 \end{bmatrix} + \frac{p'}{q'} \begin{bmatrix} 0 & 1 \\ 0 & 1 \end{bmatrix} \overset{\eref{cond1}}= \frac{p}{q} \begin{bmatrix} 1 & 0 \\ 1 & 0 \end{bmatrix} + \frac{p}{q} \begin{bmatrix} 0 & 1 \\ 0 & 1 \end{bmatrix}  ~, \\
c_2&=\frac{p^2}{q^2} \begin{bmatrix} 2 & 0 \\ 2 & 0 \end{bmatrix} + \frac{p'^2}{q'^2} \begin{bmatrix} 0 & 2 \\ 0 & 2 \end{bmatrix} + \frac{p p'}{q q' }  \begin{bmatrix} 1 & 1 \\ 1 & 1 \end{bmatrix} + p q p' q'  \begin{bmatrix} 1 & 1 \\ 0 & 0 \end{bmatrix} + \frac{1}{p q p' q' } \begin{bmatrix} 0 & 0 \\ 1 & 1 \end{bmatrix}\\
& \quad  -   \begin{bmatrix} 2 & 0 \\ 0 & 0 \end{bmatrix}  -   \begin{bmatrix} 0 & 2 \\ 0 & 0 \end{bmatrix} -   \begin{bmatrix} 0 & 0 \\ 2 & 0 \end{bmatrix} -   \begin{bmatrix} 0 & 0\\ 0 & 2 \end{bmatrix} + \frac{p^2}{q^2} - \frac{q^2}{p^2}  -4 + \underbrace{\frac{p q'}{q p'}}_{\overset{\eref{cond1}}{=}1}~.
\end{split}
\ee
Upon setting $p, q, p', q', y_1, \ldots, y_4$ to $1$, the unrefined index of this theory for $(k_1,k_2)=(2,2)$ is 
\be \label{unrefPhi}
1 + 8 x + 27 x^2 + 24 x^3 - 14 x^4+\ldots~.
\ee
From \eref{unrefS} and \eref{unrefPhi}, we see the unrefined indices of theory \eref{finalPhiglue} and theory \eref{finalSglue} are equal to each other.

Let us consider \eref{coeffindexPhiglue} in more detail.  Notice that the coefficient $c_2$ contains a negative term $- \frac{q^2}{p^2}$.  If we {\it assume} that theory \eref{finalPhiglue} flows to a superconformal fixed point, the negative terms in $c_2$ must correspond to a conserved current.  Let us proceed with this assumption. The $-\frac{q^2}{p^2}$ term should correspond to a $U(1)$ conserved current and should appear in the index as $1$ (since its character is $1$). Therefore our assumption on the conformality forces us to set $p=q$.  It follows from \eref{cond1} that $p'=q'$.  Therefore \eref{coeffindexPhiglue} can be rewritten as
\be
\begin{split}
c_1&= \begin{bmatrix} 1 & 0 \\ 1 & 0 \end{bmatrix} +  \begin{bmatrix} 0 & 1 \\ 0 & 1 \end{bmatrix} ~, \\
c_2&=\begin{bmatrix} 2 & 0 \\ 2 & 0 \end{bmatrix} +  \begin{bmatrix} 0 & 2 \\ 0 & 2 \end{bmatrix} +  \begin{bmatrix} 1 & 1 \\ 1 & 1 \end{bmatrix} + p^2 p'^2  \begin{bmatrix} 1 & 1 \\ 0 & 0 \end{bmatrix} + \frac{1}{p^2 p'^2 } \begin{bmatrix} 0 & 0 \\ 1 & 1 \end{bmatrix}\\
& \quad  -   \begin{bmatrix} 2 & 0 \\ 0 & 0 \end{bmatrix}  -   \begin{bmatrix} 0 & 2 \\ 0 & 0 \end{bmatrix} -   \begin{bmatrix} 0 & 0 \\ 2 & 0 \end{bmatrix} -   \begin{bmatrix} 0 & 0\\ 0 & 2 \end{bmatrix} -3~.
\end{split}
\ee
For the coefficient $c_3$, we report the result only for $y_1=y_2=y_3=y_4=1$:
\be
c_3 = 8+ 16 \left( p^2 p'^2 +\frac{1}{p^2 p'^2} \right) - 8 \left(\frac{p^2}{p'^2} +\frac{p'^2}{p^2}  \right)~.
\ee
It can be see from the negative terms in $c_2$ that the theory has a global symmetry $SU(2)^4 \times U(1)^3$.  Although $SU(2)^4$ is manifest in the quiver, not all three $U(1)$ symmetries is manifest.  Since we have two fugacities $p$ and $p'$ appearing in the index, only two $U(1)$ symmetries is manifest.  We conjecture that the other $U(1)$ global symmetry emerges at the superconformal fixed point in the infrared. 

In fact, it is important to emphasise that the indices of \eref{finalSglue} and \eref{finalPhiglue} are equal if we set $p=q$ and $p'=q'$:
\be \label{matchindices}
\CI^{N=2}_{\eref{finalPhiglue}}(x; p= q, p'= q', y_1, \ldots, y_4) = \CI^{N=2}_{\eref{finalSglue}}(x; p= q, p'= q', y_1, \ldots, y_4)~.
\ee
We have checked this relation up to order $x^6$ for various $(k_1,k_2)$.  We conjecture that theories \eref{finalPhiglue} and \eref{finalSglue} are dual to each other, in the sense that they flow to the same fixed point in the infrared.  For \eref{finalSglue}, the global symmetry $SU(2)^4 \times U(1)^3$ is manifest in the quiver description, and it is therefore possible to refine all of the corresponding fugacities in the index.  For \eref{finalPhiglue}, the global symmetry is also $SU(2)^4 \times U(1)^3$, but among all global fugacities, it is possible to refine only two $U(1)$ fugacities in the index, since the other $U(1)$ is emergent in the infrared.  This interpretation is consistent with the relation \eref{matchindices}.  An immediate consequence of this conjecture is that we have six dual descriptions, namely
\be
 \eref{finalPhiglue}  ~\overset{\text{for $N=2$}}{\longleftrightarrow} ~ \eref{finalSglue} ~.
\ee 

Let us end this subsection by briefly discussing the case of $N=3$.  We find that the indices of models \eref{finalSglue} and \eref{finalPhiglue} are not equal, and so the two theories are not dual.  In particular, for $N=3$ and $(k_1,k_2)=(2,2)$, their unrefined indices are
\be
\begin{split}
\eref{finalPhiglue}: &\qquad 1 + 18 x + 136 x^2 + 562 x^3+ \ldots \\
\eref{finalSglue}: &\qquad 1 + 18 x + 154 x^2 + 832 x^3+\ldots ~.
\end{split}
\ee

\section{Conclusion and perspectives} \label{sec:conclude}
We study 3d $\CN=2$ gauge theories associated with $S$-duality walls in the 4d $\CN=2$ $SU(N)$ gauge theory with $2N$ flavours.  Motivated by \cite{Kim:2017toz}, we propose a prescription in gluing theories associated with multiple duality walls as well as self-gluing for arbitrary number of walls.  The analog of the geometric view point of \cite{Kim:2017toz}, involving Riemann surfaces, is presented using the skeleton diagram.   Using supersymmetric indices, we find a number of dualities between different theories, some of them hold only for $N=2$ and many of them are true for all $N\geq 2$.  In particular, we find that for an even number of walls, if all external legs of the skeleton diagrams are closed, the theories associated with the same topology of the skeleton diagram (for given rank and CS levels of the gauge groups) are dual to each other, independent of the way we glue the basic building block \eref{basicblock}.

The gluing performed in this paper can also be viewed as a generalisation of the $S$-fold theory \cite{Terashima:2011qi, Gang:2015wya, Gang:2018wek, Gang:2018huc, Assel:2018vtq} associated with duality walls in the 4d $\CN=4$ super-Yang-Mills to a theory with lower amounts of supersymmetry, which is the 4d $\CN=2$ gauge theory in our case. 

This work has led to a number of open problems that deserve a further investigation in the future.  First of all, it would be interesting to understand the geometric origin, such as compactification of a higher dimensional theory, for our theories and, in particular, the skeleton diagrams.  Secondly, in certain theories presented in this paper, we assume that they flow to superconformal fixed points and deduce some properties from the indices, such as an emergent $U(1)$ symmetry.  It would be nice to better understand such an assumption and, if it is true, the property of such conformal fixed points.  Another important future work is to understand the holographic dual of the theories discussed in this paper along the line of \cite{Assel:2018vtq, Bobev:2019jbi}.  Finally, we would like to understand properties of the moduli space of vacua of the 3d $\mathcal{N}=2$ theories in this paper along the line of \cite{Garozzo:2018kra}, as well as to generalise our result to 4d $\CN=2$ gauge theory with orthogonal, symplectic and exceptional gauge groups in analogy to those studied in \cite{Garozzo:2019hbf}.

\acknowledgments
We thank Antonio Amariti, Sara Pasquetti, Valentin Reys, Alessandro Tomasiello and Masahito Yamazaki for valuable discussions. NM acknowledges the Pollica Summer Workshop, where this work was initiated. The workshop was supported in part by the Simons Foundation (Simons Collaboration on the Non-perturbative Bootstrap) and in part by the INFN.  MS is partially supported by the ERC-STG grant 637844-HBQFTNCER, by the University of Milano-Bicocca grant 2016-ATESP-0686, and by the INFN.  He also acknowledges the following workshop and conferences: String Math 2019; Elliptic integrable systems, special functions and quantum field theory; and Mathematics and Physics of Knots. He would like to express his gratitude to Ambrogio Sacchi and Lorenza Longhi for their continuous support and in particular for providing the laptop computer, which was fundamental for some of the computations in this paper.

\appendix
\section{3d supersymmetric index} \label{app:3dindex}
In this appendix, we briefly review some basic facts about the 3d supersymmetric index and and explain the conventions used in the paper, which follow those adopted in our previous paper \cite{Garozzo:2019ejm} and as in \cite {Aharony:2013dha, Aharony:2013kma}. The index is defined as a trace over states on $\mathbb{S}^2 \times \BR$ \cite{Bhattacharya:2008zy,Bhattacharya:2008bja, Kim:2009wb,Imamura:2011su, Kapustin:2011jm, Dimofte:2011py}:
\be
\cI(x, \bm{\mu})\,=\,\Tr \left[ (-1)^{2J_3} x^{\Delta+J_3}\prod_i \mu_i^{T_i} \right]\,,
\ee
where $\Delta$ is the energy in units of the $\mathbb{S}^2$ radius (for superconformal field theories, $\Delta$ is related to the conformal dimension), $J_3$ is the Cartan generator of the Lorentz $SO(3)$ isometry of $\mathbb{S}^2$, and $T_i$ are charges under non-$R$ global symmetries. The index only receives contributions from the states that satisfy:
\be
\Delta-R-J_3 =  0\,,
\ee
where $R$ is the $R$-charge.

The 3d supersymmetric index can also be computed as the supersymmetric partition function on $\mathbb{S}^2 \times  \mathbb{S}^1$ using localization, provided that the superconformal $R$-charge is chosen for all the matter fields
\be\label{indexpartitionfunction}
\cI(x; \{\bm{\mu},\bm{n}\})\,=\,\sum_{\bm{m}}\frac{1}{|\mathcal{W}_{\bm{m}}|}\oint\frac{d\bm{z}}{2\pi i \bm{z}} Z_{\text{cl}}\,Z_{\text{vec}}\,Z_{\text{mat}}\,,
\ee
where we denoted by $\bm{z}$ the fugacities parameterising the maximal torus of the gauge group, and by $\bm{m}$ the corresponding GNO magnetic fluxes on $S^2$. The integration contour is taken to be the unit circle $\mathbb{T}$ for each integration variable and the prefactor $|\mathcal{W}_{\bm{m}}|$ is the dimension of the Weyl group of the residual gauge symmetry in the monopole background labelled by the configuration of magnetic fluxes $\bm{m}$.  We also use $\{\bm{\mu},\bm{n}\}$ to denote possible fugacities and fluxes for the background vector multiplets associated with global symmetries, respectively. The different contributions to the integrand of \eqref{indexpartitionfunction} are:
\begin{enumerate}[$\bullet$]
\item the contribution from the classical action of CS and BF interactions
\be
Z_{\text{cl}}\,=\,\prod_{i=1}^{\text{rk}G}\omega^{m_i}z_i^{k\,m_i+\mathfrak{n}}\,,
\ee
where $\text{rk}G$ is the rank of the gauge group $G$ and we denoted with $k$ the CS level and with $\omega$ and $\mathfrak{n}$ the fugacity and the background flux for the global symmetry;

\item the contribution of the $\mathcal{N}=2$ vector multiplet
\be
Z_{\text{vec}}\,=\,\prod_{\alpha\in\frak{g}}x^{-\frac{|\alpha(\bm{m})|}{2}}(1-(-1)^{\alpha(\bm{m})}\bm{z}^{\alpha}x^{|\alpha(\bm{m})|})\,
\ee
where $\alpha$ are roots in the gauge algebra $\frak g$;

\item the contribution of an $\mathcal{N}=2$ chiral field transforming in some representation $\mathcal{R}$ and $\mathcal{R}_F$ of the gauge and the flavour symmetry respectively and with $R$-charge $r$
\begin{eqnarray}
&&Z_{mat}\,=\,\prod_{\rho \in \mathcal{R}}\,\prod_{\tilde \rho \in \mathcal{R}_F}\left(\bm{z}^{\rho}\,\bm{\mu}^{\tilde \rho}\,x^{r-1} \right)^{-\frac{|\rho(\bm{m})+\tilde{\rho}(\bm{n})|}{2}}\times\nonumber \\
&&\qquad\qquad\qquad\times\frac{((-1)^{\rho(\bm{m})+\tilde{\rho}(\bm{n})}\,\bm{z}^{-\rho}\,\bm{\mu}^{-\tilde \rho}\,x^{2-r+|\rho(\bm{m})+\tilde{\rho}(\bm{n})|};x^2)_\infty}{((-1)^{\rho(\bm{m})+\tilde{\rho}(\bm{n})}\,\bm{z}^{\rho}\,\bm{\mu}^{\tilde \rho}\,x^{r+|\rho(\bm{m})+\tilde{\rho}(\bm{n})|};x^2)_\infty}\,,
\end{eqnarray}
where $\rho$ and $\tilde \rho$ are the weights of $\mathcal{R}$ and $\mathcal{R}_F$ respectively.
\end{enumerate}

Let us now apply the index to 3d superconformal field theories.  In which case, the index keeps track of the short multiplets of the theory, up to recombination.  It proves useful to compute the index perturbatively by expanding the integrand in the fugacity $x$ and taking the gauge projection $\oint\frac{d\bm{z}}{2\pi i \bm{z}}$ at each order. Turning off the background fluxes for the global symmetries, we obtain a result which is a power series in $x$
\be
\cI(x,\{\vec \mu, \vec n =0 \})\,=\,\sum_{p=0}^\infty \chi_{p}(\bm{\mu})\,x^{p}\,
\ee
where $\chi_p(\bm{\mu})$ is the character of some representation of the global symmetry of the theory.  As demonstrated in \cite{Razamat:2016gzx} (see also \cite{Gadde:2009dj, Beem:2012yn}), one can study the contribution of superconformal multiplets to each order of $x$ in the power series.  Since the classification of the shortening conditions for 3d superconformal algebras is known \cite{Dolan:2008vc, Cordova:2016emh}, it is possible to obtain useful information about the superconformal theory in question using the power series of the index.  In this paper, we mainly focus on the coefficient of $x$ and $x^2$.  The coefficients of $x$ correspond to the $\CN=2$ relevant operators, contributing with only a positive sign.  The coefficient of $x^2$ receives a contribution from the $\CN=2$ marginal operators, contributing with a positive sign, and the conserved currents, contributing with a negative sign.


\bibliographystyle{ytphys}
\bibliography{ref}
\end{document}